\documentclass[12pt]{article}
\usepackage{axodraw,a4wide,epsfig}
  \usepackage{graphicx}
  \usepackage{epsfig}
  \usepackage{amssymb}
  \usepackage{subfigure}

= 0mm \oddsidemargin = -5mm

\catcode`@=11
\newcount\@tempcntc
\def\@citex[#1]#2{\if@filesw\immediate\write\@auxout{\string\citation{#2}}\fi
  \@tempcnta\z@\@tempcntb\m@ne\def\@citea{}\@cite{\@for\@citeb:=#2\do
    {\@ifundefined
       {b@\@citeb}{\@citeo\@tempcntb\m@ne\@citea\def\@citea{,}{\bf ?}\@warning
       {Citation `\@citeb' on page \thepage \space undefined}}%
    {\setbox\z@\hbox{\global\@tempcntc0\csname b@\@citeb\endcsname\relax}%
     \ifnum\@tempcntc=\z@ \@citeo\@tempcntb\m@ne
       \@citea\def\@citea{,}\hbox{\csname b@\@citeb\endcsname}%
     \else
      \advance\@tempcntb\@ne
      \ifnum\@tempcntb=\@tempcntc
      \else\advance\@tempcntb\m@ne\@citeo
      \@tempcnta\@tempcntc\@tempcntb\@tempcntc\fi\fi}}\@citeo}{#1}}
\def\@citeo{\ifnum\@tempcnta>\@tempcntb\else\@citea\def\@citea{,}%
  \ifnum\@tempcnta=\@tempcntb\the\@tempcnta\else
   {\advance\@tempcnta\@ne\ifnum\@tempcnta=\@tempcntb \else \def\@citea{--}\fi
    \advance\@tempcnta\m@ne\the\@tempcnta\@citea\the\@tempcntb}\fi\fi}
\catcode`@=12

\begin{document}

\newcommand{\be}{\begin{equation}}
\newcommand{\ee}{\end{equation}}
\newcommand{\bfm}[1]{\mbox{\boldmath$#1$}}
\newcommand{\bff}[1]{\mbox{\scriptsize\boldmath${#1}$}}
\newcommand{\al}{\alpha}
\newcommand{\bt}{\beta}
\newcommand{\lm}{\lambda}
\newcommand{\bea}{\begin{eqnarray}}
\newcommand{\eea}{\end{eqnarray}}
\newcommand{\gm}{\gamma}
\newcommand{\Gm}{\Gamma}
\newcommand{\dl}{\delta}
\newcommand{\Dl}{\Delta}
\newcommand{\ep}{\epsilon}
\newcommand{\vep}{\varepsilon}
\newcommand{\kp}{\kappa}
\newcommand{\Lm}{\Lambda}
\newcommand{\om}{\omega}
\newcommand{\pa}{\partial}
\newcommand{\nn}{\nonumber}
\newcommand{\dd}{\mbox{d}}
\newcommand{\grtsim}{\mbox{\raisebox{-3pt}{$\stackrel{>}{\sim}$}}}
\newcommand{\lessim}{\mbox{\raisebox{-3pt}{$\stackrel{<}{\sim}$}}}
\newcommand{\uk}{\underline{k}}
\newcommand{\gsim}{\;\rlap{\lower 3.5 pt \hbox{$\mathchar \sim$}} \raise 1pt \hbox {$>$}\;}
\newcommand{\lsim}{\;\rlap{\lower 3.5 pt \hbox{$\mathchar \sim$}} \raise 1pt \hbox {$<$}\;}
\newcommand{\Li}{\mbox{Li}}
\newcommand{\bc}{\begin{center}}
\newcommand{\ec}{\end{center}}

\newcommand{\xrr}{x_{rr}}
\newcommand{\xr}{x_r}
\newcommand{\yr}{y_r}
\newcommand{\zr}{z_r}
\newcommand{\xb}{x_b}
\newcommand{\yb}{y_b}
\newcommand{\zb}{z_b}
\newcommand{\hypF}{{}_2\mbox{F}_1}

\begin{titlepage}
\nopagebreak {\flushright{
        \begin{minipage}{5cm}
         ALBERTA-THY-24-07 \\
         IFIC/07-77 \\
         ZU-TH 31/07
        \end{minipage}        }

}
\renewcommand{\thefootnote}{\fnsymbol{footnote}}
\vskip 3cm
\begin{center}
\boldmath {\Large\bf Calculation of the \\[3pt]
Two-Loop Heavy-Flavor Contribution  \\[9pt]
to Bhabha Scattering}\unboldmath \vskip 1.0cm {\large
R.~Bonciani$\rm \, ^{a, \,}$\footnote{Email: {\tt
Roberto.Bonciani@ific.uv.es}}}, {\large A.~Ferroglia$\rm \, ^{b,
\,}$\footnote{Email: {\tt Andrea.Ferroglia@physik.unizh.ch}}}, and
{\large A.A.~Penin$\rm \, ^{c, \, d, \,}$\footnote{Email: {\tt
apenin@phys.ualberta.ca}}}
\vskip .7cm {\it $\rm ^a$ Departamento de F\'{\i}sica Te\`orica,
IFIC, CSIC -- Universidad de
Valencia, \\
E-46071 Valencia, Spain} \vskip .3cm {\it $\rm ^b$ Institut f{\"u}r
Theoretische Physik,
Universit{\"a}t Z\"urich, \\
CH-8057 Zurich, Switzerland} \vskip .3cm
{\it $\rm ^c$ Department of Physics, University Of Alberta, \\
Edmonton, AB T6G 2J1, Canada} \vskip .3cm
{\it $\rm ^d$ Institute for Nuclear Research of Russian Academy of Sciences, \\
117312 Moscow, Russia}
\end{center}
\vskip .5cm

\begin{abstract}
We describe in detail the calculation of the two-loop corrections to the
QED Bhabha scattering cross section  due to the vacuum polarization by
heavy fermions. Our approach eliminates one mass scale from the most
challenging part of the calculation and allows  us to obtain the corrections
in a closed analytical form. The  result is valid for arbitrary values of
the heavy fermion mass  and the Mandelstam invariants, as long as
$s,t,u \gg m_e^2$. \\[2mm]
\end{abstract}
\vfill
\end{titlepage}
 \newpage

\section{Introduction}

High energy electron-positron or {\it Bhabha} scattering \cite{Bha}
is among the most important and carefully studied processes in
particle physics. It provides a very efficient tool for luminosity
determination at electron-positron colliders and thus mediates the
process of extracting physical information from the raw experimental
data \cite{reviews}. The small-angle Bhabha scattering is
particularly effective as a luminosity monitor at high-energy
colliders.\footnote{At LEP, the luminometers were located at an
angle between $1.4^\circ$ and $2.9^\circ$.  At the future
International Linear Collider  (ILC), they will be placed  between
$0.7^\circ$ and $2.3^\circ$ \cite{Moenig}.} The large-angle Bhabha
scattering is used to measure the luminosity at colliders operating
at the  center-of-mass energy, $\sqrt{s}$, of a few GeV, such as
BABAR/PEP-II, BELLE/KEKB, BES/BEPC, KLOE/DA$\Phi$NE, and CMD,
SND/VEPP-2M \cite{Car}.\footnote{For example, in KLOE experiment the
luminosity measurement is based on the events with scattering
angles between $55^{\circ}$ and $125^{\circ}$ \cite{Denig}.} Moreover,
it will be also used to disentangle the luminosity spectrum at the ILC
\cite{Too,Heu}.
Bhabha scattering involves stable charged leptons both in the initial
and the final states and, therefore, it can be measured experimentally
with very high precision. At LEP, the experimental error in the
luminosity measurement has been reduced to 0.4 permille \cite{LEP}
and it is expected to be even smaller at the ILC: the goal of the
TESLA forward calorimeter collaboration  is to reach the experimental
accuracy of 0.1 permille in the first year of run \cite{TESLA}.
Finally, at the low-energy accelerators DA$\Phi$NE and VEPP-2M the
cross section of the large-angle scattering is measured with the
accuracy of about 1 permille \cite{Alo,Eid}.
In the phenomenologically most interesting cases of low energy
or small angle scattering, the Bhabha cross section is QED
dominated, with the electroweak and hadronic effects being strongly
suppressed. Therefore, it can be reliably computed in perturbative
QED, with the accuracy  limited only by uncalculated high order
corrections.

These properties make in such a way that Bhabha scattering is
an ideal ``standard candle'' for electron-positron colliders.

Realistic simulations of the Bhabha events, which take into account the detector
geometry and experimental cuts, are performed by means of sophisticated Monte
Carlo  generators, such as BHLUMI \cite{BHLUMI}, BABAYAGA \cite{Car,BABAYAGA},
BHAGENF \cite{BHAGENF}, BHWIDE \cite{BHWIDE}, MCGPJ \cite{MCGPJ}, and SABSPV
\cite{SABSPV}. To match the experimental needs, the two-loop QED corrections
must be included into the theoretical analysis and incorporated into the event
generators.
Since the theoretical accuracy  directly affects the luminosity
determination and may jeopardize the high-precision physics program
at electron-positron colliders, remarkable efforts were devoted
to the study of the radiative corrections. The one-loop corrections
have been known in the full electroweak theory  for a long time
\cite{Bhabha1loop}. The two-loop electroweak corrections are still
elusive. However, recently the calculation of  the  two-loop QED
corrections was completed. These corrections can be divided
into three main categories: (i) the pure photonic corrections, (ii)
the corrections involving the electron vacuum polarization, {\it
i.e.} with at least one closed electron loop, and (iii) the
corrections involving the vacuum polarization by heavy flavors
(leptons or quarks).
The first results for the  photonic corrections were obtained
in the limit of small scattering angles
\cite{russians,Fadin:1993ha,Arbuzov:1995vj,Jad1}, in  the massless
electron approximation \cite{Bern}, and for the terms enhanced by
powers of the large logarithm $\ln(s/m_e^2)$ \cite{Bas}. Finally,
the photonic corrections to the differential cross section were
obtained in \cite{Pen} in the leading order of the small electron
mass expansion through the {\it infrared matching} to the massless
approximation. This result is  sufficient for all phenomenological
applications at present and future colliders \cite{BF}  and was
recently confirmed within a slightly different framework
\cite{BecMel} (see also \cite{Mitov:2006xs}).\footnote{The full
dependence of the pure photonic corrections on the electron mass
$m_e$ is not known at the moment. The corresponding calculation
involves the two-loop box diagrams with three scales: $s$, $t$ and
$m_e$, which are not yet available, though many relevant results have
been already obtained
\cite{Smirnov:2001cm,BMR1,BMR2,DK,BFMRvBbox,CGR}.}
The corrections involving a closed electron loop were obtained
in \cite{BMR,BFMRvB1,BFMRvB2} by direct diagrammatic calculation,
retaining the full dependence on $m_e$. The calculation was
performed   by using the Laporta algorithm \cite{Lap} for the
reduction of the Feynman diagrams to the master integrals (MIs)
\cite{Tka}, which were  subsequently  evaluated
\cite{BMR1,BMR2,DK,BFMRvBbox,CGR} by means of  the differential
equation method \cite{DiffEq}. The result was obtained in
analytical form  in terms of harmonic polylogarithms
\cite{HPLs,NUMHPL,Daniel1,Weinzierl}.
The corrections due to the vacuum polarization by heavy fermion of
mass $m_f \gg m_e$ were first evaluated  in the limit
$m_f^2 \ll s,t,u $ by two different methods \cite{BecMel,Act}. The
calculation of \cite{BecMel} is based on the expansion in the
electron mass within the effective theory approach, while the
calculation  of \cite{Act} is  diagrammatic  and based on the
reduction to the MIs evaluated in the asymptotic regime
\cite{Czakon:2006pa}.
The condition $m_f^2 \ll s,t,u $, however,  does not hold for
$\tau$-lepton, $c$- and $b$-quarks in the practically interesting
energy range of about a few GeV, as well as for the top quark at
typical ILC energies $500~{\rm GeV} \lsim \sqrt{s} \lsim 1000~{\rm
GeV}$. In a recent letter \cite{hfbha} we announced the result for
the two-loop heavy-flavor contribution which is valid for any ratio
of the heavy fermion  mass to the  Mandelstam invariants, provided
$s,t,u \gg m_e^2$.  The calculation was performed in the small
electron mass limit. We used the general theory of infrared and
collinear divergencies to separate the singular dependence of the
corrections on the vanishing  electron mass.  The most difficult
part of the calculation was then carried out with a strictly massless
electron. This critically reduced the complexity of the problem and
made it solvable by the method of
\cite{BMR,BFMRvB1,BFMRvB2}.\footnote{When this work was in
preparation a numerical result for the  two-loop heavy-flavor
contribution was obtained by means of the dispersion relation
approach \cite{Actis:2007}.}

In this paper we provide a detailed account of our calculation \cite{hfbha}  and
we present the complete analytical result for the correction to the Bhabha
cross section. The paper  is organized  as follows. In Section~\ref{nota} we
introduce our notations and conventions. In Section~\ref{cancellation} we
discuss the infrared and collinear  structure of the corrections and outline the
strategy of the calculation. In Section~\ref{intro} we describe the calculation
of  two previously unknown four-point two-loop master integrals. In
Section~\ref{crosssection} we present the analytical result for the correction
to the cross section. The numerical analysis is given in Section~\ref{num}.
Section~\ref{conclusions} contains our conclusions. Some technical aspects of
the calculation including the auxiliary functions, generalized harmonic
polylogarithms (GHPLs), and the asymptotic behavior  of the corrections  are
discussed in the appendices.

\section{Notation and Conventions \label{nota}}

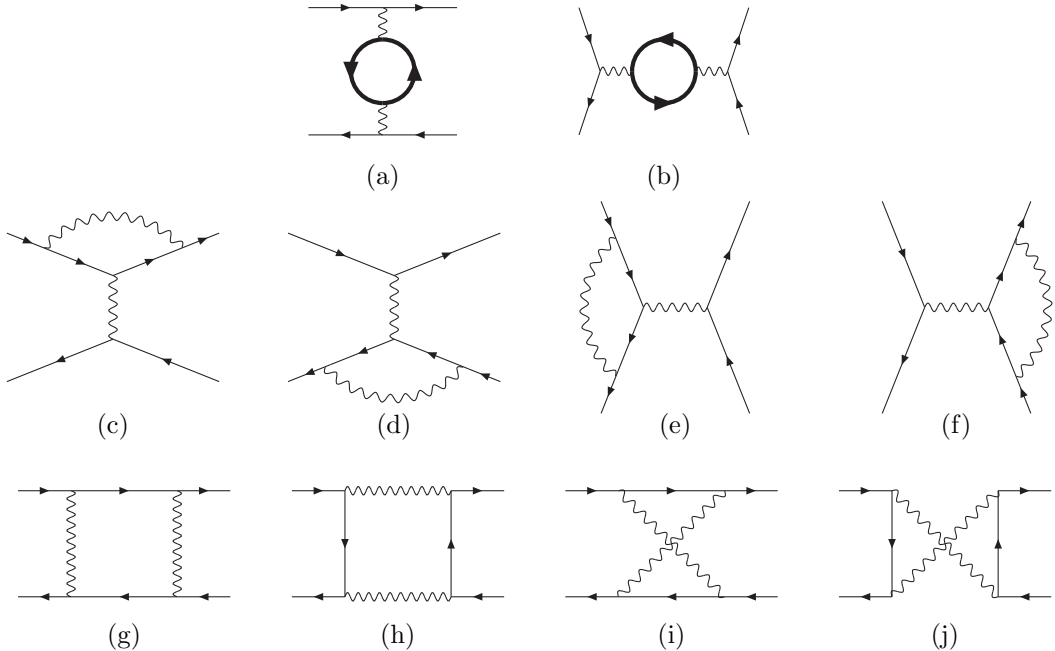
\begin{figure}
\vspace*{.5cm}
\[ \hspace*{-3mm}
\vcenter{
\hbox{
  \begin{picture}(0,0)(0,0)
\SetScale{0.8}
  \SetWidth{.5}
\ArrowLine(-35,30)(0,30)
\ArrowLine(0,30)(35,30)
\ArrowLine(0,-30)(-35,-30)
\ArrowLine(35,-30)(0,-30)
\Photon(0,30)(0,15){2}{3}
\Photon(0,-15)(0,-30){2}{3}
  \SetWidth{2}
\ArrowArc(0,0)(15,90,270)
\ArrowArc(0,0)(15,270,90)
\Text(0,-45)[cb]{{\footnotesize (a)}}
\end{picture}}
}
\hspace{3.6cm}
  \vcenter{
\hbox{
  \begin{picture}(0,0)(0,0)
\SetScale{0.8}
  \SetWidth{.5}
\ArrowLine(-30,0)(-40,-30)
\ArrowLine(-40,30)(-30,0)
\Photon(-30,0)(-15,0){2}{3}
\Photon(15,0)(30,0){2}{3}
\ArrowLine(30,0)(40,30)
\ArrowLine(40,-30)(30,0)
  \SetWidth{2}
\ArrowArc(0,0)(15,0,180)
\ArrowArc(0,0)(15,180,0)
\Text(0,-45)[cb]{{\footnotesize (b)}}
\end{picture}}
}
\]
\vspace*{1.7cm}
\[
  \vcenter{
\hbox{
  \begin{picture}(0,0)(0,0)
  \SetScale{0.8}
  \SetWidth{.5}
\ArrowLine(-50,35)(-30,27)
\ArrowLine(-30,27)(0,15)
\ArrowLine(0,15)(35,29)
\ArrowLine(35,29)(50,35)
\Photon(0,15)(0,-15){2}{5}
\PhotonArc(0,0)(43,41,139){2}{10}
\ArrowLine(0,-15)(-50,-35)
\ArrowLine(50,-35)(0,-15)
\Text(0,-50)[cb]{{\footnotesize (c)}}
\end{picture}}
}
\hspace{3.6cm}
  \vcenter{
\hbox{
\begin{picture}(0,0)(0,0)
  \SetScale{.8}
  \SetWidth{.5}
\ArrowLine(-50,35)(0,15)
\ArrowLine(0,15)(50,35)
\Photon(0,15)(0,-15){2}{5}
\PhotonArc(0,0)(43,-139,-41){2}{10}
\ArrowLine(-30,-27)(-50,-35)
\ArrowLine(0,-15)(-30,-27)
\ArrowLine(35,-29)(0,-15)
\ArrowLine(50,-35)(35,-29)
\Text(0,-50)[cb]{{\footnotesize (d)}}
\end{picture}}
}
\hspace{3.6cm}
  \vcenter{\hbox{
\begin{picture}(0,0)(0,0)
  \SetScale{.8}
  \SetWidth{.5}
\ArrowLine(35,-50)(15,0)
\ArrowLine(15,0)(35,50)
\PhotonArc(0,0)(43,131,229){2}{10}
\Photon(15,0)(-15,0){2}{5}
\ArrowLine(-15,0)(-27,-30)
\ArrowLine(-27,-30)(-35,-50)
\ArrowLine(-35,50)(-27,30)
\ArrowLine(-27,30)(-15,0)
\Text(0,-50)[cb]{{\footnotesize (e)}}
\end{picture}}}
\hspace{3.6cm}
  \vcenter{\hbox{
\begin{picture}(0,0)(0,0)
  \SetScale{0.8}
  \SetWidth{.5}
\ArrowLine(27,-30)(15,0)
\ArrowLine(35,-50)(27,-30)
\ArrowLine(27,30)(35,50)
\ArrowLine(15,0)(27,30)
\PhotonArc(0,0)(43,-49,49){2}{10}

\Photon(15,0)(-15,0){2}{5}
\ArrowLine(-15,0)(-35,-50)
\ArrowLine(-35,50)(-15,0)
\Text(0,-50)[cb]{{\footnotesize (f)}}
\end{picture}}}
\]
\vspace*{1.9cm}
\[
\vcenter{\hbox{
  \begin{picture}(0,0)(0,0)
\SetScale{.8}
  \SetWidth{.5}
\ArrowLine(-50,25)(-25,25)
\ArrowLine(-25,25)(25,25)
\ArrowLine(25,25)(50,25)
\Photon(-25,25)(-25,-25){2}{10}
\Photon(25,25)(25,-25){2}{10}
\ArrowLine(-25,-25)(-50,-25)
\ArrowLine(25,-25)(-25,-25)
\ArrowLine(50,-25)(25,-25)
\Text(0,-40)[cb]{{\footnotesize (g)}}
\end{picture}}}
\hspace{3.5cm}
\vcenter{\hbox{
  \begin{picture}(0,0)(0,0)
\SetScale{.8}
  \SetWidth{.5}
\ArrowLine(-50,25)(-25,25)
\ArrowLine(25,25)(50,25)
\Photon(-25,-25)(25,-25){2}{10}
\Photon(-25,25)(25,25){2}{10}
\ArrowLine(-25,-25)(-50,-25)
\ArrowLine(50,-25)(25,-25)
\ArrowLine(-25,25)(-25,-25)
\ArrowLine(25,-25)(25,25)
\Text(0,-40)[cb]{{\footnotesize (h)}}
\end{picture}}}
\hspace{3.5cm}
\vcenter{\hbox{
  \begin{picture}(0,0)(0,0)
\SetScale{.8}
  \SetWidth{.5}
\ArrowLine(-50,25)(-25,25)
\ArrowLine(-25,25)(25,25)
\ArrowLine(25,25)(50,25)
\Photon(-25,25)(25,-25){2}{10}
\Photon(25,25)(-25,-25){2}{10}
\ArrowLine(-25,-25)(-50,-25)
\ArrowLine(25,-25)(-25,-25)
\ArrowLine(50,-25)(25,-25)
\Text(0,-40)[cb]{{\footnotesize (i)}}
\end{picture}}}
%
\hspace{3.5cm}
\vcenter{\hbox{
  \begin{picture}(0,0)(0,0)
\SetScale{.8}
  \SetWidth{.5}
\ArrowLine(-50,25)(-25,25)
\ArrowLine(25,25)(50,25)
\Photon(-25,25)(25,-25){2}{10}
\Photon(-25,-25)(25,25){2}{10}
\ArrowLine(-25,-25)(-50,-25)
\ArrowLine(50,-25)(25,-25)
\ArrowLine(-25,25)(-25,-25)
\ArrowLine(25,-25)(25,25)
\Text(0,-40)[cb]{{\footnotesize (j)}}
\end{picture}}}
\]
\vspace*{.6cm}
\caption{\it One-loop diagrams.}
\label{fig1ltot}
\end{figure}

In this Section we briefly summarize our notation and conventions
which follow \cite{BFMRvB1,BFMRvB2}. We consider the photon mediated
process
\be e^{-}(p_1) + e^{+}(p_2) \rightarrow e^{-}(p_3) + e^{+}(p_4) \,,
\ee
where  $p_i^2=-m_e^2$. In the following, we will neglect the electron mass,
which is much smaller than any of the other mass scales involved in the
problem. $m_e$ will be set consistently to zero everywhere, except where it
acts as a regulator for the collinear singularities.
The kinematics of the process is  described in terms of the
Mandelstam invariants $s$, $t$ and $u$:
\bea
s &=& - P^2 \equiv -(p_1 + p_2)^2  = 4 E^2 \, ,
\label{sdef} \\
t &=& - Q^2 \equiv - (p_1 - p_3)^2  =
- 4 E^2  \sin^2{\frac{\theta}{2}} \, ,
\label{tdef} \\
u &=& - R^2 \equiv - (p_1 - p_4)^2  =
- 4 E^2  \cos^2{\frac{\theta}{2}}  \, ,
\label{udef}
\eea
where $s + t + u = 0$, $E$ is the particle energy in the center-of-mass frame,
and $\theta$ is the scattering angle. The Bhabha scattering differential cross
section is given by a series in the fine-structure constant $\alpha$:
\be \frac{d \sigma}{d \Omega} = \frac{d \sigma_{0}}{d \Omega} +
\left( \frac{\alpha}{\pi} \right) \frac{d \sigma_{1}}{d \Omega} +
\left( \frac{\alpha}{\pi} \right)^2 \frac{d \sigma_{2}}{d \Omega}
 + {\mathcal O}
\left(\alpha^3 \right) \, ,
\label{sigser}
\ee
where
\bea \frac{d \sigma_{0}}{d\Omega} & = & \frac{\alpha^2}{s} \Bigg[
\frac{1}{s^2} \left(s t + \frac{s^2}{2} + t^2 \right) +
\frac{1}{t^2} \left( s t + \frac{t^2}{2} + s^2 \right)   +
\frac{1}{s t} (s + t)^2  \Bigg] +{\cal O}(m_e^2/s) 
\label{TL}
\eea
is the Born cross section. In this paper, we consider only the radiative
corrections $d\sigma_{i}/d\Omega$ that involve the vacuum polarization by
heavy fermions. The first order correction to the cross section comes from
the interference of the diagrams (a) and (b) in Fig.~\ref{fig1ltot}
with the Born amplitude and it reads
\bea
\frac{d \sigma_{1}}{d \Omega} &=& \frac{\alpha^2}{s} Q_f^2 N_c
\Bigg[ \frac{1}{s^2}\left(s t + \frac{s^2}{2} + t^2\right)
2 \mbox{Re}\Pi^{(1l,0)}_0(s)
+ \frac{1}{t^2}\left(s t + \frac{t^2}{2} + s^2\right)
2 \Pi^{(1l,0)}_0(t) \nn\\
& & + \frac{1}{s t} (s + t)^2
 \left(\mbox{Re}\Pi^{(1l,0)}_0(s) +  \Pi^{(1l,0)}_0(t)\right)
\Bigg] +{\cal O}(m_e^2/s)\, .
\label{amp1lS}
\eea
The  expression of the one-loop vacuum polarization functions $\Pi_0^{(1l,0)}$
is given in Appendix~\ref{AF}; $Q_f$ is the electric charge of the heavy
fermion, $N_c$ (number of colors) is equal to $1$ for leptons and $3$ for
quarks, and we adopt the on-shell scheme for the renormalization of $\alpha$
and of the fermion mass. Note that Eq.~(\ref{amp1lS}) is infrared finite and has
a regular behavior in the small electron mass limit.

\section{Structure of the Second-Order Corrections and Calculation Method 
   \label{cancellation}}

The two-loop virtual corrections are infrared divergent. These {\it
soft} divergencies are canceled in the inclusive cross section when
one adds the photonic bremsstrahlung \cite{Kin}. We regulate all
the soft divergencies by dimensional regularization in $D$
space-time dimensions. The standard approach to deal with the
bremsstrahlung is to split it into a soft part, which accounts for
the emission of the photons with the energy below a given cut-off
$\omega \ll m_e$, and a hard part corresponding to the emission of
the photons with the energy larger than $\omega$.  The infrared
finite hard part is then computed numerically using Monte-Carlo
methods with physical cuts dictated by the experimental setup.  At
the same time, the soft part is computed analytically and combined
with the virtual corrections ensuring the cancellation of the soft
divergencies in Eq.~(\ref{sigser}). Thus we consider the
second-order contribution to the cross section given by the sum of
two terms:
\be
\frac{d \sigma_{2}}{d \Omega}=\frac{d \sigma_{2}^V}{d \Omega}+
\frac{d \sigma_{2}^S}{d \Omega} \, ,
\label{vps}
\ee
which correspond to the two-loop virtual correction,\footnote{We do
not consider the trivial correction given by two heavy-fermion
loop insertions which are are usually treated by Dyson resummation.}
and the one-loop correction to the single soft photon emission which
factorizes into the product of the first-order contributions
\cite{YFS}.

The calculation of the virtual corrections is a highly nontrivial
problem since in principle it involves the two-loop  box diagrams
depending on four mass scales: $s,~t,~m_f$, and $m_e$. These
diagrams are beyond the reach of the available calculational
techniques. However, in practice the electron mass is much smaller
than any of the other mass scales involved in the problem and the
calculation can be significantly simplified  by exploiting this
scale hierarchy. The small electron mass limit is not trivial
because a finite electron mass regulates the {\it collinear}
divergencies, giving a logarithmic dependence of the second
order correction on $m_e$. One way to perform a systematic expansion
in the small electron mass is to use the expansion by regions
approach \cite{Smi}. However, if we are interested only in the
leading order term in $m_e^2/s$, the problem can be solved in an
elegant way without the expansion of the individual diagrams
\cite{hfbha}. The main idea of the method is to use the general
theory of collinear divergencies to identify a set of simple
diagrams responsible for the singular behavior of the corrections on
the electron mass. Then, we compute the remaining corrections with
a strictly massless electron, effectively removing one mass scale from
the most complicated part of the calculation.

Let us describe the approach of \cite{hfbha} in more detail. The
second order contribution to the cross section
can be split in the sum of two terms
according to the asymptotic dependence on $m_e$:
\begin{equation}
\frac{d \sigma_{2}}{d \Omega} =
\left[\delta^{(2)}_{1}\ln\left({s\over m_e^2} \right) +
\delta^{(2)}_{0} + {\mathcal O}(m_e^2/s)\right]\frac{d \sigma_{0}}{d
\Omega}\, . \label{2lexp}
\end{equation}
The logarithmic term in Eq.~(\ref{2lexp}) is a remnant of the
collinear divergence  regulated by the electron mass.
The quantities $ \delta^{(2)}_{1}$ and $\delta^{(2)}_{0}$ in
Eq.~(\ref{2lexp}) depend on $s$, $t$, and $m_f$ only. The collinear
divergencies, and hence the singular dependence of the corrections on
$m_e$, have a peculiar structure which was extensively studied in the
context of QCD. In particular, in a physical (Coulomb or axial)
gauge the collinear divergencies factorize and can be absorbed in
the external field renormalization \cite{FreTay}.
Due to the factorization, the singular dependence on $m_e$ is the
same for the Bhabha amplitude and the square of the vector form
factor \cite{Pen}. This attributes the total logarithmic corrections
to the two-loop Bhabha scattering amplitude to the one-particle
reducible  diagrams (s)--(v) and the one-particle irreducible diagrams
(g)--(j) of Fig.~\ref{fig2ltot}. Moreover, due to the on-shell
renormalization condition, the vacuum polarization does not change
the photon propagator near the mass shell where the collinear
divergencies are located. As a result, the irreducible diagrams are
infrared finite even for $m_e=0$ and the singular terms are entirely
contained in the reducible diagrams.  In calculating the cross
section one has to take into account also the contributions coming
from the interference of the one-loop corrections to the amplitude
and the soft emission. Both contributions have a factorized form and
can be easily evaluated for $m_e \ne 0$. Thus, it is straightforward
to obtain the coefficient of the logarithmic term in
Eq.~(\ref{2lexp}), which reads
\be
\delta^{(2)}_{1} = \left[ 2\ln\left(\frac{4 \omega^2}{s}\right) + 3
\right] \frac{d \sigma_{1}}{d \sigma_{0}} \, .
\label{del21}
\ee
At the same time the sum of the remaining two-loop one-particle
irreducible  diagrams  has a regular behavior in the small electron
mass limit and can be computed with $m_e=0$. The two-loop vacuum
polarization given by the diagrams (a)--(f) in Fig.~\ref{fig2ltot}
does not develop collinear singularities, because the
corresponding photon is far off-shell. Hence, the sum of the two-loop
box diagrams (k)--(r) in Fig.~\ref{fig2ltot} is free of
collinear divergencies as well.  Let us emphasize that this property
in general holds only for the sum of the diagrams. The individual
diagrams computed in a covariant gauge do exhibit the collinear
divergencies for $m_e=0$. This, however, does not pose any additional
problem since we work in dimensional regularization. In this case,
the collinear divergencies show up as extra poles in $(D-4)$,
which are not related to the soft emission and disappear in the sum
of the one-particle irreducible diagrams.  The cancellation of the
collinear singularities of the box diagrams in the Feynman gauge is
schematically shown in Figs.~\ref{cancellationatoneloop} and
\ref{cancellationattwoloop} for the one- and two-loop cases, respectively.
Let us  demonstrate this cancellation explicitly in the case of the one-loop
graphs. Each one-loop box diagram for $m_e=0$ exhibits a double
pole in $(D-4)$, arising from the overlapping of soft and collinear
divergencies. In particular, for the
diagram (g) in Fig.~\ref{fig1ltot} one finds
\be
\frac{d \sigma_1^{V}}{d \Omega} \Bigg|_{(\mbox{g})} =
\frac{\alpha^2}{s} \frac{1}{(D-4)^2} \left[ \frac{m_f^2}{s}
B_1^{(1l,-2)}(s,t) + \frac{m_f^2}{t} B_2^{(1l,-2)}(s,t) \right] +
{\mathcal O}(1/(D-4)) \, ,
\ee
while for the diagram (i) one obtains
\be
\frac{d \sigma_1^{V}}{d \Omega} \Bigg|_{(\mbox{i})} =
\frac{\alpha^2}{s} \frac{1}{(D-4)^2} \left[ \frac{m_f^2}{s}
B_3^{(1l,-2)}(u,t) - \frac{m_f^2}{t} B_2^{(1l,-2)}(u,t) \right] +
{\mathcal O}(1/(D-4)) \, .
\ee
The explicit expressions of the auxiliary functions $B_i^{(1l,-2)}$
($i=1,2,3$) for $x > 0$  are collected in Appendix~\ref{AF}. It is
easy to check that
\bea
 B_1^{(1l,-2)}(s,t) + B_3^{(1l,-2)}(u,t)  &=& 0 \, , \nn \\
 B_2^{(1l,-2)}(s,t) - B_2^{(1l,-2)}(u,t)  &=& 0 \, ,
\eea
so that the double pole disappears in the sum of the diagrams. The
residual single pole in $(D-4)$ is of soft nature and it is
canceled after adding the soft-photon emission. The cancellation of
the collinear poles of the two-loop box diagrams is completely
analogous to the one-loop case.

Since the sum of the box diagrams has a smooth limit $m_e\to 0 $,
the result does not depend on whether this limit or the limit 
$\epsilon \to 0$ is taken first. In other words,
the absence of collinear divergencies makes in such a way that
the expression of the sum of the box diagrams cannot depend on which
collinear regulator (electron mass or dimensional regularization) is 
employed in the calculation.
All the ``true'' two-loop diagrams contribute only to the
non-logarithmic term in Eq.~(\ref{2lexp}) and, thus, can be evaluated for
$m_e =0$.  The two-loop problem with massless electron falls in the
same complexity class as the one considered in
\cite{BMR,BFMRvB1,BFMRvB2} and can be solved by similar approach. In
the reduction of the two-loop box diagrams, however, two completely
new MIs appear. The calculation of these MIs is described in the
next section.

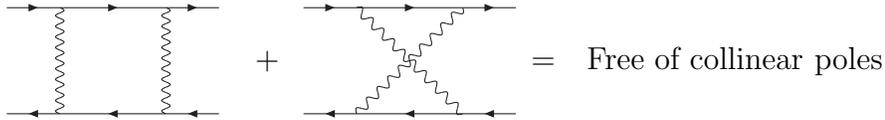
\begin{figure}
\[
\hspace*{0.8cm}
\vcenter{
\hbox{
  \begin{picture}(0,0)(0,0)
\SetScale{.8}
  \SetWidth{.5}
\ArrowLine(-50,25)(-25,25)
\ArrowLine(-25,25)(25,25)
\ArrowLine(25,25)(50,25)
\Photon(-25,25)(-25,-25){2}{10}
\Photon(25,25)(25,-25){2}{10}
\ArrowLine(-25,-25)(-50,-25)
\ArrowLine(25,-25)(-25,-25)
\ArrowLine(50,-25)(25,-25)
\end{picture}}}
\hspace*{1.8cm}
+
\hspace*{1.5cm}
\vcenter{
\hbox{
  \begin{picture}(0,0)(0,0)
\SetScale{.8}
  \SetWidth{.5}
\ArrowLine(-50,25)(-25,25)
\ArrowLine(-25,25)(25,25)
\ArrowLine(25,25)(50,25)
\Photon(-25,25)(25,-25){2}{10}
\Photon(25,25)(-25,-25){2}{10}
\ArrowLine(-25,-25)(-50,-25)
\ArrowLine(25,-25)(-25,-25)
\ArrowLine(50,-25)(25,-25)
\end{picture}}
}
\hspace*{1.5cm}
 = \hspace*{0.3cm} \mbox{Free of collinear poles}
\]
\vspace*{3mm} \caption{\it Cancellation of the collinear poles among
one-loop box diagrams calculated by setting $m_e = 0$ from the
start.} \label{cancellationatoneloop}
\end{figure}
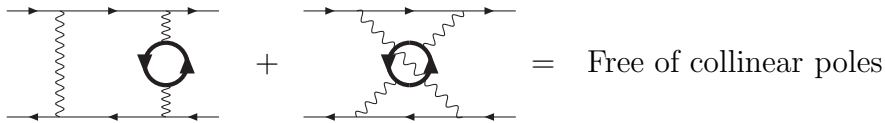
\begin{figure}
\[
\hspace*{0.8cm}
\vcenter{
\hbox{
  \begin{picture}(0,0)(0,0)
\SetScale{.8}
  \SetWidth{.5}
\ArrowLine(-50,25)(-25,25)
\ArrowLine(-25,25)(25,25)
\ArrowLine(25,25)(50,25)
\Photon(-25,25)(-25,-25){2}{10}
\Photon(25,25)(25,10){2}{4}
\Photon(25,-25)(25,-10){2}{4}
\ArrowLine(-25,-25)(-50,-25)
\ArrowLine(25,-25)(-25,-25)
\ArrowLine(50,-25)(25,-25)
  \SetWidth{2}
\ArrowArc(25,0)(10,-90,-270)
\ArrowArc(25,0)(10,-270,-90)
\end{picture}}}
\hspace*{1.8cm}
+
\hspace*{1.5cm}
\vcenter{
\hbox{
  \begin{picture}(0,0)(0,0)
\SetScale{.8}
  \SetWidth{.5}
\ArrowLine(-50,25)(-25,25)
\ArrowLine(-25,25)(25,25)
\ArrowLine(25,25)(50,25)
\Photon(-25,25)(25,-25){2}{10}
\Photon(25,25)(7,7){2}{4}
\Photon(-25,-25)(-7,-7){2}{4}
\ArrowLine(-25,-25)(-50,-25)
\ArrowLine(25,-25)(-25,-25)
\ArrowLine(50,-25)(25,-25)
  \SetWidth{2}
\ArrowArc(0,0)(10,-90,-270)
\ArrowArc(0,0)(10,-270,-90)
\end{picture}}
}
\hspace*{1.5cm}
 = \hspace*{0.3cm} \mbox{Free of collinear poles}
\]
\vspace*{3mm} \caption{\it Cancellation of the collinear poles among
two-loop box diagrams calculated by setting $m_e = 0$ from the
start.} \label{cancellationattwoloop}
\end{figure}

\section{The Master Integrals \label{intro}}

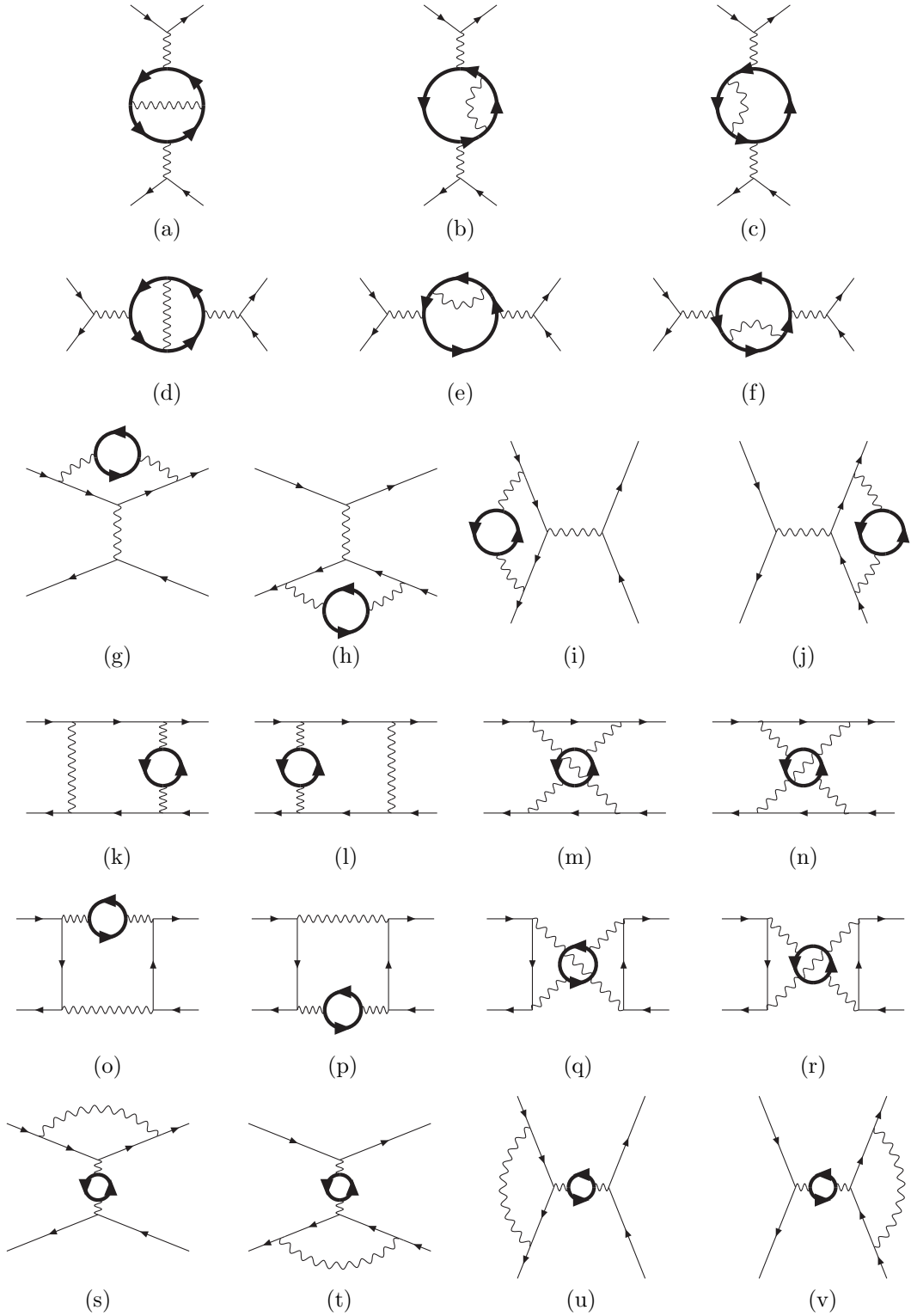
\begin{figure}
\vspace*{.3cm}
\[\vcenter{
\hbox{
  \begin{picture}(0,0)(0,0)
\SetScale{.8}
  \SetWidth{.5}
\ArrowLine(20,-55)(0,-40)
\ArrowLine(0,-40)(-20,-55)
\ArrowLine(0,40)(20,55)
\ArrowLine(-20,55)(0,40)
\Photon(0,-40)(0,-20){2}{4}
\Photon(0,20)(0,40){2}{4}
\Photon(20,0)(-20,0){2}{8}
\Text(0,-60)[cb]{{\footnotesize (a)}}
  \SetWidth{2}
\ArrowArc(0,0)(20,-90,0)
\ArrowArc(0,0)(20,0,90)
\ArrowArc(0,0)(20,90,180)
\ArrowArc(0,0)(20,180,270)
\end{picture}}
}
\hspace{4.4cm}
  \vcenter{
\hbox{
  \begin{picture}(0,0)(0,0)
\SetScale{.8}
  \SetWidth{.5}
\ArrowLine(20,-55)(0,-40)
\ArrowLine(0,-40)(-20,-55)
\Photon(0,-40)(0,-20){2}{4}
\Photon(0,20)(0,40){2}{4}
\ArrowLine(0,40)(20,55)
\ArrowLine(-20,55)(0,40)
\PhotonArc(20,0)(15,-248,-112){2}{5}
\Text(0,-60)[cb]{{\footnotesize (b)}}
  \SetWidth{2}
\ArrowArc(0,0)(20,90,270)
\ArrowArc(0,0)(20,-60,60)
\ArrowArc(0,0)(20,60,90)
\ArrowArc(0,0)(20,-90,-60)
\end{picture}}
}
\hspace{4.4cm}
  \vcenter{
\hbox{
  \begin{picture}(0,0)(0,0)
\SetScale{.8}
  \SetWidth{.5}
\ArrowLine(20,-55)(0,-40)
\ArrowLine(0,-40)(-20,-55)
\Photon(0,-40)(0,-20){2}{4}
\Photon(0,20)(0,40){2}{4}
\ArrowLine(0,40)(20,55)
\ArrowLine(-20,55)(0,40)
\PhotonArc(-20,0)(15,-68,68){2}{5}
\Text(0,-60)[cb]{{\footnotesize (c)}}
  \SetWidth{2}
\ArrowArc(0,0)(20,-90,90)
\ArrowArc(0,0)(20,90,120)
\ArrowArc(0,0)(20,120,240)
\ArrowArc(0,0)(20,240,270)
\end{picture}}
}\]
\vspace*{1.8cm}
\[\vcenter{
\hbox{
  \begin{picture}(0,0)(0,0)
\SetScale{.8}
  \SetWidth{.5}
\ArrowLine(-55,20)(-40,0)
\ArrowLine(-40,0)(-55,-20)
\ArrowLine(40,0)(55,20)
\ArrowLine(55,-20)(40,0)
\Photon(-40,0)(-20,0){2}{4}
\Photon(20,0)(40,0){2}{4}
\Photon(0,20)(0,-20){2}{8}
\Text(0,-40)[cb]{{\footnotesize (d)}}
  \SetWidth{2}
\ArrowArc(0,0)(20,0,90)
\ArrowArc(0,0)(20,90,180)
\ArrowArc(0,0)(20,180,270)
\ArrowArc(0,0)(20,270,360)
\end{picture}}
}
\hspace{4.4cm}
  \vcenter{
\hbox{
  \begin{picture}(0,0)(0,0)
\SetScale{.8}
  \SetWidth{.5}
\ArrowLine(-55,20)(-40,0)
\ArrowLine(-40,0)(-55,-20)
\ArrowLine(40,0)(55,20)
\ArrowLine(55,-20)(40,0)
\Photon(-40,0)(-20,0){2}{4}
\Photon(20,0)(40,0){2}{4}
\PhotonArc(0,20)(15,-158,-22){2}{5}
\Text(0,-40)[cb]{{\footnotesize (e)}}
  \SetWidth{2}
\ArrowArc(0,0)(20,0,30)
\ArrowArc(0,0)(20,30,150)
\ArrowArc(0,0)(20,150,180)
\ArrowArc(0,0)(20,180,360)
\end{picture}}
}
\hspace{4.4cm}
  \vcenter{
\hbox{
  \begin{picture}(0,0)(0,0)
\SetScale{.8}
  \SetWidth{.5}
\ArrowLine(-55,20)(-40,0)
\ArrowLine(-40,0)(-55,-20)
\Photon(-40,0)(-20,0){2}{4}
\Photon(20,0)(40,0){2}{4}
\ArrowLine(40,0)(55,20)
\ArrowLine(55,-20)(40,0)
\PhotonArc(0,-20)(15,22,158){2}{5}
\Text(0,-40)[cb]{{\footnotesize (f)}}
  \SetWidth{2}
\ArrowArc(0,0)(20,0,180)
\ArrowArc(0,0)(20,180,210)
\ArrowArc(0,0)(20,210,330)
\ArrowArc(0,0)(20,330,360)
\end{picture}}
}\]
\vspace*{1.7cm}
%

\[
  \vcenter{
\hbox{
  \begin{picture}(0,0)(0,0)
  \SetScale{0.8}
  \SetWidth{.5}
\ArrowLine(-50,35)(-30,27)
\ArrowLine(-30,27)(0,15)
\ArrowLine(0,15)(35,29)
\ArrowLine(35,29)(50,35)
\Photon(0,15)(0,-15){2}{5}
\PhotonArc(0,0)(43,41,74){2}{4}
\PhotonArc(0,0)(43,106,139){2}{4}
\ArrowLine(0,-15)(-50,-35)
\ArrowLine(50,-35)(0,-15)
\Text(0,-60)[cb]{{\footnotesize (g)}}
  \SetWidth{2}
\ArrowArc(0,43)(12,-10,190)
\ArrowArc(0,43)(12,190,350)
\end{picture}}
}
\hspace{3.4cm}
  \vcenter{
\hbox{
\begin{picture}(0,0)(0,0)
  \SetScale{.8}
  \SetWidth{.5}
\ArrowLine(-50,35)(0,15)
\ArrowLine(0,15)(50,35)
\Photon(0,15)(0,-15){2}{5}
\PhotonArc(0,0)(43,-74,-41){2}{4}
\PhotonArc(0,0)(43,-139,-106){2}{4}
\ArrowLine(-30,-27)(-50,-35)
\ArrowLine(0,-15)(-30,-27)
\ArrowLine(35,-29)(0,-15)
\ArrowLine(50,-35)(35,-29)
\Text(0,-60)[cb]{{\footnotesize (h)}}
  \SetWidth{2}
\ArrowArc(0,-43)(12,-10,190)
\ArrowArc(0,-43)(12,190,350)
\end{picture}}
}
\hspace{3.4cm}
  \vcenter{\hbox{
\begin{picture}(0,0)(0,0)
  \SetScale{.8}
  \SetWidth{.5}
\ArrowLine(35,-50)(15,0)
\ArrowLine(15,0)(35,50)
\PhotonArc(0,0)(43,131,164){2}{4}
\PhotonArc(0,0)(43,196,229){2}{4}
\Photon(15,0)(-15,0){2}{5}
\ArrowLine(-15,0)(-27,-30)
\ArrowLine(-27,-30)(-35,-50)
\ArrowLine(-35,50)(-27,30)
\ArrowLine(-27,30)(-15,0)
\Text(0,-60)[cb]{{\footnotesize (i)}}
  \SetWidth{2}
\ArrowArc(-43,0)(12,70,280)
\ArrowArc(-43,0)(12,280,430)
\end{picture}}}
\hspace{3.4cm}
  \vcenter{\hbox{
\begin{picture}(0,0)(0,0)
  \SetScale{0.8}
  \SetWidth{.5}
\ArrowLine(27,-30)(15,0)
\ArrowLine(35,-50)(27,-30)
\ArrowLine(27,30)(35,50)
\ArrowLine(15,0)(27,30)
\PhotonArc(0,0)(43,16,49){2}{4}
\PhotonArc(0,0)(43,-49,-16){2}{4}
\Photon(15,0)(-15,0){2}{5}
\ArrowLine(-15,0)(-35,-50)
\ArrowLine(-35,50)(-15,0)
\Text(0,-60)[cb]{{\footnotesize (j)}}
  \SetWidth{2}
\ArrowArc(43,0)(12,70,280)
\ArrowArc(43,0)(12,280,430)
\end{picture}}}
\]
\vspace*{2.2cm}
\[\vcenter{\hbox{
  \begin{picture}(0,0)(0,0)
\SetScale{.8}
  \SetWidth{.5}
\ArrowLine(-50,25)(-25,25)
\ArrowLine(-25,25)(25,25)
\ArrowLine(25,25)(50,25)
\Photon(-25,25)(-25,-25){2}{10}
\Photon(25,25)(25,10){2}{4}
\Photon(25,-25)(25,-10){2}{4}
\ArrowLine(-25,-25)(-50,-25)
\ArrowLine(25,-25)(-25,-25)
\ArrowLine(50,-25)(25,-25)
\Text(0,-45)[cb]{{\footnotesize (k)}}
  \SetWidth{2}
\ArrowArc(25,0)(10,-90,-270)
\ArrowArc(25,0)(10,-270,-90)
\end{picture}}}
\hspace{3.4cm}
\vcenter{\hbox{
  \begin{picture}(0,0)(0,0)
\SetScale{.8}
  \SetWidth{.5}
\ArrowLine(-50,25)(-25,25)
\ArrowLine(-25,25)(25,25)
\ArrowLine(25,25)(50,25)
\Photon(25,25)(25,-25){2}{10}
\Photon(-25,25)(-25,10){2}{4}
\Photon(-25,-25)(-25,-10){2}{4}
\ArrowLine(-25,-25)(-50,-25)
\ArrowLine(25,-25)(-25,-25)
\ArrowLine(50,-25)(25,-25)
\Text(0,-45)[cb]{{\footnotesize (l)}}
  \SetWidth{2}
\ArrowArc(-25,0)(10,-90,-270)
\ArrowArc(-25,0)(10,-270,-90)
\end{picture}}}
\hspace{3.4cm}
\vcenter{\hbox{
  \begin{picture}(0,0)(0,0)
\SetScale{.8}
  \SetWidth{.5}
\ArrowLine(-50,25)(-25,25)
\ArrowLine(-25,25)(25,25)
\ArrowLine(25,25)(50,25)
\Photon(-25,25)(25,-25){2}{10}
\Photon(25,25)(7,7){2}{4}
\Photon(-25,-25)(-7,-7){2}{4}
\ArrowLine(-25,-25)(-50,-25)
\ArrowLine(25,-25)(-25,-25)
\ArrowLine(50,-25)(25,-25)
\Text(0,-45)[cb]{{\footnotesize (m)}}
  \SetWidth{2}
\ArrowArc(0,0)(10,-90,-270)
\ArrowArc(0,0)(10,-270,-90)
\end{picture}}}
\hspace{3.4cm}
\vcenter{\hbox{
  \begin{picture}(0,0)(0,0)
\SetScale{.8}
  \SetWidth{.5}
\ArrowLine(-50,25)(-25,25)
\ArrowLine(-25,25)(25,25)
\ArrowLine(25,25)(50,25)
\Photon(25,25)(-25,-25){2}{10}
\Photon(-25,25)(-7,7){2}{4}
\Photon(25,-25)(7,-7){2}{4}
\Text(0,-45)[cb]{{\footnotesize (n)}}
\ArrowLine(-25,-25)(-50,-25)
\ArrowLine(25,-25)(-25,-25)
\ArrowLine(50,-25)(25,-25)
  \SetWidth{2}
\ArrowArc(0,0)(10,-90,-270)
\ArrowArc(0,0)(10,-270,-90)
\end{picture}}}
\]
\vspace*{1.8cm}
\[\vcenter{\hbox{
  \begin{picture}(0,0)(0,0)
\SetScale{.8}
  \SetWidth{.5}
\ArrowLine(-50,25)(-25,25)
\ArrowLine(25,25)(50,25)
\Photon(-25,-25)(25,-25){2}{10}
\Photon(-25,25)(-10,25){2}{4}
\Photon(25,25)(10,25){2}{4}
\ArrowLine(-25,-25)(-50,-25)
\ArrowLine(50,-25)(25,-25)
\ArrowLine(-25,25)(-25,-25)
\ArrowLine(25,-25)(25,25)
\Text(0,-50)[cb]{{\footnotesize (o)}}
  \SetWidth{2}
\ArrowArc(0,25)(10,-180,0)
\ArrowArc(0,25)(10,0,180)
\end{picture}}}
\hspace{3.5cm}
\vcenter{\hbox{
  \begin{picture}(0,0)(0,0)
\SetScale{.8}
  \SetWidth{.5}
\ArrowLine(-50,25)(-25,25)
\ArrowLine(25,25)(50,25)
\Photon(-25,25)(25,25){2}{10}
\Photon(-25,-25)(-10,-25){2}{4}
\Photon(25,-25)(10,-25){2}{4}
\ArrowLine(-25,-25)(-50,-25)
\ArrowLine(50,-25)(25,-25)
\ArrowLine(-25,25)(-25,-25)
\ArrowLine(25,-25)(25,25)
\Text(0,-50)[cb]{{\footnotesize (p)}}
  \SetWidth{2}
\ArrowArc(0,-25)(10,-180,0)
\ArrowArc(0,-25)(10,0,180)
\end{picture}}}
\hspace{3.5cm}
\vcenter{\hbox{
  \begin{picture}(0,0)(0,0)
\SetScale{.8}
  \SetWidth{.5}
\ArrowLine(-50,25)(-25,25)
\ArrowLine(25,25)(50,25)
\Photon(-25,25)(25,-25){2}{10}
\Photon(-25,-25)(-7,-7){2}{4}
\Photon(25,25)(7,7){2}{4}
\ArrowLine(-25,-25)(-50,-25)
\ArrowLine(50,-25)(25,-25)
\ArrowLine(-25,25)(-25,-25)
\ArrowLine(25,-25)(25,25)
\Text(0,-50)[cb]{{\footnotesize (q)}}
  \SetWidth{2}
\ArrowArc(0,0)(10,-180,0)
\ArrowArc(0,0)(10,0,180)
\end{picture}}}
\hspace{3.5cm}
\vcenter{\hbox{
  \begin{picture}(0,0)(0,0)
\SetScale{.8}
  \SetWidth{.5}
\ArrowLine(-50,25)(-25,25)
\ArrowLine(25,25)(50,25)
\Photon(25,25)(-25,-25){2}{10}
\Photon(-25,25)(-7,7){2}{4}
\Photon(25,-25)(7,-7){2}{4}
\ArrowArc(0,0)(10,-90,-270)
\ArrowArc(0,0)(10,-270,-90)
\ArrowLine(-25,-25)(-50,-25)
\ArrowLine(50,-25)(25,-25)
\ArrowLine(-25,25)(-25,-25)
\ArrowLine(25,-25)(25,25)
\Text(0,-50)[cb]{{\footnotesize (r)}}
  \SetWidth{2}
\ArrowArc(0,0)(10,-90,-270)
\ArrowArc(0,0)(10,-270,-90)
\end{picture}}}
\]
%
%
\vspace*{2.2cm}
\[
  \vcenter{
\hbox{
  \begin{picture}(0,0)(0,0)
  \SetScale{0.8}
  \SetWidth{.5}
\ArrowLine(-50,35)(-30,27)
\ArrowLine(-30,27)(0,15)
\ArrowLine(0,15)(35,29)
\ArrowLine(35,29)(50,35)
\Photon(0,15)(0,7){2}{2}
\Photon(0,-15)(0,-7){2}{2}
\PhotonArc(0,0)(43,41,139){2}{10}
\ArrowLine(0,-15)(-50,-35)
\ArrowLine(50,-35)(0,-15)
\Text(0,-55)[cb]{{\footnotesize (s)}}
  \SetWidth{2}
\ArrowArc(0,0)(7,-90,90)
\ArrowArc(0,0)(7,90,270)
\end{picture}}
}
\hspace{3.6cm}
  \vcenter{
\hbox{
\begin{picture}(0,0)(0,0)
  \SetScale{.8}
  \SetWidth{.5}
\ArrowLine(-50,35)(0,15)
\ArrowLine(0,15)(50,35)
\PhotonArc(0,0)(43,-139,-41){2}{10}
\Photon(0,15)(0,7){2}{2}
\Photon(0,-15)(0,-7){2}{2}
\ArrowLine(-30,-27)(-50,-35)
\ArrowLine(0,-15)(-30,-27)
\ArrowLine(35,-29)(0,-15)
\ArrowLine(50,-35)(35,-29)
\Text(0,-55)[cb]{{\footnotesize (t)}}
  \SetWidth{2}
\ArrowArc(0,0)(7,-90,90)
\ArrowArc(0,0)(7,90,270)
\end{picture}}
}
\hspace{3.6cm}
  \vcenter{\hbox{
\begin{picture}(0,0)(0,0)
  \SetScale{.8}
  \SetWidth{.5}
\ArrowLine(35,-50)(15,0)
\ArrowLine(15,0)(35,50)
\PhotonArc(0,0)(43,131,229){2}{10}
\Photon(-15,0)(-7,0){2}{2}
\Photon(15,0)(7,0){2}{2}
\ArrowLine(-15,0)(-27,-30)
\ArrowLine(-27,-30)(-35,-50)
\ArrowLine(-35,50)(-27,30)
\ArrowLine(-27,30)(-15,0)
\Text(0,-55)[cb]{{\footnotesize (u)}}
  \SetWidth{2}
\ArrowArc(0,0)(7,0,180)
\ArrowArc(0,0)(7,180,360)
\end{picture}}}
\hspace{3.6cm}
  \vcenter{\hbox{
\begin{picture}(0,0)(0,0)
  \SetScale{0.8}
  \SetWidth{.5}
\ArrowLine(27,-30)(15,0)
\ArrowLine(35,-50)(27,-30)
\ArrowLine(27,30)(35,50)
\ArrowLine(15,0)(27,30)
\PhotonArc(0,0)(43,-49,49){2}{10}
\Photon(-15,0)(-7,0){2}{2}
\Photon(15,0)(7,0){2}{2}
\ArrowLine(-15,0)(-35,-50)
\ArrowLine(-35,50)(-15,0)
\Text(0,-55)[cb]{{\footnotesize (v)}}
  \SetWidth{2}
\ArrowArc(0,0)(7,0,180)
\ArrowArc(0,0)(7,180,360)
\end{picture}}}
\]
\vspace*{1.6cm}
\caption{\it Two-loop diagrams containing a heavy-flavor loop.}
\label{fig2ltot}
\end{figure}

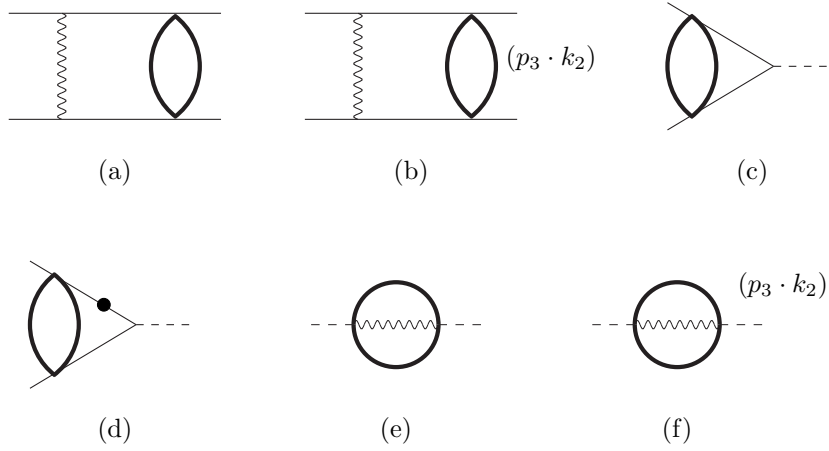
\begin{figure}
\vspace*{.3cm}
\[\vcenter{\hbox{
  \begin{picture}(0,0)(0,0)
\SetScale{.8}
  \SetWidth{.5}
\Line(-50,25)(-25,25)
\Line(-25,25)(25,25)
\Line(25,25)(50,25)
\Photon(-25,25)(-25,-25){2}{10}
%
\Line(-25,-25)(-50,-25)
\Line(25,-25)(-25,-25)
\Line(50,-25)(25,-25)
\Text(0,-45)[cb]{{\footnotesize (a)}}
  \SetWidth{2}
\CArc(47,0)(30,127,233)
\CArc(10,0)(30,-53,53)
\end{picture}}}
\hspace{3.8cm}
\vcenter{\hbox{
  \begin{picture}(0,0)(0,0)
\SetScale{.8}
  \SetWidth{.5}
\Line(-50,25)(-25,25)
\Line(-25,25)(25,25)
\Line(25,25)(50,25)
\Photon(-25,25)(-25,-25){2}{10}
%
\Line(-25,-25)(-50,-25)
\Line(25,-25)(-25,-25)
\Line(50,-25)(25,-25)
\Text(0,-45)[cb]{{\footnotesize (b)}}
\Text(53,-2)[cb]{{\footnotesize $(p_3 \cdot k_2)$}}
  \SetWidth{2}
\CArc(47,0)(30,127,233)
\CArc(10,0)(30,-53,53)
\end{picture}}}
\hspace{4.4cm}
\vcenter{\hbox{
  \begin{picture}(0,0)(0,0)
\SetScale{.8}
  \SetWidth{.5}
\Line(-40,30)(10,0)
\Line(-40,-30)(10,0)
\DashLine(10,0)(35,0){4}
\Text(0,-45)[cb]{{\footnotesize (c)}}
  \SetWidth{2}
\CArc(-10,0)(30,127,233)
\CArc(-47,0)(30,-53,53)
\end{picture}}}
\]
\vspace*{2.0cm}
\[ \hspace*{-1.0cm}
\vcenter{\hbox{
  \begin{picture}(0,0)(0,0)
\SetScale{.8}
  \SetWidth{.5}
\Line(-40,30)(10,0)
\Line(-40,-30)(10,0)
\DashLine(10,0)(35,0){4}
\Text(0,-45)[cb]{{\footnotesize (d)}}
\GCirc(-5.2,9.4){3}{0}
  \SetWidth{2}
\CArc(-10,0)(30,127,233)
\CArc(-47,0)(30,-53,53)
\end{picture}}}
\hspace{3.6cm}
\vcenter{\hbox{
  \begin{picture}(0,0)(0,0)
\SetScale{0.8}
  \SetWidth{.4}
\DashLine(-40,0)(-20,0){4}
\DashLine(20,0)(40,0){4}
\Photon(-20,0)(20,0){2}{8}
\Text(0,-45)[cb]{{\footnotesize (e)}}
  \SetWidth{2}
\CArc(0,0)(20,0,180)
\CArc(0,0)(20,180,360)
\end{picture}}}
\hspace{3.6cm}
\vcenter{\hbox{
  \begin{picture}(0,0)(0,0)
\SetScale{0.8}
  \SetWidth{.4}
\DashLine(-40,0)(-20,0){4}
\DashLine(20,0)(40,0){4}
\Photon(-20,0)(20,0){2}{8}
\Text(40,10)[cb]{{\footnotesize $(p_3 \cdot k_2)$}}
\Text(0,-45)[cb]{{\footnotesize (f)}}
  \SetWidth{2}
\CArc(0,0)(20,0,180)
\CArc(0,0)(20,180,360)
\end{picture}}}
\]
\vspace*{1.2cm}
\caption{\it The set of 6 two-loop Master Integrals involved in the
calculation.}
\label{MIs}
\end{figure}

The two-loop heavy-fermion correction to the Bhabha scattering
amplitude is given by the Feynman diagrams shown in
Fig.~\ref{fig2ltot}. We express the square modulus of the amplitude
in terms of scalar integrals.  The ultraviolet, soft, and collinear
divergencies of the integrals are treated by dimensional
regularization. By means of the Laporta algorithm \cite{Lap} the
scalar integrals are reduced to six MIs
diagrammatically shown in Fig.~\ref{MIs}. Four of them,
Fig.~\ref{MIs}~(c)--(f), were already known \cite{BMR2,MIs}.
The integrals Fig.~\ref{MIs}~(a) and (b) represent the main
computational result of the present paper. Below we describe their
calculation.


%

For the pair of MIs of the box topology we choose the integrals
Fig.~\ref{MIs}~(a) and (b) with the following momentum routing:
\vspace*{2mm}
\bea
M_{1}(D,m_f^2,P^2,Q^2) & = & \hspace{10mm} \hbox{
  \begin{picture}(0,0)(0,0)
\SetScale{.6}
  \SetWidth{.5}
\Line(-50,30)(50,30)
\Line(-50,-20)(50,-20)
\Photon(-25,30)(-25,-20){2}{10}
  \SetWidth{2}
\CArc(47,5)(30,127,233)
\CArc(10,5)(30,-53,53)
\end{picture}}
 \hspace{12mm} = \int  {\mathfrak{D}^D k_{1}}  {\mathfrak{D}^D k_{2}}
\, \frac{1}{{\mathcal D}_{1} {\mathcal D}_{3} {\mathcal D}_{4}
{\mathcal D}_{5} {\mathcal D}_{6} } \, ,
\label{MInt1} \\
\nn\\
\nn\\
M_{2}(D,m_f^2,P^2,Q^2) & = & \hspace{10mm}
\hbox{
  \begin{picture}(0,0)(0,0)
\SetScale{.6}
  \SetWidth{.5}
\Line(-50,30)(50,30)
\Line(-50,-20)(50,-20)
\Photon(-25,30)(-25,-20){2}{10}
\Text(50,-1)[cb]{{\footnotesize $(p_3 \cdot k_2)$}}
  \SetWidth{2}
\CArc(47,5)(30,127,233)
\CArc(10,5)(30,-53,53)
\end{picture}}
 \hspace{25mm} = \int  {\mathfrak{D}^D k_{1}}  {\mathfrak{D}^D k_{2}}
\, \frac{p_{3} \cdot k_{2}}{{\mathcal D}_{1} {\mathcal D}_{3}
{\mathcal D}_{4} {\mathcal D}_{5} {\mathcal D}_{6} } \, ,
\label{MInt2}
\eea
\vspace*{2mm}
where ${\mathcal D}_{1} = k_1^2$, ${\mathcal D}_{2} =
(p_1-k_1)^2$, ${\mathcal D}_{3} = (p_2+k_1)^2$, ${\mathcal D}_{4} =
k_2^2+m_f^2$, ${\mathcal D}_{5} = (p_1-p_3-k_1+k_2)^2+m_f^2$, and
where the integration measure is defined as
\be
\int{\mathfrak
D}^Dk = \frac{1}{C(D)} \left( \frac{\mu^2}{m_f^2}
\right)^{\frac{(D-4)}{2}} \int \frac{d^D k}{(2\pi)^{(D-2)}} \, .
\label{measure}
\ee
\noindent $C(D)$ is a function of the space-time dimension $D$:
\be
C(D) = (4 \pi)^{\frac{(4-D)}{2}} \Gamma \left( 3 - \frac{D}{2} \right) \, ,
\ee
with $C(4)=1$. In Eq.~(\ref{measure}) $\mu$ stands for the 't Hooft
scale of dimensional regularization and we set $\mu = m_f$ in the
rest of the paper. The integration measure in Eq.~(\ref{measure}) is chosen
in such a way that the one-loop massive tadpole becomes
\be
\int{\mathfrak D}^Dk \ \frac{1}{k^2+m_f^2} =
              \frac {m_f^2}{(D-2)(D-4)} \, .
\label{Tadpole}
\ee
%
The MIs $M_1$ and $M_2$ satisfy two systems of first-order linear
differential equations \cite{DiffEq} in the external kinematical
invariants $s$ and $t$. With our choice of MIs, the $s$-system is
completely decoupled, {\it i.e.} every MI satisfies a single
first-order linear differential equation. By contrast, the
$t$-system consists of two coupled equations and, therefore, is
equivalent to a second-order linear differential equation for one of
the MIs.\footnote{As it is  shown below, the second-order equation
in $t$ turns out to be particularly simple because the system of the
first-order equations in $s$ is decoupled.} The solution of the
system requires two initial conditions. Our MIs are functions of
$m_f^2$, $s$ and $t$. They are regular for $t\to 0$ and it is
possible to determine the initial conditions at $t=0$ for any value
of $s$. Therefore, it is more convenient to solve the system of the
differential equations in $t$ then the decoupled first-order
equations in $s$.
We define the following dimensionless variables:
\be
x = \frac{P^2}{m_f^2} = -\frac{s}{m_f^2} \, , \quad
y = \frac{Q^2}{m_f^2} = -\frac{t}{m_f^2} \, .
\label{xy}
\ee
In terms of these variables  the $t$-system takes the form:
\bea
\! \! \! \frac{dM_1}{dy} \!\! \! & = & \!\! \! \left[ \frac{D-5}{2 y} - \frac{1}{2(y\!+\!4)}
- \frac{D-4}{2 (y\!+\!x)} \right] M_1
+ \frac{3D\!-\!10}{2 m_f^2} \left[ \frac{1}{y} - \frac{1}{y\!+\!4}
 \right] M_2 \!+ \! \Omega_1(D,m_f^2,x,y) ,
\label{Syst1} \\
\! \! \! \frac{dM_2}{dy} \!\! \! & = & \!\! \! \frac{m_f^2}{2(y+4)} M_1
- \left[ \frac{D-4}{2} \left( \frac{1}{y} + \frac{1}{y+x}
 \right) - \frac{3D-10}{2(y+4)} \right] M_2 + \Omega_2(D,m_f^2,x,y) \, .
\label{Syst2}
\eea
Note that the second MI with the scalar product on the numerator is
dimensionless, while the first one has the mass dimension of
$m_f^{-2}$. The functions $\Omega_i(D,m_f^2,x,y)$ are linear combinations of
the MIs (c)--(f) of Fig.~\ref{MIs}, the product of a massless
one-loop bubble and a massive tadpole, and the product of two
tadpoles:
\bea
\Omega_1 & = & \hspace*{3.5mm}
\frac{x(x-4)}{y (4 + y) (x + y)} \hspace*{12mm}
\hbox{
  \begin{picture}(0,0)(0,0)
\SetScale{.6}
  \SetWidth{.5}
\Line(-40,33)(10,3)
\Line(-40,-27)(10,3)
\DashLine(10,3)(35,3){4}
\GCirc(-5.2,12.4){3}{0}
  \SetWidth{2}
\CArc(-10,3)(30,127,233)
\CArc(-47,3)(30,-53,53)
\end{picture}} \nn\\
& & -
\frac{1}{2 m_f^4 (D-4) y^2 (y+4)^2 (y + x)}
(-560 + 384 D - 64 D^2 + 200 x - 140 D x + 24 D^2 x  \nn\\
& & \hspace*{6mm}
+ 72 y - 76 D y +
    16 D^2 y + 10 D x y - 3 D^2 x y + 48 y^2 - 40 D y^2 + 8 D^2 y^2)
  \hspace*{12mm}
\hbox{
  \begin{picture}(0,0)(0,0)
\SetScale{0.6}
  \SetWidth{.4}
\DashLine(-40,3)(-20,3){4}
\DashLine(20,3)(40,3){4}
\Photon(-20,3)(20,3){2}{8}
  \SetWidth{2}
\CArc(0,3)(20,0,180)
\CArc(0,3)(20,180,360)
\end{picture}} \nn\\
& & +
\frac{12 (D-2) (-28 + 8 D + 10 x - 3 D x - 6 y + 2 D y)}{
m_f^6 (D-4) y^2 (y+4)^2 (y + x)}
  \hspace*{12mm}
\hbox{
  \begin{picture}(0,0)(0,0)
\SetScale{0.6}
  \SetWidth{.4}
\DashLine(-40,3)(-20,3){4}
\DashLine(20,3)(40,3){4}
\Photon(-20,3)(20,3){2}{8}
\Text(40,5)[cb]{{\footnotesize $(p_3 \cdot k_2)$}}
  \SetWidth{2}
\CArc(0,3)(20,0,180)
\CArc(0,3)(20,180,360)
\end{picture}} \nn\\
& & + \frac{2(D-2)(D-3)}{m_f^4 (D-4) x y (y+4)} \hspace*{12mm}
\hbox{
  \begin{picture}(0,0)(0,0)
\SetScale{0.6}
  \SetWidth{.4}
\DashLine(-40,3)(-25,3){4}
\DashLine(5,3)(20,3){4}
\CArc(-10,3)(15,0,180)
\CArc(-10,3)(15,180,360)
\Text(20,-2)[cb]{{\footnotesize $\times$}}
  \SetWidth{2}
\CArc(60,3)(10,0,180)
\CArc(60,3)(10,180,360)
\end{picture}} \nn\\
& & -
\frac{(D-2)}{8 m_f^6 (D-5) (D-4)
    (D-3) y^2 (y+4)^2 (y+x)}
     (5600 - 4960 D + 1408 D^2 - 128 D^3  \nn\\
& & \hspace*{6mm}
  -  2000 x + 1800 D x - 520 D^2 x + 48 D^3 x + 1184 y -
     1096 D y + 328 D^2 y - 32 D^3 y  \nn\\
& & \hspace*{6mm}
  - 296 x y +
     304 D x y - 98 D^2 x y + 10 D^3 x y - 24 y^2 +
     20 D y^2 - 4 D^2 y^2 - 24 x y^2  \nn\\
& & \hspace*{6mm}
  + 26 D x y^2 -
     9 D^2 x y^2 + D^3 x y^2)
 \hspace*{12mm}
\hbox{
  \begin{picture}(0,0)(0,0)
\SetScale{0.6}
  \SetWidth{.4}
  \SetWidth{2}
\CArc(-10,3)(10,0,180)
\CArc(-10,3)(10,180,360)
\CArc(10,3)(10,0,180)
\CArc(10,3)(10,180,360)
\end{picture}}
\hspace*{10mm} , \\ 
& & \nn\\
\Omega_2 & = & \hspace*{3.5mm}
\frac{(D-4)x}{8 y (y + x)}
 \hspace*{12mm}
\hbox{
  \begin{picture}(0,0)(0,0)
\SetScale{.6}
  \SetWidth{.5}
\Line(-40,33)(10,3)
\Line(-40,-27)(10,3)
\DashLine(10,3)(35,3){4}
  \SetWidth{2}
\CArc(-10,3)(30,127,233)
\CArc(-47,3)(30,-53,53)
\end{picture}}
\hspace*{10mm} +
\frac{m_f^2 x (x-4)}{4 (y+4) (y + x)}
 \hspace*{12mm}
\hbox{
  \begin{picture}(0,0)(0,0)
\SetScale{.6}
  \SetWidth{.5}
\Line(-40,33)(10,3)
\Line(-40,-27)(10,3)
\DashLine(10,3)(35,3){4}
\GCirc(-5.2,12.4){3}{0}
  \SetWidth{2}
\CArc(-10,3)(30,127,233)
\CArc(-47,3)(30,-53,53)
\end{picture}} \nn\\
& & -
\frac{1}{4 m_f^2 (D-4) y^2 (y+4)^2 (y + x)}
 (-360 x + 224 D x - 32 D^2 x - 480 y + 312 D y - 48 D^2 y  \nn\\
& & \hspace*{6mm}
  + 16 x y - 34 D x y +
    8 D^2 x y - 8 y^2 - 20 D y^2 + 6 D^2 y^2 + 24 x y^2 - 21 D x y^2 +
    4 D^2 x y^2  \nn\\
& & \hspace*{6mm}
+ 48 y^3 - 44 D y^3 + 9 D^2 y^3)
 \hspace*{12mm}
\hbox{
  \begin{picture}(0,0)(0,0)
\SetScale{0.6}
  \SetWidth{.4}
\DashLine(-40,3)(-20,3){4}
\DashLine(20,3)(40,3){4}
\Photon(-20,3)(20,3){2}{8}
  \SetWidth{2}
\CArc(0,3)(20,0,180)
\CArc(0,3)(20,180,360)
\end{picture}} \nn\\
& &  \nn\\
& & 
- \frac{6 (D-2)(18 x \! - \! 4 D x \! + \! 24 y \! - \! 6 D y \! + \! 4 x y \!
- \! D x y \! + 10 y^2 - 3 D y^2)
}{m_f^4 (D-4) y^2 (y+4)^2 (y + x)}
 \hspace*{12mm}
\hbox{
  \begin{picture}(0,0)(0,0)
\SetScale{0.6}
  \SetWidth{.4}
\DashLine(-40,3)(-20,3){4}
\DashLine(20,3)(40,3){4}
\Photon(-20,3)(20,3){2}{8}
\Text(40,5)[cb]{{\footnotesize $(p_3 \cdot k_2)$}}
  \SetWidth{2}
\CArc(0,3)(20,0,180)
\CArc(0,3)(20,180,360)
\end{picture}} \nn\\
& & +
\frac{(D-3) (D-2)}{2 m_f^2 (D-4) x (y+4)}
 \hspace*{12mm}
\hbox{
  \begin{picture}(0,0)(0,0)
\SetScale{0.6}
  \SetWidth{.4}
\DashLine(-40,3)(-25,3){4}
\DashLine(5,3)(20,3){4}
\CArc(-10,3)(15,0,180)
\CArc(-10,3)(15,180,360)
\Text(20,-2)[cb]{{\footnotesize $\times$}}
  \SetWidth{2}
\CArc(60,3)(10,0,180)
\CArc(60,3)(10,180,360)
\end{picture}} \nn\\
& & -
\frac{D-2}{32 m_f^4 (D-5) (D-4) (D-3) y^2 (y+4)^2
    (y + x)}
  (7200 x - 5920 D x + 1536 D^2 x  \nn\\
& & \hspace*{6mm}
  - 128 D^3 x + 9600 y - 8160 D y +
     2208 D^2 y - 192 D^3 y + 1360 x y - 1032 D x y  \nn\\
& & \hspace*{6mm}
  + 232 D^2 x y - 16 D^3 x y +
     3424 y^2 - 2944 D y^2 + 808 D^2 y^2 - 72 D^3 y^2 - 176 x y^2  \nn\\
& & \hspace*{6mm}
  + 200 D x y^2 -
     72 D^2 x y^2 + 8 D^3 x y^2 + 16 y^3 - 8 D y^3 - 24 x y^3 + 26 D x y^3 -
     9 D^2 x y^3  \nn\\
& & \hspace*{6mm}
  + D^3 x y^3)
 \hspace*{12mm}
\hbox{
  \begin{picture}(0,0)(0,0)
\SetScale{0.6}
  \SetWidth{.4}
  \SetWidth{2}
\CArc(-10,3)(10,0,180)
\CArc(-10,3)(10,180,360)
\CArc(10,3)(10,0,180)
\CArc(10,3)(10,180,360)
\end{picture}}
\hspace*{15mm}
 \, .
\eea
All the possible singularities of the integrals $M_1$ and $M_2$ are
those appearing in  their coefficients in  Eqs.~(\ref{Syst1},\ref{Syst2}).
Thus, the singularities  are  potentially located at $y=0$, $y=-4$ and
$y=-x$. The point $y=-4$ is a singular point for the integrals. It
corresponds to the three (two massive and one massless) particle cut
in the $t$ channel. By contrast,  $y=0$ is a regular point. We can
use this information  in order to determine
the initial conditions. In fact, multiplying Eqs.~(\ref{Syst1},~\ref{Syst2}) 
by $y$ and taking into account that $y\, dM_{1,2} / dy
|_{y \to 0}\to 0$, in the limit $y \to 0$ we find
\bea
M_1(D,x,y=0) & = &
- \frac{1}{2 m_f^2 x} \, \frac{1}{(D-4)^3}
- \frac{G(0,x)}{4 m_f^2 x}  \, \frac{1}{(D-4)^2}
+ \frac{\zeta(2) - G(0,0,x)}{8 m_f^2 x}
  \, \frac{1}{(D-4)} \nn\\
& &
+ \frac{1}{16 m_f^2 x} \bigl[ 8 - 2\zeta(3)
  + \bigl( \zeta(2) - 4 \bigr) G(0,x) - G(0,0,0,x)
  + G(\mu,\mu,0,x) \bigr]  \nn\\
& &
  + \frac{4-x}{8 m_f^2 x \sqrt{x(4-x)}} G(\mu,0,x)
+ {\mathcal O}(D-4) \, ,
\label{Init1} \\
M_2(D,x,y=0) & = & \frac{1}{16} \, \frac{1}{(D-4)^2}
- \frac{\bigl( 2 - G(0,x) \bigr)}{32}
  \, \frac{1}{(D-4)}
+ \frac{1}{64 x} \bigl[ 11x - \zeta(2)x
  - 5x G(0,x)  \nn\\
& &
+ x G(0,0,x) \!
  + \! 2 G(\mu,\mu,0,x) \bigr] \!
+ \frac{4-x}{64 \sqrt{x(4\!-\!x)}} G(\mu,0,x)
+ {\mathcal O}(D-4) \, ,
\label{Init2}
\eea
where the functions $G$ are generalized harmonic polylogarithms (GHPLs)
\cite{AB2} described in Appendix~\ref{GHPLs}.
The system of Eqs.~(\ref{Syst1},\ref{Syst2}) is equivalent to a
second-order linear differential equation for one of the MIs. For
the integral $M_1$, after trivial manipulations, we find
\bea
\hspace*{-10mm}
& & \frac{d^2M_1}{dy^2}
+ \left[ \frac{3}{2 y} - \frac{3D-13}{2(y+4)}
+ \frac{D-4}{y+x}
\right] \frac{dM_1}{dy}
- (D-4) \Biggl[ \frac{D-5}{4y^2}
- \left( \frac{3D-17}{16} + \frac{3}{4x}
\right) \frac{1}{y}  \nn\\
\hspace*{-10mm}
& & \hspace*{11mm}
- \frac{D-6}{4(y+x)^2}
+ \left(
 \frac{3}{4x} - \frac{3D-13}{4(x-4)}
\right) \frac{1}{y+x}
+ \left(
\frac{3D-17}{16} + \frac{3D-13}{4(x-4)}
\right) \frac{1}{y+4} \,
\Biggr] \, M_1  \nn\\
\hspace*{-10mm}
& & \hspace*{11mm}
+ \Omega(D,m_f^2,x,y) = 0 \, .
\label{Secord}
\eea
The function $\Omega(D,m_f^2,x,y)$ contains the MIs (c)--(f) of 
Fig.~\ref{MIs}, some products of one-loop integrals and the product 
of two tadpoles. 

Let us discuss briefly the structure of the  Eq.~(\ref{Secord}).
Note that we are not interested in a solution of the differential
equation valid for arbitrary value of the space-time parameter $D$.
Instead,  we look for a solution in the form of a Laurent series in
$(D-4)$. Since  the coefficient of $M_1$ is proportional to $(D-4)$,
Eq.~(\ref{Secord}) at each order in $(D-4)$ is reduced to a
first-order linear differential equation for the derivative $M_1' =
dM_1 / dy$:
\bea
M_1'(D,m_f^2,x,y) & = & \sum_{i=-3}^N M_{1,i}' \, (D-4)^i
+ {\mathcal O}(D-4)^{N+1}\, , \\
\frac{dM_{1,i}'}{dy} & = & - \left[ \frac{3}{2 y} + \frac{1}{2(y+4)}
\right] \, M_{1,i}' + \Psi_i(m_f^2,x,y) \, .
\label{Laurent}
\eea
The functions $\Psi_i(m_f^2,x,y)$ in Eq.~(\ref{Laurent}) are defined
as follows ($M_{1,j}=0$ for $j<-3$):
\bea
\Psi_i(m_f^2,x,y) & = & - \Omega_i(D,m_f^2,x,y)
+ \Biggl[
\frac{3}{2(y+4)}
- \frac{1}{y+x} \Biggr] M_{1,i-1}'
- \Biggl[ \frac{1}{4y^2}
   + \left( \frac{3}{4x} - \frac{5}{16} \right) \frac{1}{y} \nn\\
& &
   - \frac{1}{2(y+x)^2}
   - \left( \frac{3}{4x} + \frac{1}{4(x-4)} \right) \frac{1}{y+x}
   + \left( \frac{5}{16} + \frac{1}{4(x-4)} \right) \frac{1}{y+4}
\Biggr] M_{1,i-1}  \nn\\
& &
+ \Biggl[
     \frac{1}{4y^2}
   - \frac{3}{16y}
   - \frac{1}{4(y+x)^2}
   - \frac{3}{4(x-4)} \, \frac{1}{y+x} \nn\\
& &
   + \left( \frac{3}{16} + \frac{3}{4(x-4)} \right) \frac{1}{y+4}
\Biggr] M_{1,i-2} \, .
\eea
Note that $\Psi_i(m_f^2,x,y)$ contains  $M'_{1,i-1}$, $M_{1,i-1}$
and $M_{1,i-2}$.
Moreover, $\Psi_i(m_f^2,x,y)$ contains the $i$-th order term of the
Laurent expansion of $\Omega(D,m_f^2,x,y)$:
\bea 
\Omega(D,m_f^2,x,y) &=& \sum_{i=-2}^0 \Omega_i(D,m_f^2,x,y) \,
(D-4)^i
+ {\mathcal O}(D-4) \, , \nn\\
& = & \frac{2 x^2 + 10 x y - x^2 y + 4 y^2}{4 m_f^2 x y^2 (4 + y) (x + y)^2} \,
\frac{1}{(D-4)^2}
\nn\\
&  & - \frac{1}{8 m_f^2 y^2 (x + y)^2}
\Biggl[
\frac{8 x + 20 y - 2 x y + y^2}{(4 + y)}
- \frac{2 x^2 + 10 x y - x^2 y + 4 y^2}{x (4 + y)} G(0,x) \nn\\
& &
- \frac{2 x + 6 y - x y}{\sqrt{y(y+4)}} G( - \mu,y)
\Biggr]
\frac{1}{(D-4)} \nn\\
& &
+ \frac{1}{16 m_f^2 y^2 (x + y)^2 x (4 + y)}
\Biggl\{
   16 x^2 + 40 x y - 4 x^2 y + 2 x y^2 - 2 x^2 \zeta(2)  \nn\\
& &
 - 10 x y \zeta(2) + x^2 y \zeta(2) - 4 y^2 \zeta(2)
 - x( 4 x + 16 y - 3 x y) G(0,x)
 + (2 x^2  \nn\\
& & + 10 x y - x^2 y + 4 y^2) G(0,0,x)
+ 2x(2 x - 2 y + x y) G( - \mu, - \mu,y) \nn\\
& &
+ \frac{xy(x-4)^2}{\sqrt{x(4-x)}} G(\mu,0,x)
- \frac{x(y+4)}{\sqrt{y(y+4)}} \Bigl[
(12 x + 24 y - x y + 2 y^2) G( - \mu,y) \nn\\
& &
- ( 2 x + 6 y - x y) G(0, - \mu,y)
- 3 ( 2 x + 6 y - x y) G(-4, - \mu,y)
\Bigr]
\Biggr\} \nn\\
& & + {\mathcal O}(D-4)\, .
\eea
The formal solution of Eq.~(\ref{Laurent}) reads
\be
M_{1,i}' = \frac{1}{y \sqrt{y(y+4)}} \left[
\int^y r \sqrt{r(r+4)} \Psi_i \, dr + J_i \right] \, ,
\label{eqder}
\ee
where $J_i$ are integration constants, which  are fixed by the
regularity condition for  the derivative $M_{1,i}'$ at $y=0$.
Actually, in every order in $(D-4)$ we find $J_i = 0$. This means
that we have to discard the homogeneous solution and keep only the
particular solution of the inhomogeneous Eq.~(\ref{Laurent}). Then,
we integrate Eq.~(\ref{eqder}):
\be
M_{1,i} = \int^y M_{1,i}'(m_f^2,x,r) \, dr + J_{1,i} \, .
\label{eqMI}
\ee
$J_{1,i}$ are again integration constants which  are
determined by the initial conditions Eq.~(\ref{Init1}). Once we
have the master integral $M_1$, the calculation of $M_2$ is
straightforward. From Eq.~(\ref{Syst1}), we express $M_2$ in terms
of $M_1$ and $dM_1/dy$. The final analytical  expression for $M_1$
and $M_2$ reads
\be
M_1(D,m_f^2,x;y) = \sum_{i=-3}^0 M_{1,i}(m_f^2,x;y) \, (D-4)^i
+ {\mathcal O}(D-4) \, ,
\label{MI1}
\ee
where
\bea
M_{1,-3} & = & - \frac{1}{2m_f^2 x} \, , \\
M_{1,-2} & = & \frac{1}{4 m_f^2 x} \left[ 2 - G(0;x)
- \frac{y+4}{\sqrt{y(y+4)}} G( - \mu;y) \right] \, , \\
M_{1,-1} & = & \frac{1}{8 m_f^2 x} \Biggl\{
- 4 + \zeta(2) + 2 G(0;x) - G(0,0;x) + 2 G( - \mu, - \mu;y) \nn\\
& &
+ \frac{y+4}{\sqrt{y(y+4)}} \Bigl[
2 G( - \mu;y) - 3 G(-4, - \mu;y) - G(0;x) G( - \mu;y)
\Bigr]
\Biggr\}
\, , \\
M_{1,0} & = &
- \frac{(x-4)(y+4)}{16m_f^2 \sqrt{x(4-x)} \sqrt{y(y+4)}}
G(\mu,0;x) G( - x - \mu;y) \nn\\
& &
+ \frac{4+y}{16m_f^2 x \sqrt{y(y+4)}} \bigl[
       -    G( - x,0, - \mu;y)
       +  6 G( - \mu, - \mu, - \mu;y)
       -  ( 4 - \zeta(2) ) G( - \mu;y) \nn\\
& &
       + G(0;x) (
       -  3 G(-4, - \mu;y)
       +    G( - x, - \mu;y)
       +  2 G( - \mu;y)
       -    G(0, - \mu;y) ) \nn\\
& &
       +  6 G(-4, - \mu;y)
       -  9 G(-4,-4, - \mu;y)
       +    G(0,0, - \mu;y)
       -    G(0,0;x)G( - \mu;y)
\bigr] \nn\\
& &
+ \frac{1}{16m_f^2 x} \bigl[
         8
          - 2 \zeta(2)
          - 2 \zeta(3)
       - 4 G( - \mu, - \mu;y)
       + 6 G( - \mu,-4, - \mu;y)
       +  G(\mu,\mu,0;x) \nn\\
& &
       - 4 G(0, - \mu, - \mu;y)
       -  ( 4 - \zeta(2) ) G(0;x)
       + 2 G(0;x)G( - \mu, - \mu;y)
       + 2 G(0,0;x) \nn\\
& &
       - G(0,0,0;x)
\bigr] \, ,
\eea
and
\be
M_2(D,m_f^2,x;y) = \sum_{i=-3}^0 M_{2,i}(m_f^2,x;y) \, (D-4)^i
+ {\mathcal O}(D-4) \, ,
\label{MI2}
\ee
where
\bea
M_{2,-3} & = &
       - \frac{y}{8 x} \, , \\
M_{2,-2} & = &
         \frac{1}{16 x} \left[ x + y( 4 - G(0;x) )
       - \frac{y(y+4)}{\sqrt{y(y+4)}} G( - \mu;y) \right]
\, , \\
M_{2,-1} & = &
         \frac{1}{32 x} \Biggl[
            \zeta(2) y \!
          - \! 5x \!
          - \! 14y
       +  2 \Bigl( \!
           2 \!
          + \! y \!
          + \! \frac{x}{y}
          \Bigr) G( - \mu, - \mu;y)
       +  (
           x \! + \! 4y
          ) G(0;x)
       - y G(0,0;x)
\Biggr] \nn\\
& &
+ \frac{4 +y}{32 x \sqrt{y(y+4)}} \bigl[
         (x + 4 y
          ) G( - \mu;y)
       -  3 y G(-4, - \mu;y)
       -  y G(0;x)G( - \mu;y)
\bigr] \, ,
\\
M_{2,0} & = &
- \frac{y(y+4)(x-4)}{64 \sqrt{x(4-x)} \sqrt{y(y+4)}}
G(\mu,0;x) G( - x - \mu;y)
- \frac{x-4}{64 \sqrt{x(4-x)}} G(mu,0;x) \nn\\
& &
+ \frac{y(y+4)}{64 x \sqrt{y(y+4)}} \biggl[
       - G( - x,0, - \mu;y)
       + \biggl( \zeta(2) -14 - \frac{5x}{y} \biggr) G( - \mu;y) \nn\\
& &
       + \biggl( 12 + \frac{3x}{y} \biggr) G(-4, - \mu;y)
       + 6 G( - \mu, - \mu, - \mu;y)
       - 9 G(-4,-4, - \mu;y) \nn\\
& &
       + \frac{x}{y}  G(0, - \mu;y)
       + G(0,0, - \mu;y)
       + G(0;x) \bigl(
         G( - x, - \mu;y)
       - 3 G(-4, - \mu;y) \nn\\
& &
       + 4 G( - \mu;y)
       - G(0, - \mu;y) \bigr)
       - G(0,0;x)G( - \mu;y)
\biggr]
+ \frac{1}{64 x} \biggl[
            19x + 46y - \zeta(2) x  \nn\\
& &
       - 4\zeta(2) y - 2\zeta(3) y
       - 2 \biggl( 6 + 3 \frac{x}{y} + x + 4y \biggr) G( - \mu, - \mu;y) \nn\\
& &
       + 6 \biggl( 2 + \frac{x}{y} + y \biggr) G( - \mu,-4, - \mu;y)
       + \frac{2x}{y} G( - \mu,0, - \mu;y)
       + ( 2 + y ) G(\mu,\mu,0;x)\nn\\
& &
       - 2 \biggl( 4 + \frac{x}{y} + 2 y \biggr) G(0, - \mu, - \mu;y)
       + 2 ( 2 + y ) G(0;x)G( - \mu, - \mu;y)\nn\\
& &
       + ( \zeta(2) y - 5 x - 14 y ) G(0;x)
       + ( x + 4y ) G(0,0;x)
       - y G(0,0,0;x)
\biggr] \, ,
\eea
The result is expressed in terms of GHPLs of two variables, $x$ and $y$. More
details on these functions can be found in Appendix \ref{GHPLs}.
We checked that Eqs.~(\ref{MI1},\ref{MI2}) do satisfy the system of
linear differential equations in $s$, and the initial condition
of Eq.~(\ref{Init2}) for $M_2$ is recovered in the limit $x \to 0$.

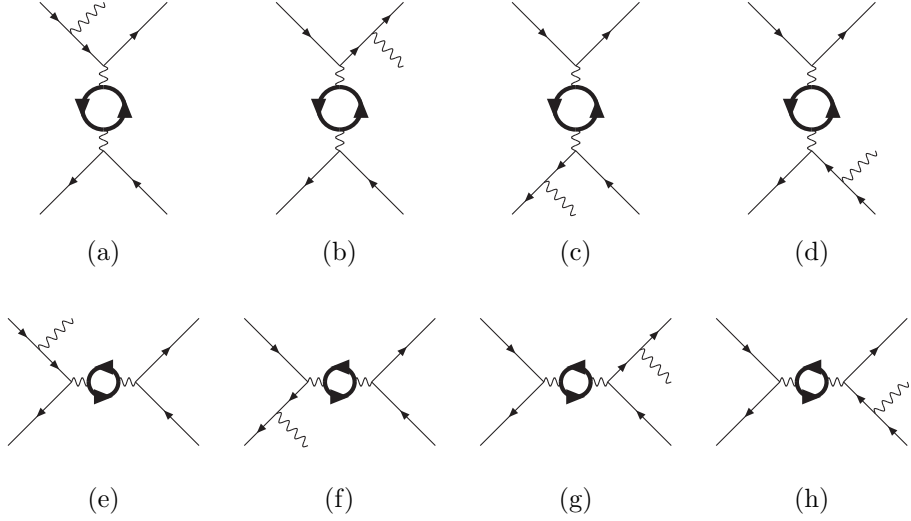
\begin{figure}
\vspace*{.3cm}
\[\vcenter{
\hbox{
  \begin{picture}(0,0)(0,0)
\SetScale{.8}
  \SetWidth{.5}
\ArrowLine(0,20)(30,50)
\ArrowLine(-30,50)(-15,35)
\ArrowLine(-15,35)(0,20)
\ArrowLine(30,-50)(0,-20)
\ArrowLine(0,-20)(-30,-50)
\Photon(0,-20)(0,-10){2}{2}
\Photon(0,10)(0,20){2}{2}
\Photon(-15,35)(0,50){2}{4}
\Text(0,-60)[cb]{{\footnotesize (a)}}
  \SetWidth{2}
\ArrowArc(0,0)(10,-90,90)
\ArrowArc(0,0)(10,90,270)
\end{picture}}
}
\hspace{3cm}
  \vcenter{
\hbox{
  \begin{picture}(0,0)(0,0)
\SetScale{.8}
  \SetWidth{.5}
\ArrowLine(0,20)(15,35)
\ArrowLine(15,35)(30,50)
\ArrowLine(-30,50)(0,20)
\ArrowLine(30,-50)(0,-20)
\ArrowLine(0,-20)(-30,-50)
\Photon(0,-20)(0,-10){2}{2}
\Photon(0,10)(0,20){2}{2}
\Photon(15,35)(30,20){2}{4}
\Text(0,-60)[cb]{{\footnotesize (b)}}
  \SetWidth{2}
\ArrowArc(0,0)(10,-90,90)
\ArrowArc(0,0)(10,90,270)
\end{picture}}
}
\hspace{3cm}
  \vcenter{
\hbox{
  \begin{picture}(0,0)(0,0)
\SetScale{.8}
  \SetWidth{.5}
\ArrowLine(0,20)(30,50)
\ArrowLine(-30,50)(0,20)
\ArrowLine(-15,-35)(-30,-50)
\ArrowLine(0,-20)(-15,-35)
\ArrowLine(30,-50)(0,-20)
\Photon(0,-20)(0,-10){2}{2}
\Photon(0,10)(0,20){2}{2}
\Photon(-15,-35)(0,-50){2}{4}
\Text(0,-60)[cb]{{\footnotesize (c)}}
  \SetWidth{2}
\ArrowArc(0,0)(10,-90,90)
\ArrowArc(0,0)(10,90,270)
\end{picture}}
}
\hspace{3cm}
  \vcenter{
\hbox{
  \begin{picture}(0,0)(0,0)
\SetScale{.8}
  \SetWidth{.5}
\ArrowLine(0,20)(30,50)
\ArrowLine(-30,50)(0,20)
\ArrowLine(0,-20)(-30,-50)
\ArrowLine(30,-50)(15,-35)
\ArrowLine(15,-35)(0,-20)
\Photon(0,-20)(0,-10){2}{2}
\Photon(0,10)(0,20){2}{2}
\Photon(15,-35)(30,-20){2}{4}
\Text(0,-60)[cb]{{\footnotesize (d)}}
  \SetWidth{2}
\ArrowArc(0,0)(10,-90,90)
\ArrowArc(0,0)(10,90,270)
\end{picture}}
}
\]
\vspace*{2.2cm}
\[\vcenter{
\hbox{
  \begin{picture}(0,0)(0,0)
\SetScale{.8}
  \SetWidth{.5}
\ArrowLine(-45,30)(-30,15)
\ArrowLine(-30,15)(-15,0)
\ArrowLine(-15,0)(-45,-30)
\ArrowLine(15,0)(45,30)
\ArrowLine(45,-30)(15,0)
\Photon(-15,0)(-7,0){2}{2}
\Photon(7,0)(15,0){2}{2}
\Photon(-30,15)(-15,30){2}{4}
\Text(0,-50)[cb]{{\footnotesize (e)}}
  \SetWidth{2}
\ArrowArc(0,0)(7,0,180)
\ArrowArc(0,0)(7,180,360)
\end{picture}}
}
\hspace{3cm}
  \vcenter{
\hbox{
  \begin{picture}(0,0)(0,0)
\SetScale{.8}
  \SetWidth{.5}
\ArrowLine(-45,30)(-15,0)
\ArrowLine(-15,0)(-30,-15)
\ArrowLine(-30,-15)(-45,-30)
\ArrowLine(15,0)(45,30)
\ArrowLine(45,-30)(15,0)
\Photon(-15,0)(-7,0){2}{2}
\Photon(7,0)(15,0){2}{2}
\Photon(-30,-15)(-15,-30){2}{4}
\Text(0,-50)[cb]{{\footnotesize (f)}}
  \SetWidth{2}
\ArrowArc(0,0)(7,0,180)
\ArrowArc(0,0)(7,180,360)
\end{picture}}
}
\hspace{3cm}
  \vcenter{
\hbox{
  \begin{picture}(0,0)(0,0)
\SetScale{.8}
  \SetWidth{.5}
\ArrowLine(-45,30)(-15,0)
\ArrowLine(-15,0)(-45,-30)
\ArrowLine(15,0)(30,15)
\ArrowLine(30,15)(45,30)
\ArrowLine(45,-30)(15,0)
\Photon(-15,0)(-7,0){2}{2}
\Photon(7,0)(15,0){2}{2}
\Photon(30,15)(45,0){2}{4}
\Text(0,-50)[cb]{{\footnotesize (g)}}
  \SetWidth{2}
\ArrowArc(0,0)(7,0,180)
\ArrowArc(0,0)(7,180,360)
\end{picture}}
}
\hspace{3cm}
  \vcenter{
\hbox{
  \begin{picture}(0,0)(0,0)
\SetScale{.8}
  \SetWidth{.5}
\ArrowLine(-45,30)(-15,0)
\ArrowLine(-15,0)(-45,-30)
\ArrowLine(15,0)(45,30)
\ArrowLine(45,-30)(30,-15)
\ArrowLine(30,-15)(15,0)
\Photon(-15,0)(-7,0){2}{2}
\Photon(7,0)(15,0){2}{2}
\Photon(30,-15)(45,0){2}{4}
\Text(0,-50)[cb]{{\footnotesize (h)}}
  \SetWidth{2}
\ArrowArc(0,0)(7,0,180)
\ArrowArc(0,0)(7,180,360)
\end{picture}}
}
\]
\vspace*{1.2cm}
\caption{\it Diagrams contributing to the real corrections to the NNLO heavy
 flavor cross section.}
\label{2lreal}
\end{figure}
%

\section{The Two-Loop Heavy-Flavor Correction \label{crosssection}}

In this Section we present the analytical result for  the
contribution of the different classes of the two-loop diagrams shown
in Fig.~\ref{fig2ltot} and the corresponding soft-photon emission
contribution to the differential cross section. We keep the
notations as close as possible to  \cite{BFMRvB1,BFMRvB2}. The
ultraviolet divergencies  are renormalized in the on-shell scheme.

It is convenient to  split the two-loop virtual correction in
Eq.~(\ref{vps}) into the sum of five terms
\be
\frac{d \sigma_2^V}{d \Omega} =
   \frac{d \sigma_{2}^{V}}{d \Omega} \Big|_{ (2l,S)}
 \!+ \frac{d \sigma_{2}^{V}}{d \Omega} \Big|_{ (2l,V)}
 \!+ \frac{d \sigma_{2}^{V}}{d \Omega} \Big|_{ (2l,B)}
 \!+ \frac{d \sigma_{2}^{V}}{d \Omega} \Big|_{ (2l,R)}
 \!+ \frac{d \sigma_{2}^{V}}{d \Omega} \Big|_{ (S,V)}
 \!+ \frac{d \sigma_{2}^{V}}{d \Omega} \Big|_{ (S,B)} \, ,
 \label{virtualCS}
\ee
which correspond to the contribution of the two-loop self-energy
diagrams, two-loop vertex diagrams, two-loop box diagrams,  two-loop
reducible diagrams, and to the interference of one-loop vertex and
one-loop box diagrams with the one-loop self-energy diagrams,
respectively. We  drop the list of arguments of the various
contributions to the cross section. All the terms  in
Eq.~(\ref{virtualCS}) depend on $s$, $t$, and $m_f$. As it is
explained in Section~\ref{cancellation}, only the fourth and fifth
terms on the r.~h.~s. of Eq.~(\ref{virtualCS})
depend logarithmically on the electron mass $m_e$.

\subsection{Two-Loop Vacuum Polarization Corrections}

The contribution of the  diagrams Fig.~\ref{fig2ltot}~(a)--(f) can
be obtained by replacing $\Pi^{(1l)}_0$ with the two-loop vacuum
polarization function $\Pi^{(2l)}_0$  in Eq.~(\ref{amp1lS}):
\bea
\frac{d \sigma_{2}^{V}}{d \Omega} \Big|_{ (2l,S)}
&=& \frac{\alpha^2}{s} Q_f^4 N_c\Bigg\{ \frac{1}{s^2} \left[ s t +
\frac{s^2}{2} + t^2 \right] 2 \mbox{Re}\Pi^{(2l)}_0(s)
+ \frac{1}{t^2} \left[ s t + \frac{t^2}{2} + s^2 \right]
2 \Pi^{(2l)}_0(t) \nn\\
& & + \frac{1}{s t} (s + t)^2
 \left(\mbox{Re}\Pi^{(2l)}_0(s) +  \Pi^{(2l)}_0(t)\right)
\Bigg\} \, .
\label{amp2lS}
\eea
The explicit expression of the renormalized function  $\Pi^{(2l)}_0$
in terms of GHPLs is given in Eq.~(\ref{A1lS1}) of  Appendix
\ref{AF}. Note that   the two-loop vacuum polarization corrections
are proportional to the fourth power of the heavy fermion charge,
while all the other corrections  are proportional to $Q_f^2$.

\subsection{Two-Loop Vertex Corrections}

The two-loop vertex diagram are shown in
Fig.~\ref{fig2ltot}~(g)--(j).  These diagrams are infrared finite
and can be evaluated for $m_e=0$. The analytical result for the
two-loop  vertex correction reads
\be
\frac{d \sigma_{2}^{V}}{d \Omega} \Big|_{(2l,V)} = 2
\frac{\alpha^2}{s} Q_f^2 N_c \Biggl[ \frac{1}{s^2} V^{(2l)}_2(t,s) +
\frac{1}{t^2} V^{(2l)}_2(s,t) + \frac{1}{s t} \left( V_1^{(2l)}(s,t)
\! + \! V_1^{(2l)} (t,s)\right) \Biggr] \,.
\label{tot2lvertex}
\ee
The functions $V_1^{(2l)}$ and $V_2^{(2l)}$ in this equation are
related to the two-loop Dirac form factor\footnote{The Pauli form
factor vanishes in the limit $m_e \to 0$.} $F_1^{(2l)} (-p^2)$ as
follows:
\bea
V_1^{(2l)}(s,t) & = & c_{1,1}(s,t) \mbox{Re}F_1^{(2l)}(t)\, ,
\label{V1V22lU} \\
V_2^{(2l)}(s,t) & = & c_{2,1}(s,t) \mbox{Re}F_1^{(2l)}(t)\, ,
\label{V1V22l}
\eea
where the coefficients $c_{i,1}(s,t)$  read
\bea
c_{1,1}(s,t) &=& (s+t)^2 \, , \quad c_{2,1}(s,t) = 2 \left(s t +s^2 +
\frac{t^2}{2}\right) \, .
\label{coeffs1121}
\eea
The explicit expressions of the  renormalized form factor is given
in Eq.~(\ref{2lF1}) of Appendix~\ref{AF}.

\subsection{Two-Loop Box Corrections}

There are eight two-loop box diagrams  shown in
Fig.~\ref{fig2ltot}~(k)--(r). If we transform the external momenta
$p_4 \leftrightarrow -p_1$ and $p_2 \leftrightarrow - p_3$ in the
diagram~(l), it becomes identical to the diagram~(k). Since this
transformation does not  change the Mandelstam variables $s$ and
$t$, the contributions of diagrams (k) and~(l) to the differential
cross section  are equal. The same is true for the pairs of
diagrams:  (m)-(n), (o)-(p), and (q)-(r). Thus the contribution of
the two-loop box diagrams can be written as follows:
\bea
\frac{d \sigma_{2}^{V}}{d \Omega}
\Big|_{(2l,B)} \hspace*{-2mm} & = & \hspace*{-2mm}
 - 2 \frac{\alpha^2}{4s} Q_f^2 N_c
\Bigl[ \frac{m_f^2}{s} \Bigl(\mbox{Re}B_1^{(2l)}(s,t) +
\mbox{Re}B_2^{(2l)}(t,s)+ B_3^{(2l)}(u,t)
 - \mbox{Re}B_2^{(2l)}(u,s)
\Bigr) \nn\\
\hspace*{-2mm}  & &  \hspace*{-2mm}
+ \frac{m_f^2}{t} \Bigl(
\mbox{Re}B_2^{(2l)}(s,t) +
\mbox{Re}B_1^{(2l)}(t,s)
- B_2^{(2l)}(u,t) + \mbox{Re}B_3^{(2l)}(u,s)
\Bigr)
\Bigr] \, ,
\label{2lBamp}
\eea
where the overall minus sign  is due to the closed fermion loop and
the overall factor  2 reflects the identity of the diagrams
discussed above.
The Laurent expansion of the renormalized functions $B_{i}^{(2l)}$
reads
\be
B_i^{(2l)}(s,t) = \frac{1}{(D-4)^2} B_i^{(2l,-2)}(s,t)+
\frac{1}{(D-4)} B_i^{(2l,-1)}(s,t)
                  + B_i^{(2l,0)}(s,t)
                  + {\mathcal{O}} \Big( (D-4) \Big) \ .
\label{BiLaur2l}
\ee
The expressions of  the coefficients $B_i^{(2l,j)}$ for $j=-2,-1,0$,
are given in Eqs.~(\ref{B12lM2}--\ref{B32l0}) of Appendix~\ref{AF}.
As it was pointed out in Section~\ref{cancellation}, the double
poles of the auxiliary functions $B_i^{(2l,j)}$ cancel in the sum of
the planar and crossed box contributions.

\subsection{Two-Loop Reducible Corrections \label{RedCorr}}

The two-loop reducible diagrams are shown in
Fig.~\ref{fig2ltot}~(s)--(v). They reduce to the product of the
one-loop vacuum polarization function and one-loop Dirac form factor
and can be easily obtained from Eq.~(43) of \cite{BFMRvB1}. The
result is of the following form:
\be
\frac{d \sigma_{2}^{V}}{d \Omega}
\Big|_{(2l,R)} = 2 \frac{\alpha^2}{s} Q_f^2
N_c\Biggl[\frac{1}{s^2} V^R_2(t,s)  + \frac{1}{t^2} V^R_2(s,t) +
\frac{1}{s t}\left(V^R_1(s,t) + V^R_1 (t,s)\right) \Biggr] \ ,
\label{2lRtot}
\ee
where the  functions $V_i^{R}(s,t)$ have the  Laurent expansion
\be
 V_i^{(R)}(s,t) = \frac{1}{(D-4)}V_i^{(R,-1)}(s,t)
                 + V_i^{(R,0)}(s,t)
                 + {\mathcal{O}} \Big((D-4)\Big) \ ,
\label{ViLaur}
\ee
with
\bea
 V_i^{(R,-1)}(s,t) & = &  c_{i,1}(s,t) \
          \mbox{Re}\Big(\ F_1^{(1l,-1)}(t)\ \Pi^{(1l,0)}_0(t) \ \Big) \ ,
\label{2lRa} \\
 V_i^{(R,0)}(s,t) & = &  c_{i,1}(s,t) \
          \mbox{Re}\Big(\  F_1^{(1l,0)}(t)\ \Pi^{(1l,0)}_0(t)
                        + F_1^{(1l,-1)}(t)\ \Pi^{(1l,1)}_0(t) \ \Big) \nn\\
                    & & + c_{i,2}(s,t) \
          \mbox{Re}\Big(\ F_1^{(1l,-1)}(t)\ \Pi^{(1l,0)}_0(t) \ \Big) \, .
\label{2lR}
\eea
The coefficients $c_{1,1}(s,t)$ and $c_{2,1}(s,t)$ are given in
Eq.~(\ref{coeffs1121}). Moreover, we have:
\be
c_{1,2}(s,t) = \frac{1}{2} ( st   + s^2  +  t^2 ) \ , \qquad
c_{2,2}(s,t) = \frac{1}{2} t^2
\label{coeffs1222} \, .
\ee
In Eqs.~(\ref{2lRa},\ref{2lR}) the quantities $\Pi^{(1l,i)}$
($i=0,~1$) and $F_1^{(1l,i)}$   ($i=-1,~0$) are the coefficients of
the Laurent expansion of the one-loop vacuum polarization function
and one-loop Dirac form factor given in
Eqs.~(\ref{A1lS0},~\ref{A1lS1}) and Eqs.~(\ref{F1ren0},~\ref{F1ren})
of Appendix \ref{AF}, respectively. Note that the form factor should
be calculated by keeping a nonzero electron mass as collinear
regulator.

\subsection{Interference of Two One-Loop Graphs \label{1Lx1LCorr}}

Finally, we discuss the interference between the one-loop vacuum
polarization diagrams of Fig.~\ref{fig1ltot}~(a) and (b), and the
one-loop vertex and box diagrams of Fig.~\ref{fig1ltot}~(c)--(j).
The calculation is straightforward because  the one-loop vacuum
polarization factorizes with respect to the tree-level amplitude and
the interference term can be obtained from the one-loop vertex and
box corrections to the cross section. For the vertex diagrams we
obtain
\bea
\frac{d \sigma_{2}^{V}}{d \Omega} \Big|_{(S,V)} &= & 2
\frac{\alpha^2}{s}Q_f^2 N_c \mbox{Re} \Bigg[ \frac{1}{s^2}
V_2^{(1l)}(t,s)\bigg( \Pi^{(1l)}_0(s) \bigg)^*
+ \frac{1}{t^2} V_2^{(1l)}(s,t)\bigg( \Pi^{(1l)}_0(t) \bigg)^* \nn\\
& &
+ \frac{1}{st} V_1^{(1l)}(s,t)\bigg( \Pi^{(1l)}_0(s) \bigg)^*
+ \frac{1}{st} V_1^{(1l)}(t,s)\bigg( \Pi^{(1l)}_0(t) \bigg)^* \Bigg] \ ,
\label{1lS1lV}
\eea
where the functions $V_i^{(1l)}(s,t)$ are defined as
\bea
V_1^{(1l)}(s,t) &=&  \Biggl[
2 \left( s t  + \frac{1}{2} s^2  + \frac{1}{2} t^2   \right)  +
\frac{1}{2} (D-4) ( st  + s^2 +  t^2 ) \Biggr]  \mbox{Re} F_1^{(1l)} (t)
 \, ,
 \label{VsU} \\
V_2^{(1l)}(s,t) & = & \left[2 \left(s t
 + s^2   + \frac{1}{2} t^2 \right)
+ \frac{1}{2} (D - 4) t^2 \right] \mbox{Re} F_1^{(1l)} (t) \, .
\label{VsD}
\eea
Their Laurent expansions read
\be
 V_i^{(1l)}(s,t) = \frac{1}{(D-4)}V_i^{(1l,-1)}(s,t)
                 + V_i^{(1l,0)}(s,t)
                 + {\mathcal{O}} \Big((D-4)\Big) \, ,
\label{ViLaur1l}
\ee
where
\bea
 V_i^{(1l,-1)}(s,t) &=& c_{i,1}(s,t) \mbox{Re}F_1^{(1l,-1)}(t) \, ,
\label{ViLaur1U} \\
 V_i^{(1l,0)}(s,t) &=& c_{i,1}(s,t) \mbox{Re}F_1^{(1l,0)}(t)
                     + c_{i,2}(s,t) \mbox{Re}F_1^{(1l,-1)}(t) \ ,
\label{ViLaur1D}
\eea
and with the coefficients $c_{i,j}(s,t)$ given in
Eqs.~(\ref{coeffs1121},\ref{coeffs1222}). Thus, if for instance we consider the
the first term of Eq.~(\ref{1lS1lV}), we obtain
\bea
V_2^{(1l)}(t,s)\bigg( \Pi^{(1l)}_0(s) \bigg)^* &=& \frac{1}{(D-4)}
V_2^{(1l,-1)}(t,s)\bigg( \Pi^{(1l,0)}_0(s) \bigg)^*
+ V_2^{(1l,0)}(t,s)\bigg( \Pi^{(1l,0)}_0(s) \bigg)^* \nn\\
 & &
+ V_2^{(1l,-1)}(t,s)\bigg( \Pi^{(1l,1)}_0(s) \bigg)^*
+ {\mathcal{O}} \Big((D-4)\Big) \, .
\label{1lS1lVLaur}
\eea
Similar expressions hold for the other terms.

For the box diagrams we obtain
\bea
\frac{d \sigma_{2}^{V}}{d \Omega} \Big|_{(S,B)}
\hspace*{-3mm} & = & \hspace*{-2mm}
    \frac{\alpha^2}{4s} Q_f^2 N_c \mbox{Re} \Bigg[
    \frac{m_f^2}{s} \bigg( B_1^{(1l)}(s,t)\! + \! B_2^{(1l)}(t,s)\!
+ \!B_3^{(1l)}(u,t) \!- \!B_2^{(1l)}(u,s) \bigg)
\bigg( \Pi^{(1l)}_0(s) \bigg)^* \nn\\
\hspace*{-3mm} & & \hspace*{-2mm} + \frac{m_f^2}{t} \bigg(
B_2^{(1l)}(s,t) \!+ \!B_1^{(1l)}(t,s)\! -\! B_2^{(1l)}(u,t) \!+\!
B_3^{(1l)}(u,s)  \bigg) \bigg( \Pi^{(1l)}_0(t) \bigg)^* \Biggr] \, ,
\label{1lS1lB}
\eea
where the functions $B_i^{(1l)}(s,t)$ ($i=1,2,3$) have the Laurent expansion:
\be
  B_i^{(1l)}(s,t) = \frac{1}{(D-4)^2} B_i^{(1l,-2)}(s,t)
                  + \frac{1}{(D-4)} B_i^{(1l,-1)}(s,t)
                  + B_i^{(1l,0)}(s,t)
                  + {\mathcal{O}} \left((D-4)\right) \ ,
\label{BiLaur}
\ee
The explicit expressions of the coefficients of the
Laurent expansion can be found in Appendix~\ref{AF}.
Thus, for instance for the first term in Eq.~(\ref{1lS1lB}) we obtain
\bea
B_1^{(1l)}(s,t)\bigg( \Pi^{(1l)}_0(s) \bigg)^* & = & \frac{1}{(D-4)}
  B_1^{(1l,-1)}(s,t)\bigg( \Pi^{(1l,0)}_0(s) \bigg)^*
+ B_1^{(1l,0)}(s,t)\bigg( \Pi^{(1l,0)}_0(s) \bigg)^* \nn\\
& &
+ B_1^{(1l,-1)}(s,t)\bigg( \Pi^{(1l,1)}_0(s) \bigg)^*
+ {\mathcal{O}} \Big((D-4)\Big) \ ,
\label{1lS1lBLaur}
\eea
and similar expressions for  the other terms.

\subsection{Soft Photon Emission}

Let us now discuss the calculation of the second term in
Eq.~(\ref{vps}). We use the procedure applied  in \cite{BFMRvB2} to
the case of the electron vacuum polarization. It is convenient to
introduce the quantity
\be
\frac{d \sigma_1^{D}}{d \Omega}\Big|_{(1l,S)}  = \frac{d
\sigma_1}{d \Omega} + (D-4)\frac{d \sigma_1^{(D-4)}}{d \Omega} \, ,
\label{newV}
\ee
where the first term on  the right hand side (r.h.s.)  is defined in
Eq.~(\ref{amp1lS})  and  the second term  for  $m_e=0$ reads
\bea
\frac{d \sigma_{1}^{(D-4)}}{d \Omega}
\Big|_{(1l,S)} \hspace*{-2mm} &=& \hspace*{-2mm}
\frac{\alpha^2}{s}Q_f^2 N_c \Bigg\{ \frac{1}{s^2}\left[s t +
\frac{s^2}{2} + t^2\right] 2 \mbox{Re}\Pi^{(1l,1)}_0(s) +
\frac{1}{t^2}\left[s t + \frac{t^2}{2} + s^2\right]
2 \Pi^{(1l,1)}_0(t)  \nn\\
\hspace*{-2mm} & & \hspace*{-2mm}
+ \frac{1}{s t}(s + t)^2
 \left(\mbox{Re}\Pi^{(1l,1)}_0(s) +
  \Pi^{(1l,1)}_0(t)\right)
+ \frac{\mbox{Re}\Pi^{(1l,0)}_0(s) + \Pi^{(1l,0)}_0(t)}{2} \nn \\
\hspace*{-2mm} & &\hspace*{-2mm}
+ \frac{1}{2} \left[ (s +t)^2 \! - \! s t
\right] \left(\mbox{Re}\Pi^{(1l,0)}_0(s) +
  \Pi^{(1l,0)}_0(t)\right)
\Bigg\} \, .
\label{amp1lSorderep}
\eea
The contribution of the soft-photon emission is then given by
\be
\frac{d \sigma^{S}_2}{d \Omega} = \left(\frac{d \sigma_1^{D}}{d
\Omega} \right) \, 4 \sum_{j=1}^{4}
 J_{1j}(p_1 \cdot p_j, m_e^2,m_f^2) \, . \label{above2}
\ee
Here, the infrared divergent quantities $J_{1j}$ ($j =1,\cdots,4$)
are defined as follows:
\be
J_{1j} (p_1 \cdot p_j, m_e^2,m_f^2) = \epsilon_j \, \left(p_1 \cdot p_j
\right) \, I_{1j}(p_1 \cdot p_j, m_e^2,m_f^2) \, ,
\ee
where $\epsilon_j = +1$ for $j=1,4$,  $\epsilon_j = -1$ for $j=2,3$,
and
\be
I_{1j}(p_1 \cdot p_j, m_e^2,m_f^2)  = \frac{1}{\Gamma\left(3-\frac{D}{2}\right)
\pi^{(D-4)/2}}  \frac{m_f^{D-4}}{4 \pi^2} \int^{\omega} \frac{d^D k}{k_0}
\frac{1}{(p_1 \cdot k) (p_j \cdot k)} \, .
\label{Iij}
\ee
The calculation of the integrals $I_{1j}$ follows the procedure
outlined in \cite{BardinPassarino} and is described  in detail in
Appendix~A of \cite{BFMRvB2}. We need only the leading terms of the
small electron mass expansion of these integrals, which  to
${\mathcal O} \left( (D-4)^0 \right)$   read
\bea
I_{11} & = &
          \frac{1}{2 m_e^2} \left[ \frac{2}{D-4}
    + \ln\left( \frac{4 \omega^2}{s} \right)
    + \ln\left( \frac{s}{m_f^2}\right)
    - \ln\left( \frac{s}{m_e^2}\right) \right]
    + {\mathcal O}\left( m_e^0\right) \, , \\
I_{12} & = &
        \frac{1}{s} \Biggl[
          \ln\left(\frac{s}{m_e^2}\right)\left(\frac{2}{D-4}
            + \ln\left(\frac{4 \omega^2}{s}\right)
                + \ln\left(\frac{s}{m_f^2}\right)\right)
        - \frac{1}{2}\ln^2\left(\frac{s}{m_e^2}\right)
        - 2 \zeta(2) \Biggr] \nn \\
& &
    + {\mathcal O}\left( \frac{m_e^2}{s} \right) \, , \\
I_{13} & = &
        \frac{1}{t} \Biggl[
      \ln\left(-\frac{t}{m_e^2}\right) \left( \frac{2}{D-4}
            + \ln\left(\frac{4 \omega^2}{s}\right)
                + \ln \left(\frac{s}{m_f^2}\right)\right)
    - \frac{1}{2}\ln^2\left(\frac{s}{m_e^2}\right)
    - \Li_2\left(1+\frac{s}{t}\right) \nn \\
& &
        - 2 \zeta(2) \Biggr]
    + {\mathcal O}\left( \frac{m_e^2}{t} \right) \, , \\
I_{14} & = &
        \frac{1}{u} \Biggl[
      \ln\left(-\frac{u}{m_e^2}\right) \left( \frac{2}{D-4}
            + \ln\left(\frac{4 \omega^2}{s}\right)
        + \ln \left(\frac{s}{m_f^2}\right)\right)
    - \frac{1}{2} \ln^2\left(\frac{s}{m_e^2}\right)
    - \Li_2\left(1+\frac{s}{u}\right) \nn \\
& &
        - 2 \zeta(2) \Biggr]
    + {\mathcal O}\left( \frac{m_e^2}{u} \right) \, .
\eea
Note that  the term proportional to $(D-4)$ in Eq.~(\ref{newV})
gives a finite contribution to Eq.~(\ref{above2}), since $J_{1j}$
contain an infrared pole.

\section{Numerical Analysis \label{num}}

In this Section we consider the application of our result to the
phenomenologically interesting cases relevant for physics at
DA$\Phi$NE and  the ILC. We provide a detailed account of the pure
QED contribution, extend the analysis to the mixed QED-QCD
corrections, and  give an estimate of the hadronic vacuum
polarization effect. All the terms involving the logarithm of the IR
cut-off $\omega$, $\ln(4 \omega^2/s)$, are excluded from the
numerical estimates  because the corresponding contribution
critically depends on the event selection algorithm and cannot be
unambiguously estimated without imposing specific cuts on the photon
bremsstrahlung.  The actual impact of the two-loop virtual
corrections on the theoretical predictions can be determined only
after the result of the paper is consistently implemented into the
Monte-Carlo event generators. Nevertheless, the above na\"ive
procedure can be used to get a rough estimate of the magnitude and
the structure of the corrections.

\begin{figure}
\bc
\begin{picture}(0,0)%
\includegraphics{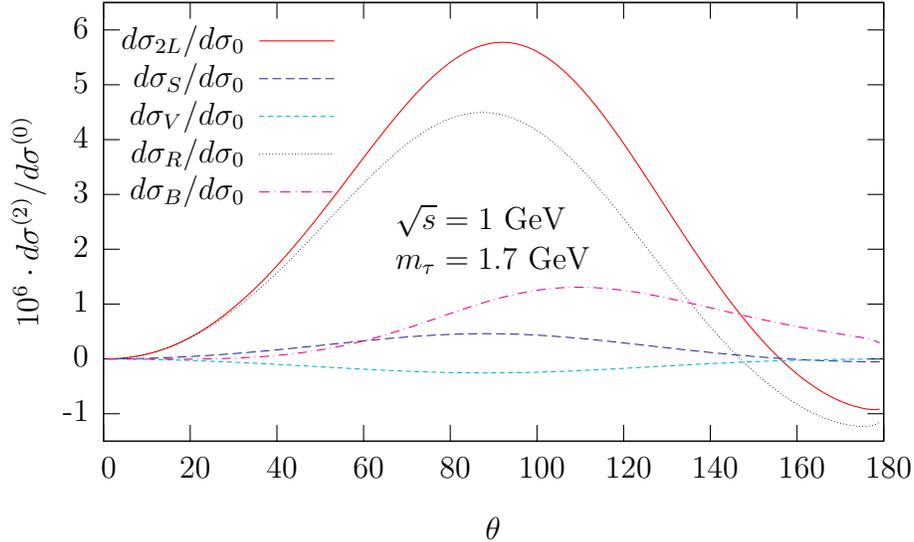}%
\end{picture}%
\begingroup
\setlength{\unitlength}{0.0200bp}%
\begin{picture}(18000,10800)(0,0)%
\put(2200,2496){\makebox(0,0)[r]{\strut{}-1}}%
\put(2200,3530){\makebox(0,0)[r]{\strut{} 0}}%
\put(2200,4564){\makebox(0,0)[r]{\strut{} 1}}%
\put(2200,5598){\makebox(0,0)[r]{\strut{} 2}}%
\put(2200,6631){\makebox(0,0)[r]{\strut{} 3}}%
\put(2200,7665){\makebox(0,0)[r]{\strut{} 4}}%
\put(2200,8699){\makebox(0,0)[r]{\strut{} 5}}%
\put(2200,9733){\makebox(0,0)[r]{\strut{} 6}}%
\put(2475,1429){\makebox(0,0){\strut{} 0}}%
\put(4108,1429){\makebox(0,0){\strut{} 20}}%
\put(5742,1429){\makebox(0,0){\strut{} 40}}%
\put(7375,1429){\makebox(0,0){\strut{} 60}}%
\put(9008,1429){\makebox(0,0){\strut{} 80}}%
\put(10642,1429){\makebox(0,0){\strut{} 100}}%
\put(12275,1429){\makebox(0,0){\strut{} 120}}%
\put(13908,1429){\makebox(0,0){\strut{} 140}}%
\put(15542,1429){\makebox(0,0){\strut{} 160}}%
\put(17175,1429){\makebox(0,0){\strut{} 180}}%
\put(9825,275){\makebox(0,0){\strut{} $\theta$ }}%
\put(7783,6115){\makebox(0,0)[l]{\strut{} $\sqrt{s}=1$~GeV }}%
\put(7783,5380){\makebox(0,0)[l]{\strut{} $m_{\tau}=1.7$~GeV }}%
\put(5140,9526){\makebox(0,0)[r]{\strut{}$d\sigma_{2L}/d\sigma_0$}}%
\put(5140,8814){\makebox(0,0)[r]{\strut{}$d\sigma_{S}/d\sigma_0$}}%
\put(5140,8102){\makebox(0,0)[r]{\strut{}$d\sigma_{V}/d\sigma_0$}}%
\put(5140,7390){\makebox(0,0)[r]{\strut{}$d\sigma_{R}/d\sigma_0$}}%
\put(5140,6678){\makebox(0,0)[r]{\strut{}$d\sigma_{B}/d\sigma_0$}}%
\put(1100,6114){\rotatebox{90}{\makebox(0,0){\strut{} $10^6 \cdot d\sigma^{(2)}/d\sigma^{(0)}$ }}}%
\end{picture}%
\endgroup
\caption{\it{Self-energy (``{\it S}''), vertex (``{\it V}''), reducible plus
one-loop times one-loop (``{\it R}''), and box (``{\it B}'') contributions to
the two-loop $\tau$-lepton correction to the  differential cross section of
Bhabha scattering at $\sqrt{s} = 1$~GeV.} \label{fig7}}
\ec
\end{figure}

As a first application, we consider the Bhabha scattering at $\sqrt{s}=1$~GeV.
The latter  is the value of  the center-of-mass energy of the KLOE experiment at
DA$\Phi$NE, which plays a crucial role in the determination of the hadronic
vacuum polarization contribution to the muon anomalous magnetic moment
\cite{Alo,Eid}. The anatomy of the heavy-flavor two-loop correction for this
choice of the center-of-mass energy is shown in Fig.~\ref{fig7}, where the case
in which the heavy fermion is the $\tau$-lepton is considered. Each curve
plotted in Fig.~\ref{fig7} represents  a specific subset of the  virtual
corrections and the corresponding soft emission. The dominant contribution
originates from the two-loop reducible corrections and from the product of the
one-loop corrections, which are considered together (see Sections \ref{RedCorr}
and \ref{1Lx1LCorr}).  This subset  is numerically dominant because it contains
all the large collinear logarithms $\ln(s/m_e^2)$. The numerical values of the
ratio of the second-order heavy-flavor  corrections to the Born QED cross
section (Eq.~(\ref{TL})) are collected in Table~\ref{tab1}.\footnote{ Note that the numerical
evaluation of our analytic formulas is done with double Fortran precision but we
do not present here all the available significant digits} We separately consider the
contributions of  muon, $\tau$-lepton, $c$-quark and $b$-quark. Since at KLOE
one is particularly interested in the large angle scattering events, we
considered the angular range between $50$ and $130$ degrees. For comparison we
also give the  value  of the electron vacuum polarization
contribution.\footnote{ The electron contribution includes the logarithmic part
of the soft-pair production. The logarithms of the soft-pair cut-off are
excluded from the numerical estimates \cite{BF}.} The following input
parameters   are used: $m_e=0.510998902$~MeV, $m_{\mu}=0.105658369$~GeV,
$m_{\tau} = 1.7$~GeV, $m_{c}=1.25$~GeV, and $m_{b}=4.7$~GeV. The contributions
of the $\tau$-lepton, $c$- and $b$-quark are suppressed with respect to the
muon at least by one order of magnitude. The total heavy-flavor contribution is
dominated by the muon loop and it reaches 0.45 permille in magnitude at
$\theta\sim 140^\circ$.
\begin{table}
\bc
{\rotatebox{90}{\makebox(0,0){\strut{} $\sqrt{s}=1$~GeV}}}
\hspace*{3mm}
\begin{tabular}{|c|c|c|c|c|c|c|}
\hline
$\theta$ & $e$ ($10^{-4}$) & $\mu$ ($10^{-4}$) & $c$ ($10^{-4}$)  &
$\tau$ ($10^{-4}$) & $b$ ($10^{-4}$) \cr
\hline
\hline
$50^{\circ}$ & 17.341004 & 1.7972877 & 0.0622677 & 0.0264013 & 0.0010328 \cr
\hline
$60^{\circ}$ & 18.407836 & 2.2267654 & 0.0861876 & 0.0367058 & 0.0014184 \cr
\hline
$70^{\circ}$ & 19.438718 & 2.6504950 & 0.1086126 & 0.0465329 & 0.0018907 \cr
\hline
$80^{\circ}$ & 20.465455 & 3.0655973 & 0.1253094 & 0.0540991 & 0.0022442 \cr
\hline
$90^{\circ}$ & 21.463240 & 3.4581845 & 0.1321857 & 0.0576348 & 0.0024428 \cr
\hline
$100^{\circ}$ & 22.366427 & 3.8070041 & 0.1268594 & 0.0560581 & 0.0024304 \cr
\hline
$110^{\circ}$ & 23.099679 & 4.0922189 & 0.1098317 & 0.0495028 & 0.0022024 \cr
\hline
$120^{\circ}$ & 23.605216 & 4.3030725 & 0.0843311 & 0.0392810 & 0.0018086 \cr
\hline
$130^{\circ}$ & 23.847394 & 4.4392717 & 0.0549436 & 0.0273145 & 0.0013297 \cr
\hline
\end{tabular}
\caption{\it{The second-order electron, muon,
$c$-quark, $\tau$-lepton, and $b$-quark QED contributions to the
differential cross section of Bhabha scattering at $\sqrt{s}=1$~GeV
in units of $10^{-4}$ of the Born cross section.} \label{tab1}}
\ec
\end{table}
Note that the energy under consideration is sufficiently below the
quarkonium threshold so that $\sqrt{4m_f^2-s}\gg \Lambda_{QCD}$ and
the heavy quarks can be treated perturbatively. Moreover, for the
heavy-quark vacuum polarization in two-loop approximation one has to
take into account the first order corrections in the strong coupling
constant $\alpha_s$ due to a gluon exchange inside the quark loop.
The resulting ${\mathcal O}(\alpha \alpha_s)$ correction to the
Bhabha cross section  can be obtained from the QED contribution,
Eq.~(\ref{virtualCS}), by adjusting the overall factor:
\bea
\frac{d \sigma_{2}^{V}}{d \Omega} \Big|^{QCD}_{ (2l,S)} \! \! \!
& = & \! \!  \frac{C_F}{Q_f^2} \, \frac{\alpha_s(m_f^2)}{\alpha} \,
\frac{d \sigma_{2}^{V}}{d \Omega} \Big|_{ (2l,S)}  \, .
\label{amp2lSqcd}
\eea
$C_F=(N_c^2-1)/(2N_c)$ is the Casimir operator of the
fundamental  representation of the $SU(N_c)$ color group,  and the
strong coupling constant is evaluated at the scale $\mu = m_f$,
using the NLO RG equation with the appropriate number of active
quarks, starting from the input value $\alpha_S(M_Z)=0.118$. The
numerical results for the  ${\mathcal O}(\alpha \alpha_s)$
corrections are listed in the first and the second columns in
Table~\ref{tab5}. At the same time, the contribution of the light
$u$-, $d$-, and $s$-quark to the vacuum polarization is
non-perturbative, due to the hadronization effects. In principle, it
requires a special treatment based on the integration of the
experimentally measured spectral density within the dispersion
relation method (see {\it e.g.} \cite{KKKS}), as it was done in
\cite{Actis:2007}. However, this contribution can be estimated by
na\"ive use of the perturbative result with effective light quark
masses. Such estimates are normally in good agreement with the
result of the rigorous analysis. To estimate the hadronic
contribution at two-loop level we use, for the three light quarks
$u$, $d$ and $s$, the value $m_u=m_d=m_s=m_{eff} \sim 180$~MeV
adopted to describe in the lowest order the hadronic contribution to
the muon anomalous magnetic moment  \cite{Pivovarov}. The numerical
results for the light-quark contribution at KLOE energies are
included in the third column of Table~\ref{tab5}. It is comparable to
the contribution of the muon ({\it c.f.} Table~\ref{tab1}). Note that at
$\theta =90^{\circ}$ our estimate reproduces  the value given in
\cite{Actis:2007} with  20\% accuracy, sufficient for
phenomenological applications at DA$\Phi$NE.

\begin{table}
\bc {\rotatebox{90}{\makebox(0,0){\strut{} $\sqrt{s}=1$~GeV }}}
\hspace*{3mm}
\begin{tabular}{|c|c|c|c|}
\hline
$\theta$ & $b$ ($10^{-4}$) & $c$ ($10^{-4}$) & $u+s+d$ ($10^{-4}$)  \cr
\hline
\hline
$50^{\circ}$  & 0.0040026 & 0.3605297 & 1.8 \cr
\hline
$60^{\circ}$  & 0.0053684 & 0.4787401 & 2.4 \cr
\hline
$70^{\circ}$  & 0.0065839 & 0.5795114 & 3.0 \cr
\hline
$80^{\circ}$  & 0.0074290 & 0.6427167 & 3.6 \cr
\hline
$90^{\circ}$  & 0.0077240 & 0.6528096 & 4.2 \cr
\hline
$100^{\circ}$ & 0.0073994 & 0.6051311 & 4.7 \cr
\hline
$110^{\circ}$ & 0.0065277 & 0.5082196 & 5.1 \cr
\hline
$120^{\circ}$ & 0.0052908 & 0.3802358 & 5.4 \cr
\hline
$130^{\circ}$ & 0.0039094 & 0.2421310 & 5.6 \cr
\hline
\end{tabular}
\caption{\it{The second-order ${\mathcal O}(\alpha \alpha_s)$
contribution of $b$ and  $c$ quarks, and the light-quark
(hadronic) contribution to the differential cross section of Bhabha
scattering at $\sqrt{s}=1$~GeV in units of $10^{-4}$ of the Born
cross section.} \label{tab5}} \ec
\end{table}

\begin{figure}
\bc
\begin{picture}(0,0)%
\includegraphics{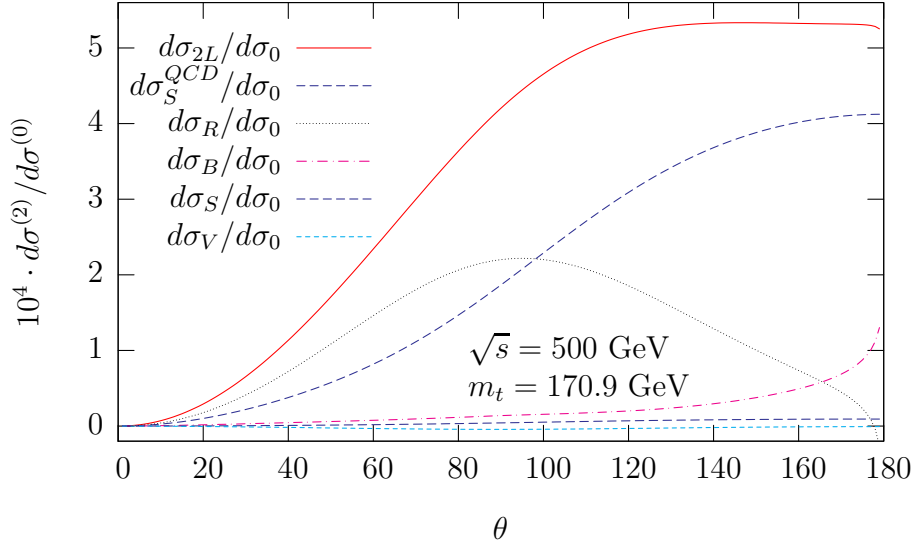}%
\end{picture}%
\begingroup
\setlength{\unitlength}{0.0200bp}%
\begin{picture}(18000,10800)(0,0)%
\put(2475,2264){\makebox(0,0)[r]{\strut{} 0}}%
\put(2475,3690){\makebox(0,0)[r]{\strut{} 1}}%
\put(2475,5116){\makebox(0,0)[r]{\strut{} 2}}%
\put(2475,6542){\makebox(0,0)[r]{\strut{} 3}}%
\put(2475,7968){\makebox(0,0)[r]{\strut{} 4}}%
\put(2475,9394){\makebox(0,0)[r]{\strut{} 5}}%
\put(2750,1429){\makebox(0,0){\strut{} 0}}%
\put(4353,1429){\makebox(0,0){\strut{} 20}}%
\put(5956,1429){\makebox(0,0){\strut{} 40}}%
\put(7558,1429){\makebox(0,0){\strut{} 60}}%
\put(9161,1429){\makebox(0,0){\strut{} 80}}%
\put(10764,1429){\makebox(0,0){\strut{} 100}}%
\put(12367,1429){\makebox(0,0){\strut{} 120}}%
\put(13969,1429){\makebox(0,0){\strut{} 140}}%
\put(15572,1429){\makebox(0,0){\strut{} 160}}%
\put(17175,1429){\makebox(0,0){\strut{} 180}}%
\put(9962,275){\makebox(0,0){\strut{} $\theta$ }}%
\put(9161,3690){\makebox(0,0)[l]{\strut{} $\sqrt{s}=500$~GeV }}%
\put(9161,2977){\makebox(0,0)[l]{\strut{} $m_t=170.9$~GeV }}%
\put(5841,9394){\makebox(0,0)[r]{\strut{}$d\sigma_{2L}/d\sigma_0$}}%
\put(5841,8682){\makebox(0,0)[r]{\strut{}$d\sigma_{S}^{QCD}/d\sigma_0$}}%
\put(5841,7970){\makebox(0,0)[r]{\strut{}$d\sigma_{R}/d\sigma_0$}}%
\put(5841,7258){\makebox(0,0)[r]{\strut{}$d\sigma_{B}/d\sigma_0$}}%
\put(5841,6546){\makebox(0,0)[r]{\strut{}$d\sigma_{S}/d\sigma_0$}}%
\put(5841,5834){\makebox(0,0)[r]{\strut{}$d\sigma_{V}/d\sigma_0$}}%
\put(1100,6114){\rotatebox{90}{\makebox(0,0){\strut{} $10^4 \cdot d\sigma^{(2)}/d\sigma^{(0)}$ }}}%
\end{picture}%
\endgroup
\caption{\it{QED and QCD self-energy (``$S$''), vertex (``$V$''), reducible
plus one-loop times one-loop (``$R$'') and box (``$B$'') contribution to the
two-loop top-quark corrections to the differential cross section of Bhabha
scattering at $\sqrt{s} = 500$~GeV.} \label{fig8bis}}
\ec
\end{figure}
\begin{table}
\bc
{\rotatebox{90}{\makebox(0,0){\strut{} $\sqrt{s}=m_Z$ }}}
\hspace*{3mm}
\begin{tabular}{|c|c|c|c|c|c|}
\hline
$\theta$ & $e$ ($10^{-3}$) & $\mu$ ($10^{-3}$) & $\tau$ ($10^{-3}$) & $t$ ($10^{-3}$)  &
$b+c+s+d+u$ ($10^{-3}$) \cr
\hline
\hline
$1^{\circ}$ & 2.1829158 & 0.3571385 & 0.0064077 & 0.0000043 & 0.93 \cr
\hline
$2^{\circ}$ & 2.6802340 & 0.5863583 & 0.0265766 & 0.0000151 & 1.4 \cr
\hline
$3^{\circ}$ & 2.9952706 & 0.7349688 & 0.0573071 & 0.0000348 & 1.8 \cr
\hline
$50^{\circ}$ & 5.5606265 & 1.7801664 & 0.9801997 & 0.0099775 & 5.2 \cr
\hline
$60^{\circ}$ & 5.7514057 & 1.8360794 & 1.0669528 & 0.0134970 & 5.5 \cr
\hline
$70^{\circ}$ & 5.9332685 & 1.8891102 & 1.1471133 & 0.0166760 & 5.7 \cr
\hline
$80^{\circ}$ & 6.1126124 & 1.9410784 & 1.2228401 & 0.0189268 & 5.9 \cr
\hline
$90^{\circ}$ & 6.2857772 & 1.9882767 & 1.2925934 & 0.0197396 & 6.1 \cr
\hline
$100^{\circ}$ & 6.4419768 & 2.0231625 & 1.3524844 & 0.0188814 & 6.2 \cr
\hline
$110^{\circ}$ & 6.5686702 & 2.0374326 & 1.3983882 & 0.0165063 & 6.3 \cr
\hline
$120^{\circ}$ & 6.6561657 & 2.0245734 & 1.4276615 & 0.0130914 & 6.3 \cr
\hline
$130^{\circ}$ & 6.6983373 & 1.9800878 & 1.4394412 & 0.0092460 & 6.2 \cr
\hline
\end{tabular}
\caption{\it The second-order electron, muon, $\tau$-lepton,
top-quark, and light-quark contributions to the Bhabha scattering
differential cross section for the Giga-Z option of the ILC and LEP1
center-of-mass energy $\sqrt{s}=M_Z$. The numbers are given in
units of $10^{-3}$ of the Born cross section. The top-quark
contribution includes the ${\mathcal O}(\alpha \alpha_s)$ term.
\label{tab2}}
\ec
\end{table}

Let us now discuss Bhabha scattering at high energies, characteristic to the
ILC. We consider two cases: the Giga-Z option with $\sqrt{s} =
M_Z$,\footnote{The consideration we make about Giga-Z are valid for LEP1 as
well.} and $\sqrt{s} = 500$~GeV.
In the first case, we consider the contributions of the leptons and
the top quark perturbatively and  give an estimate of the
non-perturbative hadronic contribution due to the five ``light''
quarks, $u$, $d$, $s$, $c$ and $b$, using the effective quark  mass
approach.   We use the following values for the effective masses:
$m_u=m_d=m_s=m_{eff} \sim 66$~MeV, $m_c=1.25$~GeV, and
$m_b=4.7$~GeV. These values were  adopted to describe the lowest
order contribution to $\alpha(M_Z)$ \cite{LEP1}. The  numerical
results are collected  in Table~\ref{tab2}. The hadronic
contribution is of the same size as the electron vacuum polarization
contribution and exceeds the one of the muon.  Note that our
estimate reproduces   the result of \cite{Actis:2007} at the
scattering angles $90^{\circ}$ and $3^{\circ}$  with 25\% and 10\%
accuracy, respectively. For the small angle scattering, which is of
primary interest, our result provides the necessary accuracy of 0.1
permille.

For $\sqrt{s} = 500$~GeV, we consider only the contributions of the
leptons and the top  quark. The large electroweak decay width of the
top quark serves as an infrared regulator and it suppresses the
hadronization effects. Therefore, the perturbative result is applicable
at energies near and above the top-antitop threshold. At the same time,
the effective mass approach  for the contribution of lighter quarks
is not reliable at this energy. The corrections become sizable due
to the logarithmically  growing terms and and a rough estimate  with
an error of about 25\%, is not accurate enough to match the
luminosity precision requirements.
Different  subsets of the second order  correction due to the
top-quark vacuum polarization to the Bhabha scattering cross section
are  plotted separately in Fig.~\ref{fig8bis}. In this figure we
include also  the ${\mathcal O}(\alpha \alpha_s)$ corrections to the
cross section described by the term in  Eq.~(\ref{amp2lSqcd}). The
numerical results for the two-loop corrections due to electron, muon,
$\tau$-lepton, and top quark vacuum polarizations for a
center-of-mass energy of $\sqrt{s} = 500$~GeV are collected in
Table~\ref{tab4}. Note that the contribution of muon and
$\tau$-lepton can be significantly reduced if one takes into account
the corresponding soft pair emission, which is justified at this
energy from an experimental point of view.

\begin{table}
\bc
{\rotatebox{90}{\makebox(0,0){\strut{} $\sqrt{s}=500$~GeV }}}
\hspace*{3mm}
\begin{tabular}{|c|c|c|c|c|}
\hline
$\theta$ & $e$ ($10^{-3}$) & $\mu$ ($10^{-3}$) & $\tau$ ($10^{-3}$) & $t$ ($10^{-3}$) \cr
\hline
\hline
$1^{\circ}$ & 3.4957072 & 0.9690710 & 0.1542329 & 0.0000575 \cr
\hline
$2^{\circ}$ & 4.1203687 & 1.2491270 & 0.3573661 & 0.0002466 \cr
\hline
$3^{\circ}$ & 4.5099086 & 1.4146106 & 0.5140242 & 0.0005763 \cr
\hline
$50^{\circ}$ & 7.5740980 & 2.3185800 & 1.8411736 & 0.1707137 \cr
\hline
$60^{\circ}$ & 7.7965875 & 2.3446744 & 1.9274750 & 0.2340996 \cr
\hline
$70^{\circ}$ & 8.0081541 & 2.3708714 & 2.0072240 & 0.2998535 \cr
\hline
$80^{\circ}$ & 8.2164081 & 2.3981523 & 2.0829886 & 0.3635031 \cr
\hline
$90^{\circ}$ & 8.4172449 & 2.4207950 & 2.1521199 & 0.4202418 \cr
\hline
$100^{\circ}$ & 8.5982864 & 2.4282953 & 2.2085332 & 0.4655025 \cr
\hline
$110^{\circ}$ & 8.7451035 & 2.4090920 & 2.2456055 & 0.4979010 \cr
\hline
$120^{\circ}$ & 8.8465287 & 2.3536259 & 2.2585305 & 0.5181602 \cr
\hline
$130^{\circ}$ & 8.8954702 & 2.2543834 & 2.2446158 & 0.5287459 \cr
\hline
\end{tabular}
\caption{\it{The second-order electron, muon, $\tau$-lepton, and
top-quark contributions to the differential cross section of Bhabha
scattering at $\sqrt{s}=500$~GeV in units of $10^{-3}$ of the
Born cross section. The top-quark contribution includes the
${\mathcal O}(\alpha \alpha_s)$ term.} \label{tab4}} \ec
\end{table}

\section{Conclusions \label{conclusions}}

In the present work we  derived the two-loop radiative corrections to Bhabha
scattering due to the  vacuum polarization by the virtual heavy-flavor
fermion-antifermion pairs. This completes the QED analysis of the process at the
two-loop level.  The result is valid for arbitrary ratio of the Mandelstam
invariants  to the heavy fermion mass, provided all these quantities are large
compared to the electron mass.  The corrections to the cross section are
expressed in terms of ordinary harmonic polylogarithms and Nielsen
polylogarithms of ratios of polynomials in $s$ and $t$ invariants. Thus we have
complete control over its analytic properties and numerical  evaluation. The
analytical result for the total two-loop QED corrections, which includes also
the photonic and the electron vacuum polarization contributions
\cite{Pen,BFMRvB1,BFMRvB2}, is now available \cite{file}.

We suggested a new approach which reduces the number of mass scales
in the most complicated part of the calculation. The approach is
based on the general properties of the infrared and collinear
divergencies and could be useful for the high-order perturbative
calculations in a wide class of processes with a clear mass
hierarchy.

The numerical impact of the perturbative  second-order heavy-flavor
corrections, including the  ${\mathcal O}(\alpha \alpha_s)$
contribution for heavy quarks, was studied for the  KLOE experiment
at DA$\Phi$NE as well as for the GigaZ and the high-energy options
of the ILC. For the first two applications we also  provided an
estimate of the hadronic vacuum polarization contribution.

Our result is crucial for the high-precision physics at
electron-positron colliders.  It removes the last piece of pure
theoretical uncertainty in luminosity determination at the
low-energy accelerators  and gives the proper account for the top
quark effects at the ILC.  The accuracy of the luminosity
determination at low-energy accelerators is now restricted only by
the precision of the Monte-Carlo event generators for the hard
photon and electron-positron pair emission. To achieve such an
accuracy for the large angle scattering at the ILC one has to
perform more careful analysis of the hadronic contribution
\cite{Actis:2007} and take into account also yet unknown two-loop
electroweak corrections.\footnote{In the case of
$e^+e^-\to\mu^+\mu^-$ annihilation the two-loop electroweak
corrections enhanced by powers of the large logarithm
$\ln(M_{W,Z}^2/s)$, which  are dominant for $\sqrt{s}\grtsim 500$~GeV,
were computed in \cite{Sud}. This analysis can be generalized
to the Bhabha scattering by adding the $t$-channel contribution.}

\subsection*{Acknowledgements}

We are grateful to J.~Vermaseren and D. Maitre for their kind assistance in the
use of {\tt FORM} \cite{FORM}, and of the {\tt Mathematica} packages {\tt HPL}
and {HypExp} \cite{Daniel1,Daniel2}.  R.~B. would like to thank D. Greynat for
useful discussions about Mellin-Barnes \cite{david}, the Galileo Galilei
Institute for Theoretical Physics for the hospitality and the INFN for partial
support. The work of R.~B. was partially
supported by Ministerio de Educaci\'on y Ciencia (MEC) under grant
FPA2004-00996, Generalitat Valenciana under grant GVACOMP2007-156, European
Commission under the grant MRTN-CT-2006-035482 (FLAVIAnet), and MEC-INFN
agreement. The work of A.~F. was supported  by the Swiss National Science
Foundation (SNF) under contract 200020-117602.
%


\appendix

\section{Generalized Harmonic Polylogarithms \label{GHPLs}}

For $m_e=0$ the two-loop corrections are  function of $s$, $t$, $u$,
and $m_f$. It is convenient to use the dimensionless variables $x$
and  $y$ given by Eq.~(\ref{xy}), and $z=R^2/m_f^2 = -u/m_f^2$.
To make the formulas as compact as possible we introduce six {\em
rescaled} dimensionless quantities. In particular, in the non
physical region $s<0$, we define:
\bea
\xr &=& \frac{\sqrt{P^2 + 4 m_f^2} - \sqrt{P^2}}{\sqrt{P^2 + 4 m_f^2} +
\sqrt{P^2}} \, , \quad
\frac{s}{m_f^2} = - \frac{(1-\xr)^2}{\xr} \, ,
\label{xr} \\
\yr &=& \frac{\sqrt{Q^2 + 4 m_f^2} - \sqrt{Q^2}}{\sqrt{Q^2 + 4 m_f^2} +
\sqrt{Q^2}} \, ,\quad
\frac{t}{m_f^2} = - \frac{(1-\yr)^2}{\yr} \, ,
\label{yr} \\
\zr &=& \frac{\sqrt{R^2 + 4 m_f^2} - \sqrt{R^2}}{\sqrt{R^2 + 4 m_f^2} +
\sqrt{R^2}} \, , \quad
\frac{u}{m_f^2} = - \frac{(1-\zr)^2}{\zr} \, ,
\label{zr} \\
\xb &=& \frac{\sqrt{P^2} - \sqrt{P^2 - 4 m_f^2}}{\sqrt{P^2 } +
\sqrt{P^2 - 4 m_f^2}} \, , \quad
\frac{s}{m_f^2} = - \frac{(1+\xb)^2}{\xb} \, ,
\label{xb} \\
\yb &=& \frac{\sqrt{Q^2} - \sqrt{Q^2- 4 m_f^2 }}{\sqrt{Q^2} +
\sqrt{Q^2- 4 m_f^2}} \, ,\quad
\frac{t}{m_f^2} = - \frac{(1+\yb)^2}{\yb} \, ,
\label{yb} \\
\zb &=& \frac{\sqrt{R^2} - \sqrt{R^2- 4 m_f^2}}{\sqrt{R^2} +
\sqrt{R^2- 4 m_f^2}} \, , \quad
\frac{u}{m_f^2} = - \frac{(1+\zb)^2}{\zb} \, .
\label{zb}
\eea
The  rescaled variable $\yr$ is  positive for $y > 0$, while the
rescaled variable $\yb$ is  positive for $y > 4$ and is a pure phase
for  $0 <y < 4$. These properties hold also for the variables
$z,~\zr,~\zb$ and $x,~\xr,~\xb$.
The analytical result for the  cross section can be expressed in
terms of a suitable set of GHPLs depending on two of the three
variables $x,~y,$ and $z$. Let us consider the pair of variables $x$
and $y$. In this case the  set of the weight functions is:
\bea
& & f_0(x) = \frac{1}{x} \, ,
\hspace*{11mm} f_{-\mu}(x) = \frac{1}{\sqrt{x (4+x)}} \, ,
\hspace*{6mm} f_{\mu}(x) = \frac{1}{\sqrt{x (4-x)}} \, , \nn \\
& & \hspace*{-2.5mm}f_{-y}(x) = \frac{1}{x+y} \, ,
\quad f_{-4}(x) = \frac{1}{x+4} \, ,
\hspace*{9.5mm} f_{-y-\mu}(x) = \frac{1}{(x+y) \sqrt{x (4+x)}} \, .
\eea
The GHPLs of weight one are defined as follows:
\be
G(0 ; x) = \ln(x) \, , \quad G(a ; x) = \int_0^x dt f_a(t) \, ;
\ee
while GHPLs of higher weight are defined by the iterated integration
\be
G(a, \cdots ; x) = \int_0^x dt f_a(t) G(\cdots; t) \, ,
\ee
with the only exception of the weight zero GHPLs, which are defined
as follows:
\be
G(\underbrace{0, \cdots, 0}_{n}; x) = \frac{1}{n!} \ln^n(x) \, .
\ee
The GHPLs  defined in this way satisfy the usual shuffle algebra
\cite{HPLs}. The subset of GHPLs which do not involve the weights
$-y$ and $-y-\mu$ can be expressed in terms of the ordinary harmonic
polylogarithms (HPLs) of the arguments $\xr$ or $\yr$, and weights
$1$, $0$, and $-1$. The table of transformations useful for our
calculation reads
\bea
G(0;y) &=&
      - H(0; y_r)
         - 2 H(1; y_r)
        \, ,  \\
 G(-\mu;y) &=&
      - H(0; y_r)
        \, ,  \\
 G(-4;y) &=&
      - 2 \ln(2)
         + 2 H(-1; y_r)
         - H(0; y_r)
        \, ,  \\
G(0,0;y) &=&
       H(0,0; y_r)
         + 2 H(0,1; y_r)
         + 2 H(1,0; y_r)
         + 4 H(1,1; y_r)
        \, ,  \\
G(-\mu,-\mu;y) &=&
       H(0,0; y_r)
        \, ,  \\
G(-4,-\mu;y) &=&
      - \zeta(2)
         - 2 H(-1,0; y_r)
         + H(0,0; y_r)
        \, ,  \\
G(-\mu,-4;y) &=&
      \zeta(2)
         + 2 \ln(2) H(0; y_r)
         - 2 H(0,-1; y_r)
         + H(0,0; y_r)
        \, ,  \\
G(-4,-4;y) &=&
       2 \ln^2(2)
         - 4 \ln(2) H(-1; y_r)
         + 4 H(-1,-1; y_r)
         - 2 H(-1,0; y_r)\nn \\
& &
         + 2 \ln(2) H(0; y_r)
         - 2 H(0,-1; y_r)
         + H(0,0; y_r)
        \, ,  \\
G(0,-\mu;y) &=&
      2 \zeta(2)
         + H(0,0; y_r)
         + 2 H(1,0; y_r)
        \, ,  \\
G(0,0,0;y) &=&
      - H(0,0,0; y_r)
         - 2 H(0,0,1; y_r)
         - 2 H(0,1,0; y_r)
         - 4 H(0,1,1; y_r)\nn \\
& &
         - 2 H(1,0,0; y_r)
         - 4 H(1,0,1; y_r)
         - 4 H(1,1,0; y_r)\nn \\
& &
         - 8 H(1,1,1; y_r)
        \, , \\
G(-\mu,-\mu,-\mu;y) &=&
      - H(0,0,0; y_r)
        \, , \\
G(-\mu,-4,-\mu;y) &=&
       3 \zeta(3)
         + \zeta(2) H(0; y_r)
         + 2 H(0,-1,0; y_r)
         - H(0,0,0; y_r)
        \, , \\
G(-4,-4,-\mu;y) &=&
       2 \zeta(3)
         - 2 \zeta(2) H(-1; y_r)
         - 4 H(-1,-1,0; y_r)
         + 2 H(-1,0,0; y_r)\nn \\ & &
         + \zeta(2) H(0; y_r)
         + 2 H(0,-1,0; y_r)
         - H(0,0,0; y_r)
        \, , \\
G(0,0,-\mu;y) &=&
       2 \zeta(3)
         - 2 \zeta(2) H(0; y_r)
         - H(0,0,0; y_r)
         - 2 H(0,1,0; y_r)\nn \\ & &
         - 4 \zeta(2) H(1; y_r)
         - 2 H(1,0,0; y_r)
         - 4 H(1,1,0; y_r)
        \, , \\
G(-\mu,0,-\mu;y) &=&
      - 4 \zeta(3)
         - 2 \zeta(2) H(0; y_r)
         - H(0,0,0; y_r)
         - 2 H(0,1,0; y_r)
        \, , \\
G(0,-\mu,-\mu;y) &=&
       2 \zeta(3)
         - H(0,0,0; y_r)
         - 2 H(1,0,0; y_r)
        \, , \\
G(-4,-\mu,-\mu;y) &=&
      - \frac{3}{2} \zeta(3)
         + 2 H(-1,0,0; y_r)
         - H(0,0,0; y_r)
        \, , \\
G(0;x) &=&
       2 H(-1; x_b)
         - H(0; x_b)\, , \\
G(\mu;x) &=&
        \pi
      + i H(0; x_b) \, , \\
G(0,0;x) &=&
       4 H(-1,-1; x_b)
         - 2 H(-1,0; x_b)
         - 2 H(0,-1; x_b)
         + H(0,0; x_b) \, , \\
G(\mu,\mu;x) &=&
      - 3 \zeta(2)
         - H(0,0; x_b)\, , \\
G(\mu,0;x) &=&
      - i \zeta(2)
         + 2 i H(0,-1; x_b)
         - i H(0,0; x_b) \, , \\
G(0,0,0;x) &=&
       8 H(-1,-1,-1; x_b)
         - 4 H(-1,-1,0; x_b)
         - 4 H(-1,0,-1; x_b) \nn \\
&&
         + 2 H(-1,0,0; x_b)
         - 4 H(0,-1,-1; x_b)
         + 2 H(0,-1,0; x_b) \nn \\
&&
         + 2 H(0,0,-1; x_b)
         - H(0,0,0; x_b)\, , \\
G(\mu,\mu,0;x) &=&
      - 2 \zeta(3)
         + \zeta(2) H(0; x_b)
         - 2 H(0,0,-1; x_b)
         + H(0,0,0; x_b) \, .
\eea
The auxiliary two-loop box $B$-functions  listed in
Appendix~\ref{AF} involve three GHPLs  which depend on two different
kinematical variables. These GHPLs can be expressed in terms of the
logarithms and Nielsen's polylogarithms depending on the rescaled
variables:
\bea
G(-x-\mu; y) & = & \int_0^{y} dw \frac{1}{(w+x) \sqrt{w (4+w)}} \, , \nn \\
&=& -\frac{\xb }{(1-\xb) (\xb+1)}\left(\ln (\xb+\yr)-\ln
   (\xb \yr+1)\right) \, ,
\label{H1} \\
G(-x,-\mu; y) & = & \int_0^{y} dw \frac{1}{w+x}
\int_0^{w} dr \frac{1}{ \sqrt{r (4+r)}}  \, , \nn \\
& = & \ln ^2(\yr)-\ln (\xb+\yr) \ln
   (\yr)-\ln (\xb \yr+1) \ln
   (\yr)+\Li_2\left(-\frac{\xb}{\yr}\right)
   \nn\\
& &
   - \Li_2(-\xb \yr)
   \, ,
\label{H2} \\
G(-x,0,-\mu; y) & = & \int_0^{y} dw \frac{1}{w+x} \int_0^{w} dq \frac{1}{q}
\int_0^{q} dr \frac{1}{ \sqrt{r (4+r)}}  \, , \nn \\
& = & \ln(1+\xb) \Biggl\{
     - 2 \ln(\yr) \bigl[ \ln(\xb \yr+1) + \ln(\xb + \yr) - \ln(\yr) \bigr]
          - 8 \zeta(2) \nn\\
& &
          - 2 \Li_2\left( \frac{\xb \yr \! + \! 1}{\xb \yr} \right) \!
          -  \! 2 \Li_2\left( \frac{\xb \! + \! \yr}{\yr} \right) \!
          -  \! 2 \Li_2\left( \frac{\xb \yr \! + \! 1}{1+\xb} \right) \!
          -  \! 2 \Li_2\left( \frac{\xb \! + \! \yr}{1+\xb} \right) \nn\\
& &
          + 2 \Li_2\left( \! \frac{\xb \yr \! + \! 1}{\yr (1 \! + \! \xb)} \! \right) \!  \!
          +  \! 2 \Li_2\left( \! \frac{\xb \! + \! \yr}{\yr (1 \! + \! \xb)} \! \right) \!  \!
          -  \! 2 \ln(\xb) \ln(\yr) \!
          + \!  4 \ln(\yr) \ln(1  \! -  \! \yr) \nn\\
& &
          + 4 \Li_2(1  \! -  \! \yr) \!
          +  \! 8 \Li_2(\yr) \!
       +  \! 4 \ln^2(1 \! + \! \xb) \ln(\yr)  \! \Biggr\} \!
       +  \! \bigl[ \ln(\xb  \! +  \! \yr) \! - \! \ln(\yr) \bigr]   \! \biggl[
            2 \zeta(2) \nn\\
& &
          + 2 \Li_2\left( \frac{\xb+\yr}{\yr} \right)
          + 2 \Li_2\left( \frac{\xb+\yr}{1+\xb} \right)
          - 2 \Li_2\left( \frac{\xb+\yr}{\yr (1+\xb)} \right)
          - 2 \Li_2(\yr)
          \biggr] \nn\\
& &
       + \ln(\xb \yr \! + \! 1)  \!  \biggl[ \!
            2 \zeta(2) \!
          +  \! 2 \Li_2\left(  \! \frac{\xb \yr \! + \! 1}{\xb \yr}  \! \right) \!
          +  \! 2 \Li_2\left(  \! \frac{\xb \yr \! + \! 1}{1+\xb}  \! \right) \!
          -  \! 2 \Li_2\left(  \! \frac{\xb \yr \! + \! 1}{\yr (1 \! + \! \xb)}  \! \right) \nn\\
& &
          + 2 \ln(\xb) \ln(\yr) \!
          +  \! \frac{1}{2} \ln^2(\yr) \!
          -  \! 2 \Li_2(\yr) \!
          \biggr] \!
          -  \! 2 \Li_3\left(  \! \frac{\xb \yr \! + \! 1}{\xb \yr}  \! \right) \!
          -  \! 2 \Li_3\left( \! \frac{\xb \! + \! \yr}{\yr} \!  \right) \nn\\
& &
          - 2 \Li_3\left( \! \frac{\xb \yr \! + \! 1}{1 \! + \! \xb}  \! \right) \!
          -  \! 2 \Li_3\left( \! \frac{\xb \! + \! \yr}{1 \! + \! \xb}  \! \right) \!
          +  \! 2 \Li_3\left( \! \frac{\xb \yr \! + \! 1}{\yr (1 \! + \! \xb)}  \! \right) \!
          +  \! 2 \Li_3\left( \! \frac{\xb \! + \! \yr}{\yr (1 \! + \! \xb)}  \! \right) \nn\\
& &
          - \ln(\yr) \biggl[
        2 \Li_2\left(  \! \frac{\xb \yr \! + \! 1}{\xb \yr}  \! \right) \!
          - \!  2 \Li_2\left(  \! \frac{\xb \! + \! \yr}{1 \! + \! \xb}  \! \right) \!
          -  \! 2 \Li_2\left(  \! \frac{\xb \yr \! + \! 1}{\yr (1 \! + \! \xb)}  \! \right)  \!
      - \Li_2\left(  \! - \frac{\yr}{\xb}  \! \right) \nn\\
& &
      - \Li_2( - \xb \yr)
      \biggr]  \!
          +  \! \frac{1}{6} \ln^2(\yr) \bigl[ 3 \ln(\xb  \! +  \! \yr)  \! -  \! 4 \ln(\yr)  \!
         -  \! 6 \ln(\xb)\bigr] \!
          +  \! 2 \Li_3\left( \!  \frac{1 \! + \! \xb}{\xb}  \! \right) \nn\\
& &
          + 2 \Li_3(1 + \xb) \!
          -  \!\Li_3\left( - \frac{\yr}{\xb} \right) \!
          +  \!\Li_3\left( - \frac{1}{\xb} \right) \!
          -  \!\Li_3( - \xb \yr) \!
          +  \!\Li_3( - \xb)
\, ,
\label{H3}
\eea
where the Nielsen's polylogarithms are related to HPLs as follows:
\be
\Li_n (a) = H(\underbrace{0,\cdots,0}_n,1; a) \, , \quad
\mbox{S}_{n,m} (a) = H(\underbrace{0,\cdots,0}_n,\underbrace{1,\cdots,1}_m;a)
\, .
\ee

\subsection{Analytical Continuation \label{acont}}

The result for the two-loop corrections in Section~\ref{crosssection} is
expressed in terms of the auxiliary functions which are given in
Appendix~\ref{AF} in the non-physical region $s<0$. The corresponding
expressions in the physical region $s>0$ can be obtained by analytical
continuation to the complex value of $P^2$:
\be
P^2 = -s-i \epsilon, \qquad \epsilon \to 0^+ \, .
\label{analytcont}
\ee
Not that in the physical region  $Q^2$ and $R^2$ are  real and
positive.
Let us consider the analytical structure of the  rescaled variables,
Eqs.~(\ref{xr}--\ref{zb}). The variables $\yr$ and $\zr$ are
positive and  vary  from 0 to 1 for $0<Q^2<\infty$ and
$0<R^2<\infty$. The variable $\xr$ varies from 0 to 1 for
$0<P^2<\infty$. For positive $s$ the variable  $\xr$ becomes
complex. In the region $0<s<4m_f^2$ it is a pure phase:
\be
\xr = \frac{\sqrt{4m_f^2-s} + i \sqrt{s}}{\sqrt{4m_f^2-s}-i
\sqrt{s}} = e^{i 2 \phi} \, , \ee where \be \phi =
\arctan{\sqrt{\frac{s}{4m_f^2-s}}} \, .
\ee
In the region $s>4m_f^2$ we have
\be \xr = - \xr' + i \epsilon \, , \ee with \be \xr' =
\frac{\sqrt{s}-\sqrt{s-4m_f^2}}{\sqrt{s}+\sqrt{s-4m_f^2}} \, .
\ee
The HPLs of the variables $\yr$ and $\zr$ are always real, while the
HPLs of $\xr$ are complex for $s>0$. In particular, their imaginary
part in the region above the heavy-flavor threshold, $s>4m_f^2$, is
defined when the analytical continuation of the logarithm is
specified:
\be
H(0;\xr) \to H(0;-\xr'+i \epsilon) = H(0;\xr') + i \pi
\, .
\label{contlog}
\ee
The case of the variables $\xb$, $\yb$, and $\zb$  is more
complicated. The variables $\yb$ and $\zb$ can be complex.  For
$0<Q^2,R^2<4m_f^2$, $\yb$ and $\zb$ are pure phases. Note that the
expressions of $t$ and $u$ in terms of $\yb$ and $\zb$ are invariant
under the inversion
\be
\yb \to \frac{1}{\yb} \, , \qquad \zb
\to \frac{1}{\zb} \, .
\ee
This means that for $0<Q^2,R^2<4m_f^2$ it does not matter whether we
give to $Q^2$ and $R^2$ a positive or a negative imaginary part.
Let us choose  $Q^2\to Q^2-i \epsilon$ and $R^2\to R^2-i \epsilon$.
Then we have
\bea
\yb = \frac{\sqrt{Q^2} + i \sqrt{4 m_f^2
-Q^2}}{\sqrt{Q^2} - i \sqrt{4 m_f^2 -Q^2}}
= e^{i 2 \psi} \, , \\
\zb = \frac{\sqrt{R^2} + i \sqrt{4 m_f^2 -R^2}}{\sqrt{R^2} - i
\sqrt{4 m_f^2 -R^2}}
= e^{i 2 \xi} \, ,
\eea
with
\bea
\psi & = & \arctan{\sqrt{\frac{4m_f^2}{Q^2}-1}}  \, , \\
\xi & = &  \arctan{\sqrt{\frac{4m_f^2}{R^2}-1}}  \, .
\eea
When $Q^2$ and $R^2$ vary from $4m_f^2$ to infinity, $\yb$ and $\zb$
are real and positive and vary  from 1 to 0.
The variable $\xb$ is a pure phase for $0<P^2<4m_f^2$. It varies
from 1 and 0 when $P^2$ varies from $4m_f^2$ to $+\infty$. In the
physical region we have
\be
\xb \to - \xb' + i \epsilon \, , \ee where \be \xb' =
\frac{\sqrt{s+4m_f^2}-\sqrt{s}}{\sqrt{s+4m_f^2}+\sqrt{s}} \, ,
\ee
{\it i.e.}  $\xb$  is negative and varies from  -1 and 0 when $s$
varies from 0 to  $\infty$.  The HPLs  of $\xb$ are complex  in the
physical region. Their imaginary part is defined in the same way as
in Eq.~(\ref{contlog}).
Finally, the analytical continuation of the three GHPLs of
Eqs.~(\ref{H1}--\ref{H3}) are defined by  the analytical properties
of the functions $\Li_2$ and $\Li_3$ \cite{Lewin}.

\subsection{Mellin-Barnes Expansion of the GHPLs}

To study the low and high  energy behavior of the two-loop
corrections one needs the expansion of three GHPLs of two
kinematical invariants, Eqs.~(\ref{H1}-\ref{H3}), in the limits $s \gg
m_f^2$ and $s \ll m_f^2$. To perform the expansions we apply the
inverse Mellin-Barnes transformation to the integral representation
of the  GHPLs. In this Section we describe  the technique and
present the result of the expansion.
We start with the GHPL of weight one in Eq.~(\ref{H1}):
\bea
G(-x-\mu; y) &=& \int_0^{y} dw \frac{1}{(w+x) \sqrt{w (4+w)}} \, .
\eea
By changing the integration variable  $w = y r$ we obtain the
following integral representation:
\bea
G(-x-\mu; y) &=& \frac{1}{x} \int_0^{1} dr \frac{1}{r}   \left(
1+\frac{y}{x} r\right)^{-1} \left( 1+ \frac{4}{y r} \right)^{-1/2} \, .
\eea
Then we apply the inverse Mellin-Barnes transformation to the square
root in the integrand:
\bea
\left( 1+ \frac{4}{y r} \right)^{-1/2}  = \frac{1}{2 \pi i}
\frac{1}{\Gamma(1/2)} \int_{-i \infty}^{i \infty} d\sigma
\Gamma\left(\frac{1}{2} + \sigma\right) \Gamma \left( -\sigma \right)
\left( \frac{4}{y r}\right)^{\sigma} \, .
\label{MBH1}
\eea
The integrand has two infinite series of poles in the $\sigma$
complex plane:
\bea
\mbox{``left hand side poles'' at} &\quad& \sigma = -n -\frac{1}{2} \, ,\quad
\mbox{for}
\quad n  = 0,1,2, \cdots \, ; \nn \\
\mbox{``right  hand side poles'' at} &\quad& \sigma = n \, , \quad \mbox{for}
\quad n  = 0,1,2, \cdots \, .
\eea
One has to choose the integration contour in Eq.~(\ref{MBH1}) so
that
\be
-\frac{1}{2}  < \mbox{Re}(\sigma) < 0 \, .
\label{integrationcontour}
\ee
Then we find
\be
G(-x-\mu; y) = \frac{1}{2 \pi i} \frac{1}{\Gamma(1/2)} \frac{1}{x}
\int_{-i \infty}^{i \infty} d\sigma \Gamma\left(\frac{1}{2} + \sigma\right)
 \Gamma \left( -\sigma \right) \left( \frac{4}{y}\right)^{\sigma}
 I(\sigma) \, .
\label{H1beforeclosing}
\ee
The integral over $r$ can be evaluated analytically
\be
I(\sigma) = \int_0^1 dr r^{-\sigma-1} \left( 1+\frac{y}{x} r\right)^{-1} =
-\frac{1}{\sigma} \, \hypF \left(1, -\sigma, 1- \sigma; -\frac{y}{x} \right) \, ,
\ee
where $\hypF$ is the hypergeometric function.
In order to obtain the asymptotic expansion in the $y \to \infty$
limit at fixed $\tau =y/x$ it is sufficient to close the integration
contour in Eq.~(\ref{H1beforeclosing}) on the r.h.s. of the $\sigma$
complex plane. One finds
\bea
G(-x-\mu; y) &=& \sum_{n=0}^{\infty}
g_n^{(1, {\mathrm L})} (\tau,y) \left(1/y\right)^{n+1} \, ,
\label{H1largey}
\eea
where the functions $g_n^{(1, {\mathrm L})}$ can be easily obtained
{\it e.g.} with the help of \cite{Daniel1}. For $n=0,~1,~2$ they
read
\bea
g_0^{(1,{\mathrm L})} (\tau,y) &=& -\tau  \left(\ln(\tau +1)-\ln(y)\right) \, , \\
g_1^{(1,{\mathrm L})} (\tau,y) &=& -2 \tau  \left[\left(\ln
(\tau +1)-\ln (y)\right)
\tau +\tau -1 \right]\, , \\
g_2^{(1,{\mathrm L})} (\tau,y) &=& -\tau  \left[6 \left(\log
(\tau +1)-\log (y)\right) \tau ^2+\left(7 \tau
   -6\right) \tau +3\right] \, .
\eea
To obtain the asymptotic expansion of this  GHPL in the limit $y \to
0$ for fixed $\tau$ it is necessary to close the integration contour
in Eq.~(\ref{H1beforeclosing}) on the left hand side (l.h.s.) of the
$\sigma$ complex plane. One obtains
\bea
G(-x-\mu; y) &=& \sum_{n=0}^{\infty} g_n^{(1,{\mathrm S})} (\tau,y) y^n \, ,
\eea
where the functions $g_n^{(1,{\mathrm S})}$ for $n=0,~1,~2$ read
\bea
g_0^{(1,{\mathrm S})} &=& \sqrt{\frac{\tau}{y}} \arctan \left(\sqrt{\tau
   }\right)\, , \\
g_1^{(1, {\mathrm S})} &=&   \frac{\arctan \left(\sqrt{\tau }\right)-\sqrt{\tau }}{8
   \sqrt{y \tau }}\, , \\
g_2^{(1, {\mathrm S})} &=&    \frac{\sqrt{\tau } (\tau -3)+3 \arctan \left(\sqrt{\tau
   }\right)}{128 \sqrt{y \tau ^3}} \, .
\label{H1smally}
\eea
The  expansion of  the remaining two GHPLs is almost identical. By
inverse Mellin-Barnes transformation the GHPL of weight two in
Eq.~(\ref{H2}) can be written as
\bea
G(-x,-\mu; y) &=&\int_0^{y} dw \frac{1}{w+x}
\int_0^{w} dr \frac{1}{ \sqrt{r (4+r)}}   \nn \\
&=& \int_0^y dw \frac{1}{w+r} \int_0^1 ds \frac{1}{s} \left( 1 +
\frac{4}{w s} \right)^{-\frac{1}{2}} \nn \\
&=& -\frac{1}{2 \pi i} \frac{1}{\Gamma(1/2)}
\int_{-i\infty}^{i\infty} d\sigma \frac{\Gamma\left(\frac{1}{2}+\sigma
\right) \Gamma(-\sigma)}{\sigma} 4^\sigma \int_0^y dw
\frac{1}{w+x} w^{-\sigma} \nn \\
&=&-\frac{1}{2 \pi i} \frac{1}{\Gamma(1/2)}
\frac{1}{x}\int_{-i\infty}^{i\infty} d\sigma
\frac{\Gamma\left(\frac{1}{2}+\sigma \right)
\Gamma(-\sigma)}{\sigma(1-\sigma)}4^\sigma y^{1-\sigma} \times \nn \\
& & \times
\hypF\left(1,1-\sigma,2-\sigma; -\frac{y}{x} \right) \, .
\eea
Again, the integration contour has to satisfy
Eq.~(\ref{integrationcontour}). By closing the integration contour
on the r.h.s. of the complex $\sigma$ plane, in the limit $y \to
\infty$ one finds
\bea  G(-x,-\mu; y) &=& \sum_{n=0}^{\infty} g_n^{(2,{\mathrm L})}
(\tau,y) \left(1/y\right)^n \, . \eea
The functions $g_n^{(1,{\mathrm L})}$ for $n=0,~1,~2$ read
\bea
g_0^{(2,{\mathrm L})} (\tau,y) &=& \ln (y) \ln (\tau +1)+\Li_2(-\tau )\, , \\
g_1^{(2,{\mathrm L})} (\tau,y) &=& 2 \tau  [\ln (y)+\ln (\tau +1)] \, , \\
g_2^{(2,{\mathrm L})} (\tau,y) &=& \tau  [3 (\ln (y) - \ln (\tau +1) )\tau -2 \tau
   +3 ] \, .
\label{H2largey}
\eea
Alternatively, by closing the integration contour on the l.h.s. of
the complex $\sigma$ plane, which corresponds to the limit $y \to
0$, one obtains
\bea G(-x,-\mu; y) &=& \sum_{n=0}^{\infty} g_n^{(2,{\mathrm S})}
(\tau,y) y^n \, , \eea
with
\bea
g_0^{(2,{\mathrm S})} &=& 2 \sqrt{\frac{y}{\tau }}
\left(\sqrt{\tau }-\arctan \left(\sqrt{\tau }\right)\right)\, , \\
g_1^{(2, {\mathrm S})} &=& -\frac{\sqrt{\frac{y}{\tau }}
\left(\sqrt{\tau } (\tau
   -3)+3 \arctan \left(\sqrt{\tau }\right)\right)}{36
   \tau } \, , \\
g_2^{(2, {\mathrm S})} &=&    \frac{\sqrt{\frac{y}{\tau }}
\left(\sqrt{\tau } (\tau
   (3 \tau -5)+15)-15 \arctan \left(\sqrt{\tau
   }\right)\right)}{1600 \tau ^2} \, .
\label{H2smally1}
\eea
For the GHPL of weight three in Eq.~(\ref{H2}) we have
\bea
G(-x,0,-\mu;y) &=& \int_0^{y} dw \frac{1}{w+x}
\int_0^{w} dq \frac{1}{q} \int_0^{q} dr \frac{1}{ \sqrt{r (4+r)}}
\,  \nn
\\
&=& \int_0^y dw \frac{1}{w+x} \int_0^w ds \frac{1}{s} \int_0^1 dq
\frac{1}{q}\left(1 +\frac{4}{s q} \right)^{-\frac{1}{2}} \nn \\
&=&\frac{1}{2 \pi i} \frac{1}{\Gamma(1/2)}
\int_{-i\infty}^{i\infty} d\sigma \Gamma\left(\sigma
+\frac{1}{2}\right) \Gamma(-\sigma) 4^\sigma \int_0^y dw
\frac{1}{w +x} \times \nn \\ && \times \int_0^w ds \, s^{-\sigma-1}
 \int_0^1 dq\, q^{-\sigma-1}\nn \\
&=& \frac{1}{2 \pi i} \frac{1}{\Gamma(1/2)} \frac{1}{x}
\int_{-i\infty}^{i\infty} d\sigma \frac{\Gamma\left(\sigma
+\frac{1}{2}\right) \Gamma(-\sigma)}{\sigma^2 (1-\sigma)} 4^\sigma
y^{1-\sigma} \times \nn \\
& &\times\hypF\left(1,1-\sigma,2-\sigma;-\frac{y}{x}\right) \, .
\label{G3}
\eea
Once again, the condition in Eq.~(\ref{integrationcontour}) has to
be satisfied by the integration contour in  Eq.~(\ref{G3}). By
closing the integration contour on the r.~h.s. of the complex
$\sigma$ plane one  obtains the asymptotic expansion in the limit $y
\to \infty$:
\bea
G(-x,0,-\mu; y) &=& \sum_{n=0}^{\infty} g_n^{(3,{\mathrm L})}
(\tau,y) \left(1/y\right)^n \, .
\eea
The first three $g_n^{(3, {\mathrm L})}$ functions are given
by
\bea
g_0^{(3,{\mathrm L})} (\tau,y) &=&
\frac{1}{6} \left(3 \ln ^2(y)+2 \pi ^2\right) \ln
   (\tau +1)+\ln (y) \Li_2(-\tau
   )-\Li_3(-\tau )\, ,  \\
g_1^{(3,{\mathrm L})} (\tau,y) &=& -2 \tau  [\ln (y)-\ln (\tau +1)+1] \, , \\
g_2^{(3,{\mathrm L})} (\tau,y) &=&  -\frac{1}{4} \tau  [ 6 ( \ln(y)
- \ln (\tau +1) )\tau -\tau +6]
\, .
\label{H3largey}
\eea
By closing the integration contour in Eq.~(\ref{G3}) on the l.h.s.
of the complex $\sigma$ plane one  finds the asymptotic behavior of
the GHPL in the $y \to 0$ limit:
\bea G(-x,0,-\mu; y) &=& \sum_{n=0}^{\infty} g_n^{(3,{\mathrm S})}
(\tau,y) y^n \, ,
\eea
where the functions $g_n^{(1,{\mathrm S})}$ for $n=0,1,2$ read
\bea
g_0^{(3,{\mathrm S})} &=& 4 \sqrt{\frac{y}{\tau }}
\left(\sqrt{\tau }-\arctan \left(\sqrt{\tau }\right)\right)\, , \\
g_1^{(3, {\mathrm S})} &=& -\frac{\sqrt{\frac{y}{\tau }}
\left(\sqrt{\tau } (\tau
   -3)+3 \arctan \left(\sqrt{\tau }\right)\right)}{54
   \tau } \, , \\
g_2^{(3, {\mathrm S})} &=&  \frac{\sqrt{\frac{y}{\tau }}
\left(\sqrt{\tau } (\tau
   (3 \tau -5)+15)-15 \arctan \left(\sqrt{\tau
   }\right)\right)}{4000 \tau ^2} \, .
\label{H2smally2}
\eea
It is interesting  that, though the expansion of the three GHPLs
discussed above in the $s \ll m_f^2$ limit involves  $\arctan
(\sqrt{\tau })$ terms, these terms completely cancel  in the
expansion of the auxiliary functions $B_1$, $B_2$, and $B_3$.

\section{Auxiliary Functions \label{AF}}

In this Appendix we collect the expressions for the auxiliary
functions used in the paper which are  valid in the non-physical
region $s<0$ ($P^2>0$). The analytical continuation to $s>0$ is
discussed in Section~\ref{acont}.
The dimensionless variables $x$ and $y$, used in the explicit formulas
below, are related to $s$ and $t$ via Eq.~(\ref{xy}).

\subsection{One-Loop Functions}

\noindent {\bf Vacuum polarization:}
\be
\Pi^{(1l)}_0 (s) = \sum_{i=0}^1 \Pi^{(1l,i)}_0 (s) (D-4)^i
+ {\mathcal O} \left( (D-4)^{2} \right) \, ,
\ee
where:
\bea
\Pi^{(1l,0)}_0 &=&
         - \frac{5}{9}
     + \frac{4}{3x}
         + \frac{1}{3}\frac{x^2+2x-8}{x\sqrt{x(x+4)}} G( - \mu;x) \, ,
\label{A1lS0} \\
\Pi^{(1l,1)}_0 &=&
           \frac{14}{27}
     - \frac{16}{9x}
         - \frac{1}{18 x \sqrt{x(x+4)}} \Bigl[
       ( 4 x - 64 + 5 x^2 ) G( - \mu;x) \nn\\
& &
     - 3 ( 2 x - 8 + x^2  ) G(-4, - \mu;x) \Bigr] \, .
\label{A1lS1}
\eea

\noindent {\bf Dirac form factor:}
\be
F_1^{(1l)}(s) = \sum_{i=-1}^0 F_1^{(1l,i)}(s)  (D-4)^i
+ {\mathcal O} (D-4) \, ,
\ee
where:
\bea
F_1^{(1l,-1)} & = & 1 - \ln\left( \frac{P^2}{m_e^2}\right) \, ,
\label{F1ren0}\\
F_1^{(1l,0)}  & = & -1 + \frac{1}{2} \zeta(2) - \frac{1}{2} G(0;x) +
\left( \frac{5}{4} + \frac{1}{2} G(0;x) \right) \ln\left(
\frac{P^2}{m_e^2}\right) - \frac{3}{4} \ln^2\left(
\frac{P^2}{m_e^2}\right)  \, .
\label{F1ren}
\eea

\noindent {\bf Box $B$-functions:}
\be
B_j^{(1l)}(s,t) = \sum_{i=-2}^0 B_j^{(1l,i)}(s,t)  (D-4)^i
+ {\mathcal O} (D-4) \, , \quad j=1,2,3 \, ,
\ee
where:
\bea
B_1^{(1l,-2)} & = &  \frac{16(x + y)^2}{y} \, , \\
B_1^{(1l,-1)} & = &  \frac{8(x^2 + x y + y^2)}{y} + \frac{8(x +
y)^2}{y} G(0;x)
\, , \\
B_1^{(1l,0)} & = &
- \frac{2 (2 x y + 8 x^2 \zeta(2) + 10 x y \zeta(2) + 5 y^2 \zeta(2))}{y}
 \nn\\
& & + \frac{2 (2 x^2 + x y + y^2)}{y} G(0;x) + 2(x+y)  G(0;y)
+ 2(2x+y)  G(0,0;x) \nn\\
& & + \frac{2 (2 x^2 + 2 x y + y^2)}{y} \left( G(0;x) G(0;y)
-G(0,0;y) \right) \, , \\
B_2^{(1l,-2)} & = &  \frac{16 (2 x^2 + 2 x y + y^2)}{y} \, , \\
B_2^{(1l,-1)} & = &  8y + \frac{8 (2 x^2 + 2 x y + y^2)}{y} G(0;x)
\, , \\
B_2^{(1l,0)} & = & - \frac{2 (16 x^2 + 10 x y + 5 y^2)
\zeta(2)}{y} + \frac{2 (4 x^2 + 2 x y + y^2)}{y} \bigl( G(0;x)
G(0;y)
- G(0,0;y) \bigr) \nn\\
& & - 2(x-y) G(0;x) + 2(x+y) G(0;y) + 2(2x+y) G(0,0;x) \, , \\
B_3^{(1l,-2)} & = & - \frac{16x^2}{y}
\, ,\\
B_3^{(1l,-1)} & = & - \frac{8 (x^2 + x y + y^2)}{y} -
\frac{8x^2}{y} G(0;x)
\, , \\
B_3^{(1l,0)} & = &
- \frac{4 (x y + y^2 - 4 x^2 \zeta(2))}{y}
- \frac{4 x^2}{y} \bigl( G(0;x) G(0;y) - G(0,0;y) \bigr) \nn\\
& & - \frac{4 (x^2 + x y + y^2)}{y} G(0;x) \, .
\eea

\subsection{Two-Loop Functions}

\noindent {\bf Vacuum polarization:}
\be
\Pi^{(2l)}_0 (s) = \Pi^{(2l,0)}_0 (s) + {\mathcal O} (D-4) \, ,
\ee
where:
\bea
\Pi^{(2l,0)}_0 & = &
\frac{x+4}{12x\sqrt{x(x+4)}} \Bigl[
    3 (- 6 + x ) G(-\mu; x) + 4 ( 2 - x ) ( 2 G(-4, -\mu; x)
    + G(0, -\mu; x) ) \Bigr] \nn\\
& &
+ \frac{1}{24 x^2}
   \Bigl[
     52 x - 5 x^2 - 8( 7 - 2 x - 3 x^2) G(-\mu, -\mu; x)  \nn\\
& &
   + ( 4 - x^2 ) ( 16 G(-4, -\mu, -\mu; x)
   + 8 G(0, -\mu, -\mu; x)
   - 16 G(-\mu, -4, -\mu; x)  \nn\\
& &
   - 8 G(-\mu, 0, -\mu; x) )
   \Bigr] \, .
\label{2lPi0}
\eea

\noindent {\bf Dirac form factor:}
\be
F_1^{(2l)}(s) = F_1^{(2l,0)}(s)  + {\mathcal O} (D-4) \, ,
\ee
where:
\bea
F_1^{(2l,0)} & = &
 \frac{1}{36 x \sqrt{x(4-x)}} \bigl[
(x-4) (46-19 x) G(\mu, 0; x) \bigr] - \frac{1}{1296 x^2} \bigl[
8568 x - 3355 x^2  \nn\\
& & - x ( 3960 - 1590 x ) G(0; x) +
    ( 1296 - 216 x^2 ) G(\mu, \mu, 0; x) \bigr] \, ,
\label{2lF1}
\eea

\noindent {\bf Box $B$-functions:}
\be
B_j^{(2l)}(s,t) = \sum_{i=-2}^0 B_j^{(2l,i)}(s,t)  (D-4)^i +
{\mathcal O} (D-4) \, , \quad j=1,2,3 \, ,
\ee
where:
\bea
B_1^{(2l,-2)} & = & -
\frac{8 (x + y)^2 (5 y-12)}{9 y^2} + \frac{8 (y-2) (y+4) (x +
y)^2}{3 y^2 \sqrt{y (y+4)}} G( - \mu;y) \, ,
\label{B12lM2} \\
B_1^{(2l,-1)} & = &
- \frac{4 (204 x^2 + 444 x y - 13 x^2 y + 204 y^2 - 41 x y^2 - 13 y^3)}{27 y^2}
 \nn\\
& & - \frac{8 (x + y)^2}{3 y} G( - \mu, - \mu;y)
- \frac{4 (x + y)^2 (5 y-12)}{9 y^2} G(0;x) \nn\\
& &
- \frac{1}{\sqrt{y(y+4)}} \biggl[
\frac{4 (y+4) (-34 x^2 - 74 x y + 2 x^2 y - 34 y^2 + 7 x y^2 + 2 y^3)}{9 y^2}
G( - \mu;y) \nn\\
& & - \frac{4 (y-2) (y+4) (x + y)^2}{3 y^2} \bigl( 3 G(-4, -
\mu;y) + G(0;x) G(-\mu;y)\bigr)
 \biggr] \, ,
\label{B12lM1} \\
B_1^{(2l,0)} & = &
\frac{2 (x-4) x (y-2) (y+4) (x + y)^2}{3 y^2\sqrt{y(y+4)} \sqrt{x(4-x)}}
G(\mu,0;x) G( - x - \mu;y) \nn\\
& &
+ \frac{1}{\sqrt{y(y+4)}} \biggl[
  \frac{2 (y-2) (y+4) (x + y)^2}{3 y^2} G( - x,0, - \mu;y)  \nn\\
& & - \frac{4 (y-2) (y+4) (x + y)^2}{y^2} G( - \mu, - \mu, -
\mu;y)
+ \frac{y+4}{135 y^2} \bigl( -1736 x^2 - 4732 x y  \nn\\
& &
+ 82 x^2 y -
    1976 y^2 - 16 x y^2 + 9 x^2 y^2 - 158 y^3 + 18 x y^3 +
    9 y^4 + 180 x^2 \zeta(2)  \nn\\
& &
+ 360 x y \zeta(2) - 90 x^2 y \zeta(2) +
    180 y^2 \zeta(2) - 180 x y^2 \zeta(2) - 90 y^3 \zeta(2) \bigr) G( - \mu;y) \nn\\
& &
- \frac{2 (y+4)}{3 y^2} (-34 x^2 - 74 x y + 2 x^2 y - 34 y^2 +
     7 x y^2 + 2 y^3) G(-4, - \mu;y)  \nn\\
& & + \frac{2 (y-2) (y+4) (x + y)^2}{3y^2} ( 3 G(-4, - \mu;y)
G(0;x)
+ 9 G(-4,-4, - \mu;y) \nn\\
& & -  G(0;x) G( - x, - \mu;y) ) - \frac{4 (4 \! + \! y) }{9
y^2}(4 x^2 \! + \! 4 x y \! - \! 5 x^2 y \! - \! 4 y^2
\! - \! 5 x y^2 \! - \! 4 y^3) G(0, - \mu;y)\nn\\
& &
- \frac{2 (4 + y)}{9 y^2} (-42 x^2 - 82 x y + 12 x^2 y - 26 y^2 +
     17 x y^2 + 10 y^3) G(0;x) G( - \mu;y) \nn\\
& &
+ \frac{2 (y \! - \! 2) (y \! + \! 4) (x  \! +  \! y)^2}{3 y^2} (
  G(0;x) G(0, \!  - \mu;y) \!
-  \! G(0,0, \!  - \mu;y) \! +  \! G(0,0;x) G( - \mu;y) )
\biggr] \nn\\
& & + \frac{2 (x-4) x (12 x+20 y-5 x y+3 y^2)}{9 y^2\sqrt{x(4-x)}}
G(\mu,0;x)
- \frac{1}{9 y^2}(144 x^2 + 288 x y - 11 x^2 y  \nn\\
& &
   + 144 y^2 - 40 x y^2 - 17 y^3) G( - \mu, - \mu;y)
- \frac{4 (x + y)^2}{y} G( - \mu,-4, - \mu;y) \nn\\
& & - \frac{2 (2 x^2 + 2 x y + y^2)}{3 y} G( - \mu,0, - \mu;y)
- \frac{2(2 x + y)}{3} ( G(0;x) G( - \mu, - \mu;y) \nn\\
& & + G(\mu,\mu,0;x) ) + \frac{2 (2 x^2 + 6 x y + 3 y^2)}{3 y}
G(0, - \mu, - \mu;y)
+ \frac{2}{27 y^2} (-180 x^2 - 372 x y  \nn\\
& & + 41 x^2 y - 156 y^2 + 121 x y^2 + 65 y^3) G(0;x)
- \frac{2 (x + y)^2 (-12 + 5 y)}{9 y^2} G(0,0;x) \nn\\
& & - \frac{(x + y)^2}{15} G(0;y) + \frac{2}{405 y^2} (4488 x^2 +
12036 x y - 1082 x^2 y +
    4488 y^2 - 2359 x y^2  \nn\\
& &
- 1082 y^3 - 540 x^2 \zeta(2) -
    1080 x y \zeta(2) + 225 x^2 y \zeta(2) - 540 y^2 \zeta(2) + 450 x y^2
    \zeta(2)  \nn\\
& &
+ 225 y^3 \zeta(2)) \, ,
\label{B12l0} \\
B_2^{(2l,-2)} & = &
\frac{8(2 x^2 + 2 x y + y^2)}{9 y^2} \Biggl[ 12 - 5 y +
\frac{3(y-2) (y+4)}{\sqrt{y(y+4)}} G( - \mu;y) \Biggr] \, ,
\label{B22lM2} \\
B_2^{(2l,-1)} & = &
- \frac{4 (480 x^2 + 480 x y - 56 x^2 y + 204 y^2 - 56 x y^2 - 13 y^3)}{27 y^2}
-\frac{4 (5 y-12)}{9 y^2} (2 x^2 + 2 x y \nn\\
& & + y^2) G(0;x) -\frac{8 (2 x^2 + 2 x y + y^2)}{3 y} G( - \mu, -
\mu;y) + \frac{y+4}{\sqrt{y(y+4)}} \Biggl[
  \frac{8}{9 y^2} (40 x^2 + 40 x y  \nn\\
& &
  - 5 x^2 y + 17 y^2 - 5 x y^2 - y^3)
   G( - \mu;y)
+ \frac{4 (y-2) (2 x^2 + 2 x y + y^2)}{3 y^2} \bigl( 3 G(-4, - \mu;y) \nn\\
& & + G(0;x) G( - \mu;y) \bigr) \Biggr] \, ,
\label{B22lM1} \\
B_2^{(2l,0)} & = &
\frac{2 (x-4) x (y-2) (y+4)(2 x^2 + 2 x y + y^2)}{3 y^2 \sqrt{y(y+4)}
\sqrt{x(4-x)}}
G(\mu,0;x) G( - x - \mu;y) \nn\\
& & + \frac{y+4}{\sqrt{y(y+4)}} \biggl[ \frac{2(y-2)(2 x^2 + 2 x y
+ y^2)}{3 y^2} \bigl( G( - x,0, - \mu;y)
- 6 G( - \mu, - \mu, - \mu;y) \bigr) \nn\\
& &
- \frac{1}{135 y^2} \bigl( 5872 x^2 + 6112 x y - 464 x^2 y +
    1976 y^2 - 224 x y^2 - 18 x^2 y^2 + 158 y^3  \nn\\
& &
    - 18 x y^3 - 9 y^4 - 360 x^2 \zeta(2) - 360 x y \zeta(2)
    + 180 x^2 y \zeta(2) - 180 y^2 \zeta(2)  \nn\\
& &
    + 180 x y^2 \zeta(2)
    + 90 y^3 \zeta(2) \bigr) G( - \mu;y)
+ \frac{4}{3 y^2}(40 x^2 + 40 x y - 5 x^2 y + 17 y^2 - 5 x y^2  \nn\\
& &
         - y^3) G(-4, - \mu;y)
+ \frac{2(y-2)(2 x^2 + 2 x y + y^2)}{3 y^2} \bigl(
  3 G(-4, - \mu;y) G(0;x) \nn\\
& & + 9 G(-4,-4, - \mu;y) - G(0;x) G( - x, - \mu;y)
+ G(0;x) G(0, - \mu;y) \nn\\
& & - G(0,0, - \mu;y) + G(0,0;x) G( - \mu;y) \bigr) - \frac{4}{9
y^2}(8 x^2 \! + \! 4 x y \! - \! 10 x^2 y \! - \! 4 y^2
          \! - \! 5 x y^2  \nn\\
& &
          - 4 y^3) G(0, - \mu;y)
+ \frac{4(48 x^2 \! + \! 44 x y \! - \! 15 x^2 y \! + \! 13 y^2
         \!  - \! 10 x y^2 \! - \! 5 y^3)}{9 y^2} G(0;x) G( - \mu;y)
\biggr] \nn\\
& & + \frac{2 (x-4) x (24 x + 16 y - 10 x y + 3 y^2)}{9
y^2\sqrt{x(4-x)}} G(\mu,0;x)
- \frac{1}{9 y^2} (288 x^2 + 288 x y - 46 x^2 y  \nn\\
& &
        + 144 y^2 - 52 x y^2
        - 17 y^3) G( - \mu, - \mu;y)
- (2 x^2 + 2 x y + y^2) \biggl(
    \frac{4}{y} G( - \mu,-4, - \mu;y) \nn\\
& &
  + \frac{1}{15} G(0;y)
  + \frac{2 (5 y-12)}{9 y^2} G(0,0;x) \biggr)
- \frac{2 (4 x^2 + 2 x y + y^2)}{3 y} G( - \mu,0, - \mu;y) \nn\\
& & - \frac{2}{3} (2 x + y) \bigl( G(\mu,\mu,0;x) +  G(0;x) G( -
\mu, - \mu;y)
  \bigr)
- \frac{2}{27 y^2} (432 x^2 + 432 x y \nn\\
& &
         - 112 x^2 y \! + \! 156 y^2 \! - \! 136 x y^2
        \!  - \! 65 y^3) G(0;x)
+ \frac{2 (4 x^2 \! + \! 6 x y \! + \! 3 y^2)}{3 y} G(0, - \mu, - \mu;y) \nn\\
& &
+ \frac{2}{405 y^2} \bigl( 16176 x^2 \! + \! 16896 x y \! - \! 3004 x^2 y \! + \!
    4488 y^2 \! - \! 3004 x y^2 \! - \! 1082 y^3 \! - \! 1080 x^2 \zeta(2)  \nn\\
& &
    - 1080 x y \zeta(2) + 450 x^2 y \zeta(2) - 540 y^2 \zeta(2) +
    450 x y^2 \zeta(2) + 225 y^3 \zeta(2) \bigr) \, ,
\label{B22l0} \\
B_3^{(2l,-2)} & = &
\frac{8 x^2 (5 y-12)}{9 y^2} - \frac{8 x^2 (y-2) (y+4)}{3
y^2\sqrt{y(y+4)}} G( - \mu;y) \, ,
\label{B32lM2} \\
B_3^{(2l,-1)} &=&
 \frac{4 (204 x^2 - 36 x y - 13 x^2 y - 36 y^2 + 15 x y^2 + 15 y^3)}{27 y^2}
+ \frac{4 x^2 (5 y-12)}{9 y^2} G(0;x) \nn\\
& & + \frac{8x^2}{3y} G( - \mu, - \mu;y) -
\frac{y+4}{\sqrt{y(y+4)}} \Biggl[
  \frac{4}{9 y^2}(34 x^2 - 6 x y - 2 x^2 y - 6 y^2 + 3 x y^2  \nn\\
& & + 3 y^3) G( - \mu;y) + \frac{4 x^2 (y-2)}{3 y^2} \bigl( 3
G(-4, - \mu;y) + G(0;x) G( - \mu;y) \bigr) \Biggr] \, ,
\label{B32lM1} \\
B_3^{(2l,0)} &=&
- \frac{2 (x-4) x^3 (y-2) (y+4)}{3 y^2\sqrt{y(y+4)} \sqrt{x(4-x)}}
G(\mu,0;x) G( - x - \mu;y) \nn\\
& &
- \frac{y+4}{\sqrt{y(y+4)}} \biggl[
  \frac{2 x^2 (y-2)}{3 y^2} \bigl( G( - x,0, - \mu;y)
- 6 G( - \mu, - \mu, - \mu;y) \bigr) \nn\\
& &
- \frac{1}{135 y^2} \bigl( 1736 x^2
   - 1020 x y
   - 82 x^2 y - 1020 y^2  + 60 x y^2 - 9 x^2 y^2 + 60 y^3 -
     180 x^2 \zeta(2)  \nn\\
& & + 90 x^2 y \zeta(2) \bigr) G( - \mu;y) + \frac{2}{3 y^2}
\bigl( 34 x^2 - 6 x y - 2 x^2 y - 6 y^2
   + 3 x y^2 + 3 y^3 \bigr) G(-4, - \mu;y) \nn\\
& &
- \frac{2 x^2 (y-2)}{3 y^2} \bigl(
  G(0;x) G( - x, - \mu;y)
- G(0;x) G(0, - \mu;y)
- 3 G(0;x) G(-4, - \mu;y) \nn\\
& & - 9 G(-4,-4, - \mu;y) + G(0,0, - \mu;y)
- G(0,0;x) G( - \mu;y) \bigr) \nn\\
& & + \frac{4 x^2 (5 y\! -\! 4)}{9 y^2} G(0, - \mu;y)\! + \!
\frac{2 (14 x^2 \! - \! 2 x y \! - \! 4 x^2 y \! - \! 2 y^2 \! +
\! x y^2 \! + \! y^3)}{3 y^2} G(0;x) G( - \mu;y)
\biggr] \nn\\
& & - \frac{2 (x-4) x (12 x - 4 y - 5 x y)}{9 y^2\sqrt{x(4-x)}}
G(\mu,0;x)
- \frac{2}{405 y^2} (4488 x^2 - 2340 x y - 1082 x^2 y  \nn\\
& &
   - 3060 y^2 + 195 x y^2 + 195 y^3 - 540 x^2 \zeta(2) +
     225 x^2 y \zeta(2))
+ \frac{1}{9 y^2}(144 x^2 - 11 x^2 y  \nn\\
& &
    + 12 x y^2 + 12 y^3) G( - \mu, - \mu;y)
+  \frac{4 x^2}{3y} ( 3 G( - \mu,-4, - \mu;y)
                    + G( - \mu,0, - \mu;y)  \nn\\
& &
                    -  G(0, - \mu, - \mu;y) )
+ \frac{2 (180 x^2 - 12 x y - 41 x^2 y - 36 y^2 + 15 x y^2
        + 15 y^3)}{27 y^2} G(0;x)  \nn\\
& & +  \frac{x^2}{15}  G(0;y) +  \frac{2 x^2 (5 y-12)}{9 y^2}
G(0,0;x) \, .
\label{B32l0}
\eea

\section{Expansions of the Cross Section}

The non-logarithmic part of the second order correction in
Eq.~(\ref{2lexp})  can be written as follows
\be
\delta^{(2)}_{0} = - 2 \left[ 1 + \ln\left({1-\xi\over \xi}\right)
\right] \ln\left( \frac{4 \omega^2}{s}\right) + Q_f^2 N_c \left({\xi\over
1-\xi+\xi^2}\right)^2f(\rho,\xi) \, ,
\label{del20}
\ee
where the first term is determined by the soft emission and
$f(\rho,\xi)$ is a function of two dimensionless variables:
$\rho=m_f^2/s$ and $\xi=-t/s$. The small-mass expansion of the
function $f(\rho,\xi)$  ($\rho = m_f^2/s$, $\xi = -t/s$) is of the
following form
\begin{equation}
f(\rho,\xi)=\sum_{n=0}^\infty\rho^nf_n(\rho,\xi)\, , \label{smexp}
\end{equation}
where $f_n(\rho,\xi)$ depend on $\rho$ only logarithmically. For
the leading term we obtain
\begin{eqnarray}
& & \hspace{-5mm} f_0(\rho,\xi) =
\frac{\left(\xi^2-\xi+1\right)^2}{\xi^2}\Biggl\{ \frac{1}{9}
\ln^3\left(\rho\right) + \ln^2\left(\rho\right)\Biggl[\frac{1}{3}
\ln(1 - \xi) + \frac{19 }{18}
-\frac{1}{3} \ln(\xi) \Biggr]\nn\\
& & \hspace{10mm} + \ln\left(\rho \right) \Biggl[\frac{191 }{27}
+\frac{8 }{3}  \Li_{2}(\xi)  \Biggr] + \frac{40}{9}\Li_{2}(\xi) +
\frac{1165}{81} \Biggr\} -\ln\left(\rho\right) \Biggl[
\nn\\
& & \hspace{10mm} + \frac{32 \xi^4-46 \xi^3+33 \xi^2+8 \xi-4}{6
\xi^2} \zeta(2) -\frac{\left(\xi^2-\xi+1\right) \left(4 \xi^2-7
\xi+4\right)}{6 \xi^2} \ln(1 - \xi)^2
\nn\\
& & \hspace{10mm} -\frac{20 \xi^4 \! - \! 31 \xi^3 \! + \! 60
\xi^2 \! - \! 31 \xi \! + \! 20}{18 \xi^2}\ln(1 - \xi)
+ \frac{20 \xi^4 \! - \! 67 \xi^3 \! + \! 141 \xi^2 \! - \! 112 \xi \! + \! 74}{18 \xi^2} \ln(\xi)  \nn\\
& & \hspace{10mm} + \frac{8 \xi^4-\xi^3-15 \xi^2+17 \xi-4}{12
\xi^2} \ln(\xi)^2 - \frac{(2 \xi-1) \left(4 \xi^3-3
\xi^2+4\right)}{6 \xi^2} \ln(\xi) \ln(1 - \xi)
\Biggr]  \nn\\
& & \hspace{10mm} + \frac{(2 \xi-1) \left(\xi^2-\xi+1\right)}{3
\xi} \zeta(3)
- \frac{(\xi-1)^2 \left(\xi^2-\xi+1\right)}{9 \xi^2}\ln^3(1 - \xi)  \nn \\
& & \hspace{10mm} -  \frac{196 \xi^4 \! - \! 311 \xi^3 \! + \! 258
\xi^2 \! + \! 13 \xi \! - \! 38}{18 \xi^2} \zeta(2) - \frac{2
\left(2 \xi^4 \! -\! 9 \xi^3\! +\! 16 \xi^2\! -\! 11 \xi\! +\!
4\right)}{3 \xi^2} \ln(1 \! - \! \xi) \zeta(2)  \nn\\
& & \hspace{10mm} + \frac{12 \xi^4 \! - \! 20 \xi^3 \! - \! \xi^2
\! + \! 24 \xi \! - \! 4}{6 \xi^2} \ln(\xi) \zeta(2) + \frac{2
(1-\xi^2) \left(\xi^2-\xi+1\right)}{3 \xi^2} \ln(1 - \xi)
\Li_{2}(\xi)  \nn\\
& & \hspace{10mm} + \frac{7 \left(16 \xi^4 \! - \! 23 \xi^3 \! +
\! 48 \xi^2 \! - \! 23 \xi \! + \! 16\right)}{54 \xi^2} \ln(1  \!
- \! \xi) \! +  \! \frac{20 \xi^4 \! - \! 58 \xi^3 \! + \! 81
\xi^2 \! - \! 58 \xi \! + \! 20}{18 \xi^2}
\ln^2(1  \! -  \! \xi)  \nn\\
& & \hspace{10mm} - \frac{4 \xi^3 \! - \! 6 \xi^2 \! + \! 7 \xi \!
- \! 4}{12 \xi}\ln(\xi) \ln^2(1  \! -  \! \xi)
+ \frac{40 \xi^4 \! - \! 50 \xi^3 \! + \! 9 \xi^2 \! + \! 37 \xi \! - \! 20}{18 \xi^2}\ln(\xi) \ln(1  \! -  \! \xi)   \nn\\
& & \hspace{10mm} - \frac{\xi^4-3 \xi^3+4 \xi^2-\xi+1}{3 \xi^2}
\ln^2(\xi) \ln(1 - \xi)
 +\frac{4 \xi^4-2 \xi^3-22 \xi^2+31 \xi-4}{36 \xi^2} \ln^3(\xi)    \nn\\
& & \hspace{10mm} - \frac{20 \xi^4+8 \xi^3-84 \xi^2+92 \xi-55}{18
\xi^2}\! \ln^2(\xi)
 -\frac{\left(\xi^2-\xi+1\right) \!\left(2 \xi^2-7 \xi+12\right)}{3 \xi^2}\!
 \ln(\xi) \Li_{2}(\xi)\nn\\
& & \hspace{10mm}
- \frac{112 \xi^4-449 \xi^3+1011 \xi^2-836 \xi+562}{54 \xi^2} \ln(\xi)   \nn\\
& & \hspace{10mm} + \frac{2 (1-\xi^2) \left(\xi^2-\xi+1\right)}{3
\xi^2}\Li_{3}(1 - \xi) +\frac{\left(\xi^2-\xi+1\right) \left(2
\xi^2-3 \xi+4\right)}{3 \xi^2} \Li_{3}(\xi) \nn\\
& & \hspace{10mm} -\left(Q_f^2-1\right)\frac{(1-\xi+\xi^2)}{\xi^2}\Biggl[
(1\!-\!\xi\!+\!\xi^2)\left(\frac{5}{12} \!-\!2 \zeta(3) \!+\! \frac{1}{2}\ln\left(\rho\right) \right) - \frac{2-\xi}{4}
\ln\left(\xi\right)\Biggr] \, , \label{f0}
\end{eqnarray}
in agreement with the result of  Refs.~\cite{BecMel,Act}. We
observe that the functions $f_i$ depend on the charge of heavy
fermion $Q_f$ through the contribution of the two-loop irreducible
self-energy diagrams, which are proportional to $Q_f^4$, while all
the other graphs that we consider in the present work are
proportional to $Q_f^2$. The next-to-leading term is new and reads
\begin{eqnarray}
f_1(\rho,\xi) & = & \frac{2 (\xi \! -\!1) \left(\xi^2 \! - \! \xi \! + \! 1\right) \left(2
\xi^2 \! + \! \xi \! + \! 2\right)}{\xi^3} \bigl[
\ln^2\left(\rho\right)  \! + \! 4 \Li_{2}(\xi)   \! +  \! 12 \bigr]
+ \frac{\ln\left(\rho\right)}{\xi^3} \Bigl\{
      (\xi \! - \! 1) (2 \xi^4 \! - \! 5\xi^3 \nn \\
& &
+ 5 \xi^2-5 \xi+2 )
+  2 (\xi^2-\xi+1)
\bigl[ \left( 4 - 2 \xi + \xi^2 - 2 \xi^3 \right) \ln(\xi)
+  ( \xi-1 )  ( 2 + \xi  \nn \\
& &
+ 2 \xi^2 ) \ln(1 - \xi)  \bigr] \Bigr\} \!
- \! \frac{\zeta(2)}{\xi^3} ( 40 \xi^5 \! - \! 54 \xi^4 \! + \! 50 \xi^3 \! - \! 17 \xi^2 \!
- \! 12 \xi \! + \! 8 )
+ \frac{(\xi \! - \! 1)}{2\xi^3} \Bigl[ 2(12 \xi^4 \! \nn\\
& &
- \! 5 \xi^3 \! + \! 13 \xi^2 \! - \! 5 \xi \! + \! 12)
+ (8 - 6\xi + 9 \xi^2 - 6 \xi^3 + 8 \xi^4)\ln (1 - \xi) \Bigr] \ln (1 - \xi)\nn \\
& &
- \frac{1}{2\xi^3} \Bigl[ 2(12 \xi^5-21 \xi^4+26 \xi^3-26 \xi^2+21 \xi-14)
- 2(4 - 6 \xi - \xi^2 + 8 \xi^3 - 10 \xi^4  \nn \\
& & + 8 \xi^5) \ln(1 - \xi)
+ (8 - 15 \xi + 12 \xi^2 - \xi^3 - 7 \xi^4 + 8 \xi^5) \ln(\xi) \Bigr] \ln(\xi)\nn \\
& &
+\left(Q_f^2-1\right) \frac{3}{\xi^3} \Biggl[ (2 \!-\! 3\xi\! +\! 4\xi^2\! -\! 4\xi^3 \!+ \!3\xi^4 \!-\!
2\xi^5) \ln\left(\rho \right) \nn \\
& & \! - \!\left(2 - 3 \xi + 3 \xi^2-
\xi^3\right) \ln\left( \xi\right)
\Biggr]\, .
\label{f1}
\end{eqnarray}
The expansion in the large-mass limit  takes the form
\begin{equation}
f(\rho,\xi)=\sum_{n=0}^\infty\rho^{-n}\bar{f}_n(\rho,\xi)\,,
\label{lmexp}
\end{equation}
where the leading $n=0$ term vanishes because of the
renormalization condition and $\bar{f}_n(\rho,x)$ depend on $\rho$
only logarithmically. For the next-to-leading term we obtain
\begin{eqnarray}
\bar{f}_1(\rho,\xi) & = &
\frac{955 \xi^3 \! - \! 3926 \xi^2 \! + \! 3926 \xi \! - \! 955}{1350 \xi}
 - \frac{12 \xi^3 \! - \! 19 \xi^2 \! + \! 14 \xi \! - \! 6}{10 \xi}\zeta(2)
 + \frac{3 \xi^3 \! + \! \xi^2 \! - \! \xi \! - \! 3}{30 \xi}\ln{(1  \! -  \! \xi)}
\nn \\
& &
 + \frac{2 \xi^3-5 \xi^2+5 \xi-2}{20 \xi}   \ln^2 {(1 - \xi)}
 + \frac{5 \xi^3-22 \xi^2+22 \xi-5}{30 \xi}  \ln\left(\rho\right)
\nn \\
& &
 - \frac{20 \xi^3 \! - \! 78 \xi^2 \! + \! 93 \xi \! - \! 58}{90 \xi}  \ln{(\xi)}
+  \ln{(1 - \xi)}   \frac{12 \xi^3-19 \xi^2+14 \xi-6}{30
\xi} \ln{\xi}
\nn \\
& &
 + \frac{1}{60} \left(-6 \xi^2+\xi+4\right) \ln^2 {(\xi)}
+   \frac{4 \left(\xi^3-2 \xi^2+2 \xi-1\right)}{5 \xi} \mbox{Li}_2(\xi)
\nn \\
& &   -\left(Q_f^2-1\right) \frac{41(\xi^3 -2 \xi^2 + 2 \xi -1)}{54
\xi } \, . \label{f1bar}
\end{eqnarray}
Finally, the next-to-next-to-leading order in the $s \ll m_f^2$
expansion reads
\bea \bar{f}_2(\rho,\xi)&=&
-\frac{177763 \xi^4-405359 \xi^3+676194 \xi^2-405359
   \xi+177763}{2116800 \xi}\nn \\
& &
+  \frac{3 \left(4 \xi^4-17 \xi^3+16 \xi^2-12 \xi+2\right)}{280 \xi} \zeta(2)
+ \frac{2 \xi^4+15 \xi^3+6 \xi^2+15 \xi+2}{840 \xi} \ln(1-\xi)\nn \\
& &
-\frac{\left(\xi^2 \! - \! 4 \xi \! + \! 1\right) \left(2 \xi^2-3\xi+2\right)}{560 \xi}\ln^2(1-\xi)
- \frac{53 \xi^4 \! - \! 141 \xi^3+222 \xi^2-141 \xi \! + \! 53}{1680 \xi}\ln(\rho)\nn \\
& &
+ \frac{33 \xi^4 \! - \! 139 \xi^3 \! + \! 213 \xi^2 \! - \! 212 \xi \! + \! 52}{1680 \xi}
\ln(\xi)
-\frac{4 \xi^4 \! - \! 17 \xi^3 \! + \! 16 \xi^2 \! - \! 12 \xi \! + \! 2}{280 \xi}\ln(\xi) \ln(1 - \xi)\nn \\
& &
+ \frac{1}{560} \left(2 \xi^3-4 \xi^2-2 \xi+1\right)\ln(\xi)^2
- \frac{\left(\xi^2-4 \xi+1\right) \left(\xi^2-\xi+1\right)}{35 \xi}\Li_2(\xi)
\nn \\
& & +\left(Q_f^2 -1 \right)\frac{449 \left(1 - 5\xi + 6 \xi^2 - 5 \xi^3 +
\xi^4\right)}{10800 \xi}\, .
 \eea
%


\end{document}